\documentclass[11pt]{article}
\usepackage{jheppub}
\pdfoutput=1
\usepackage{amsmath,bbm,array,amsfonts,graphicx,wrapfig,lscape,float,mathtools,multirow,longtable,amsthm}
\usepackage[dvipsnames]{xcolor}
\usepackage{stackrel}
\usepackage[all]{xy}
\usepackage{caption}
\usepackage{subcaption}
\usepackage[bottom]{footmisc}
\usepackage{mathrsfs}
\usepackage{tikz}
\usepackage{bm}
\usepackage{color}
\usepackage{enumerate}
\usetikzlibrary{decorations.pathreplacing}
\usepackage{diagbox}
\numberwithin{equation}{section}
\numberwithin{figure}{section}
\numberwithin{table}{section}
\usepackage{pgfplots}
\pgfplotsset{compat=1.14}
\usepackage{makecell}
\usepackage{lscape}
\usepackage[utf8]{inputenc}
\usepackage{mathtools}
\allowdisplaybreaks

\usepackage[autostyle=true]{csquotes}

\newtheorem{remark}{Remark}

\usepackage[toc,page]{appendix}

\usetikzlibrary{external}
\tikzset{external/system call={pdflatex \tikzexternalcheckshellescape 
		-halt-on-error
		-interaction=batchmode 
		-jobname "\image" "\texsource"
		&& pdftops -eps "\image.pdf"}}
\usepackage{tocloft}

	\title{A Note on Quiver Yangians and $\mathcal{R}$-Matrices}
	
	\author[a,b]{Jiakang Bao}
	
	\affiliation[a]{
		Department of Mathematics, City, University of London, EC1V 0HB, UK}
	\affiliation[b]{
		London Institute for Mathematical Sciences, Royal Institution, London W1S 4BS, UK}
	
	\emailAdd{jiakang.bao@city.ac.uk}
	
	\preprint{
		\begin{flushright}
			
		\end{flushright}
	}

	\abstract{In this note, we study possible $\mathcal{R}$-matrix constructions in the context of quiver Yangians and Yang-Baxter algebras. For generalized conifolds, We also discuss the relations between the quiver Yangians and some other Yangian algebras (and $\mathcal{W}$-algebras) in literature.
	}

\begin{document}
	\maketitle

\section{Introduction}\label{intro}
The study of BPS counting and BPS algebras \cite{Harvey:1996gc} has been very active in the past few decades. In the case of non-compact Calabi-Yau (CY) threefolds, especially when they afford a toric description, various techniques have been developed involving quivers \cite{Nakajima:1994nid,Douglas:1996sw}, brane tilings \cite{Hanany:2005ve,Franco:2005rj,Franco:2005sm,Feng:2005gw} and crystal melting \cite{Okounkov:2003sp,Iqbal:2003ds,Ooguri:2009ijd}.

In \cite{Li:2020rij,Galakhov:2020vyb}, the quiver Yangians were constructed as BPS algebras for type IIA string theory on toric CY threefolds from the crystal melting model. The realization of quiver Yangians should also incorporate the wall crossing phenomena \cite{Nagao:2010kx,Jafferis:2008uf}. The crystal configurations for different chambers have also been studied such as in \cite{Chuang:2009crq,Aganagic:2010qr,Bao:2022oyn}. In particular, the quiver Yangians were extended to the shifted quiver Yangians in \cite{Galakhov:2021xum} which provides a nice framework for the study of wall crossing and closed/open BPS states counting problems.

The quiver Yangians should have intimate relations with cohomological Hall algebras (CoHAs) \cite{Kontsevich:2010px} and certain vertex operator algebras (VOAs) \cite{Prochazka:2017qum,Prochazka:2018tlo,Eberhardt:2019xmf,Rapcak:2019wzw}, as well as other Yangian algebras in literature. It is expected that the positive part of the quiver Yangian is the CoHA for the corresponding CY threefold. On the other hand, the quiver Yangian for $\mathbb{C}^3$, which is essentially the affine Yangian $\mathtt{Y}\left(\widehat{\mathfrak{gl}}_1\right)$, is isomorphic to the universal enveloping algebra of $\mathcal{W}_{1+\infty}\cong\mathfrak{u}(1)\times\mathcal{W}_{\infty}$ \cite{Gaberdiel:2017dbk,Prochazka:2019dvu}. In particular, this should play a crucial role in the study of the higher spin symmetry structure in the tensionless limit of string theory in AdS$_3$ as the dual CFT contains a $\mathcal{W}_{\infty}$ symmetry algebra \cite{Gaberdiel:2014cha,Henneaux:2010xg,Campoleoni:2010zq,Gaberdiel:2010pz}. The supersymmetric extension, namely the $\mathcal{W}^{\mathcal{N}=2}_{1+\infty}$ algebra, was then studied in \cite{Prochazka:2017qum,Gaberdiel:2017hcn,Gaberdiel:2018nbs,Li:2019nna,Li:2019lgd}. This supersymmetric version of the $\mathcal{W}$ algebra contains two commuting $\mathcal{W}_{1+\infty}$ algebras and can be constructed from gluing the two subalgebras with extra fermionic or bosonic generators (see also \cite{Bao:2022oyn}). Hence, its representation theory can be nicely encoded by the twin plane partitions. Later in \cite{Eberhardt:2019xmf,Rapcak:2019wzw}, the matrix extensions known as the $\mathcal{W}_{M|N\times\infty}$ algebras were constructed for generalized conifolds. For all such $\mathcal{W}$ algebras, their truncations are believed to give rise to various VOAs that are associated to gauge theories supported on certain divisors in the CY$_3$. Moreover, $\mathcal{W}_{M|N\times\infty}$ should emerge from the Drinfeld double of the CoHA corresponding to the CY$_3$. Thus, the quiver Yangians for generalized conifolds are expected to be closely related to the $\mathcal{W}_{M|N\times\infty}$ algebras. See also \cite{Rapcak:2021hdh,Yamazaki:2022cdg} for summaries of recent developments on relevant topics.

Given a quiver $Q$ with superpotential $W$ associated to a toric CY, let us denote the sets of nodes and arrows as $Q_0$ and $Q_1$ respectively. Such quiver theory can be used to describe the supersymmetric quantum mechanics on the D-branes, where the BPS states arise from the D$p$-branes wrapping holomorphic $p$-cycles of the CY$_3$ in the type IIA compactification setting. The crystal melting model can then be thought of as the 3d uplift of the (periodic) quiver, where each atom in the crystal corresponds to a gauge node in the quiver while the bifundamental/adjoint arrows are chemical bonds. Moreover, the atoms associated to different gauge nodes have different ``colours''.

More concretely, we shall choose an initial atom $\mathfrak{o}$ in the periodic quiver. All the other atoms are placed at the nodes in the periodic quiver level by level along the arrows. As the paths connecting two fixed atoms should be equivalent in the crystal, we have the path algebra defined modulo F-term relations, that is, $\mathbb{C}Q/\langle\partial W\rangle$.

The molten crystal configurations which correspond to the BPS states are obtained following the crystal melting rule. An atom $\mathfrak{a}$ is in the molten crystal $\mathfrak{C}$ if there exists an arrow $I\in Q_1$ such that $I\cdot\mathfrak{a}\in\mathfrak{C}$. This equivalently states that the complement of the molten crystal is an ideal of the path algebra. As we will review shortly, the generators of the quiver Yangian have natural actions on the molten crystal configurations.

On the other hand, as the name suggests, the quiver Yangian should enjoy an $\mathcal{R}$-matrix formalism \cite{Drinfeld1985HopfAA,maulik2012quantum}. The $\mathcal{R}$-matrix can be defined by considering a set of vector spaces $\mathcal{F}_i$ and the operator-valued functions $\mathcal{R}_{\mathcal{F}_i,\mathcal{F}_j}(u)\in\text{End}(\mathcal{F}_i\otimes\mathcal{F}_j)(u)$. Here, $u$ is the spectral parameter and the $\mathcal{R}$-matrix should satisfy the Yang-Baxter (YB) equation
\begin{equation}
	\mathcal{R}_{12}(u)\mathcal{R}_{13}(u+v)\mathcal{R}_{23}(v)=\mathcal{R}_{23}(v)\mathcal{R}_{13}(u+v)\mathcal{R}_{12}(u),
\end{equation}
where $\mathcal{R}_{12}:=\mathcal{R}_{\mathcal{F}_1,\mathcal{F}_2}\otimes1_{\mathcal{F}_3}$. Henceforth, we shall slightly abuse the notation and simply write $\mathcal{R}_{\mathcal{F}_i,\mathcal{F}_j}$ as $\mathcal{R}_{ij}$. Now, consider the tensor product of the Fock spaces, $\mathcal{F}_1(u_1)\otimes\dots\otimes\mathcal{F}_n(u_n)$, and choose an auxiliary space $\mathcal{F}_0\in\{\mathcal{F}_i\}$. We can define the operator
\begin{equation}
	\mathcal{T}_0(u)=\mathcal{R}_{0n}(u-u_n)\dots\mathcal{R}_{01}(u-u_1).
\end{equation}
The YB equation then implies the $\mathcal{RTT}$ relation
\begin{equation}
	\mathcal{R}_{ij}(u-v)\mathcal{T}_i(u)\mathcal{T}_j(v)=\mathcal{T}_j(v)\mathcal{T}_i(u)\mathcal{R}_{ij}(u-v).
\end{equation}

More rigorously, following \cite{maulik2012quantum}, we should start with an integral domain $\mathbb{K}\supset\mathbb{Q}$ with $\otimes=\otimes_{\mathbb{K}}$ and $\textup{End}=\textup{End}_{\mathbb{K}}$. Then the Maulik-Okounkov (MO) Yangian acts on $F_i(u_i):=F_i\otimes\mathbb{K}[u_i]$ for some free $\mathbb{K}$-module $F_i$, or more generally on the tensor product $\bigotimes\limits_iF_i(u_i)=\bigotimes\limits_iF_i\otimes\mathbb{K}[u_1,\dots,u_n]$. Given a quiver $Q$, the modules $F_i$ can be identified with certain equivariant cohomologies of the Nakajima quiver variety.
	
The precise relation between quiver Yangians and MO Yangians is still not clear, but they should be different for the same quiver $Q$. For $\mathbb{C}\times\mathbb{C}^2/\mathbb{Z}_n$ whose quiver Yangian is $\mathtt{Y}\left(\widehat{\mathfrak{gl}}_n\right)$, as it is the tripled quiver\footnote{Given a quiver $Q$, its tripled quiver is defined as follows. We first construct its doubled quiver $\overline{Q}=(Q_0,Q_1\sqcup Q_1^*)$ where an arrow $I^*$ in the opposite direction is added for each $I\in Q_1$. Then the tripled quiver $\widehat{Q}$ is obtained by adding a self-loop $\omega_a$ to each node $a$. It has (super)potential $W=\sum\omega_a[X,X^*]$.} $\widehat{Q}$ of the affine A-type quiver $Q$, we conjecture that its quiver Yangian $\mathtt{Y}_{\widehat{Q}}$ is isomorphic to the MO Yangian of $Q$. This is consistent with the conjecture in \cite{Davison:2013nza} regarding their positive parts.

For the $\mathbb{C}^3$ case, the construction of MO $\mathcal{R}$-matrix and its connection to certain Yangian algebras have been well-studied in various literature such as \cite{maulik2012quantum,Prochazka:2019dvu,Litvinov:2020zeq}. In this note, we shall make an attempt to generalize this story although the discussions here would be very basic and there are still many problems to study for future works.

The paper is organized as follows. In \S\ref{QY}, we review some basic concepts and properties for quiver Yangians that would be important for further discussions. In \S\ref{YBRmat}, we will introduce the YB algebras for arbitrary quivers (mainly symmetric) and consider their $\mathcal{R}$-matrices. In \S\ref{coproduct}, we will mostly focus on generalized conifolds and discuss the relations of the quiver Yangians with some other Yangian algebras in literature. We will also show that certain quiver Yangians are actually generated by finitely many generators, which might shed light on the discussions on $\mathcal{R}$-matrices in \S\ref{YBRmat}. In \S\ref{Wmninfty}, we will have a brief study on $\mathcal{W}_{M|N\times\infty}$ and contemplate the intertwiners from Miura transformations. Nevertheless, the general connection/map between (the generators of) the quiver Yangians/YB algebras and the $\mathcal{W}$ algebras would still require further study. In \S\ref{outlook}, we will mention some future directions. As a special family of toric CY threefolds, in Appendix \ref{quivergencon}, we recall the construction of quivers for generalized conifolds. We give more examples of the computations of $\mathcal{R}$-matrices in Appendix \ref{exhigherlv}.

\section{Quiver Yangians}\label{QY}
Let us first briefly review the concept of quiver Yangians as introduced in \cite{Li:2020rij}.  Given a quiver $Q=(Q_0,Q_1)$ with superpotential $W$, its quiver Yangian $\mathtt{Y}_{Q,W}$ is generated by the modes $e^{(a)}_i$, $f^{(a)}_i$ and $\psi^{(a)}_j$ ($a\in Q_0$, $i\in\mathbb{N}$, $j\in\mathbb{Z}$)\footnote{In this paper, we have the convention $\mathbb{N}=\mathbb{Z}_{\geq0}$.} satisfying the relations
\begin{align}
	&\left[\psi^{(a)}_n,\psi^{(b)}_m\right]=0,\\
	&\left[e^{(a)}_n,f^{(b)}_m\right\}=\delta_{ab}\psi^{(a)}_{m+n},\\
	&\sum_{k=0}^{|b\rightarrow a|}(-1)^{|b\rightarrow a|-k}\sigma_{|b\rightarrow a|-k}^{b\rightarrow a}\left[\psi^{(a)}_ne^{(b)}_m\right]_k=\sum_{k=0}^{|a\rightarrow b|}\sigma_{|a\rightarrow b|-k}^{a\rightarrow b}\left[e^{(b)}_m\psi^{(a)}_n\right]^k,\\
	&\sum_{k=0}^{|b\rightarrow a|}(-1)^{|b\rightarrow a|-k}\sigma_{|b\rightarrow a|-k}^{b\rightarrow a}\left[e^{(a)}_ne^{(b)}_m\right]_k=(-1)^{|(a)||(b)|}\sum_{k=0}^{|a\rightarrow b|}\sigma_{|a\rightarrow b|-k}^{a\rightarrow b}\left[e^{(b)}_me^{(a)}_n\right]^k,\\
	&\sum_{k=0}^{|b\rightarrow a|}(-1)^{|b\rightarrow a|-k}\sigma_{|b\rightarrow a|-k}^{b\rightarrow a}\left[f^{(b)}_m\psi^{(a)}_n\right]^k=\sum_{k=0}^{|a\rightarrow b|}\sigma_{|a\rightarrow b|-k}^{a\rightarrow b}\left[\psi^{(a)}_nf^{(b)}_m\right]_k,\\
	&\sum_{k=0}^{|b\rightarrow a|}(-1)^{|b\rightarrow a|-k}\sigma_{|b\rightarrow a|-k}^{b\rightarrow a}\left[f^{(b)}_mf^{(a)}_n\right]^k=(-1)^{|(a)||(b)|}\sum_{k=0}^{|a\rightarrow b|}\sigma_{|a\rightarrow b|-k}^{a\rightarrow b}\left[f^{(a)}_nf^{(b)}_m\right]_k.
\end{align}
The notations require some explanation. The bracket $[\text{-},\text{-}\}$ is the super bracket, that is, anti-commutator for two fermionic modes and commutator otherwise. In a quiver, the nodes with (without) adjoint loops are bosonic (fermionic) such that $|(a)|=0$ ($|(a)|=1$). Then $e^{(a)}_i$ and $f^{(a)}_i$ have the $\mathbb{Z}_2$-grading same as the corresponding node $a$ while $\psi^{(a)}_j$ is always bosonic. We use $a\rightarrow b$ to denote the set of arrows from $a$ to $b$, and the total number is $|a\rightarrow b|$. For each edge $I\in Q_1$, we assign a weight/charge $\widetilde{\epsilon}_I$ to it, and $\sigma^{a\rightarrow b}_k$ is the $k^\text{th}$ symmetric sum of $\widetilde{\epsilon}_I$ for all $I\in a\rightarrow b$. Moreover, we have
\begin{equation}
	[A_nB_m]_k:=\sum_{l=0}^k(-1)^l\binom{k}{l}A_{n+k-l}B_{m+l},\quad[B_mA_n]^k:=\sum_{l=0}^k(-1)^l\binom{k}{l}B_{m+l}A_{n+k-l}.
\end{equation}
For toric CYs, as the superpotential can be unambiguously determined for a given quiver, we shall sometimes abbreviate $\mathtt{Y}_{Q,W}$ as $\mathtt{Y}_Q$ or even $\mathtt{Y}$ if it would not cause confusions.

To correctly recover the counting of crystal configurations/BPS states, we need to further mod out the Serre relations. A general expression of the Serre relations for any quiver Yangian is not known. For generalized conifolds, the Serre relations read
\begin{equation}
	\text{Sym}_{n_1,n_2}\left[e^{(a)}_{n_1},\left[e^{(a)}_{n_2},e^{(a\pm1)}_m\right]\right]=0,\quad\text{Sym}_{n_1,n_2}\left[f^{(a)}_{n_1},\left[f^{(a)}_{n_2},f^{(a\pm1)}_m\right]\right]=0,
\end{equation}
for $|(a)|=0$, and
\begin{equation}
	\text{Sym}_{n_1,n_2}\left[e^{(a)}_{n_1},\left[e^{(a+1)}_{m_1},\left[e^{(a)}_{n_2},e^{(a-1)}_{m_2}\right\}\right\}\right\}=0,\quad\text{Sym}_{n_1,n_2}\left[f^{(a)}_{n_1},\left[f^{(a+1)}_{m_1},\left[f^{(a)}_{n_2},f^{(a-1)}_{m_2}\right\}\right\}\right\}=0\label{Serre2gencon}
\end{equation}
for $|(a)|=1$. The Yangian algebra after the quotient of the Serre relations is also called the reduced quiver Yangian. However, in this note, as we will mainly focus on the Yangian algebra with Serre relations included, we shall simply refer to it as the quiver Yangian $\mathtt{Y}$. Thus, the quiver Yangian for the generalized conifold defined by $xy=z^Mw^N$ is essentially the affine Yangian $\mathtt{Y}\left(\widehat{\mathfrak{gl}}_{M|N}\right)$.

We can then introduce the currents
\begin{equation}
	e^{(a)}(u):=\sum_{n=0}^{\infty}\frac{e^{(a)}_n}{u^{n+1}},\quad f^{(a)}(u):=\sum_{n=0}^{\infty}\frac{f^{(a)}_n}{u^{n+1}},\quad\psi^{(a)}(u):=\sum_{n\in\mathbb{Z}}\frac{\psi^{(a)}_n}{u^{n+1}}.
\end{equation}
In the molten crystal, $e^{(a)}(u)$ ($e^{(a)}_n$) creates atoms in the configuration while $f^{(a)}(u)$ ($f^{(a)}_n$) annihilates atoms. Moreover, $\psi^{(a)}(u)$ contains all the Cartan modes $\psi^{(a)}_n$. It was shown in \cite{Li:2020rij} that for toric CYs without compact divisors (or more generally, any symmetric quivers), $\psi^{(a)}_{n<-1}=0$ and $\psi^{(a)}_{-1}=1$.

We may then write the relations in terms of the currents as
\begin{align}
	&\left[e^{(a)}(u),f^{(b)}(v)\right\}=\delta_{ab}\frac{\psi^{(a)}(u)-\psi^{(a)}(v)}{u-v}+\dots,\\
	&\overline{g}_{ba}(u-v)\psi^{(a)}(u)e^{(b)}(v)=g_{ab}(u-v)e^{(b)}(v)\psi^{(a)}(u)+\dots,\\
	&\overline{g}_{ba}(u-v)e^{(a)}(u)e^{(b)}(v)=(-1)^{|(a)||(b)|}g_{ab}(u-v)e^{(b)}(v)e^{(a)}(u)+\dots,\\
	&\overline{g}_{ba}(u-v)f^{(b)}(v)\psi^{(a)}(u)=g_{ab}(u-v)\psi^{(a)}(u)f^{(b)}(v)+\dots,\\
	&\overline{g}_{ba}(u-v)f^{(b)}(v)f^{(a)}(u)=(-1)^{|(a)||(b)|}g_{ab}(u-v)f^{(a)}(u)f^{(b)}(v)+\dots,\\
\end{align}
where
\begin{equation}
	g_{ab}(z):=\prod_{i=1}^{|a\rightarrow b|}\left(z+\widetilde{\epsilon}_{ab,i}\right),\quad\overline{g}_{ba}(z):=\prod_{i=1}^{|b\rightarrow a|}\left(z-\widetilde{\epsilon}_{ba,i}\right).
\end{equation}
The ellipses indicate the local terms in the sense of \cite{Litvinov:2020zeq} as they would not contribute when we compute the contour integrals to recover most of the mode relations\footnote{More specifically, when applying the contour integral $\frac{1}{(2\pi i)^2}\oint u^nv^m\text{d}u\text{d}v$ with $m,n\geq0$ (or taking the formal mode expansion), these terms do not contribute as they have zero residues. However, they would affect the results for relations such as $\left[\psi^{(a)}(u),e^{(b)}_0\right]$ which has $m=-1$. See for instance \cite{Litvinov:2020zeq} for some explicit examples.}. For instance, when the toric CY does not have compact 4-cycles, we have the local terms for the $\psi e$ relation as
\begin{equation}
	\begin{split}
		&\left(\sum_{k=0}^{|b\rightarrow a|}(-1)^{|b\rightarrow a|}\sigma^{b\rightarrow a}_{|b\rightarrow a|-k}\sum_{j=0}^k(-1)^j\binom{k}{j}u^{k-j}v^{j}\left(\psi^{(a)}(u)\left(\sum_{m=0}^{j-1}\frac{e^{(b)}_m}{v^{m+1}}\right)+\left(\sum_{n=-1}^{k-j-1}\frac{\psi^{(a)}_n}{u^{n+1}}\right)e^{(b)}(v)\right.\right.\\
		&\left.\left.-\sum_{m=0}^{j-1}\sum_{n=-1}^{k-j-1}\frac{\psi^{(a)}_n}{u^{n+1}}\frac{e^{(b)}_m}{v^{m+1}}\right)\right)-\left(\sum_{k=0}^{|a\rightarrow b|}\sigma^{a\rightarrow b}_{|a\rightarrow b|-k}\sum_{j=0}^k(-1)^j\binom{k}{j}u^{k-j}v^{j}\left(\left(e^{(b)}(v)\sum_{n=-1}^{k-j-1}\frac{\psi^{(a)}_n}{u^{n+1}}\right)\right.\right.\\
		&\left.\left.+\left(\sum_{m=0}^{j-1}\frac{e^{(b)}_m}{v^{m+1}}\right)\psi^{(a)}(u)-\sum_{m=0}^{j-1}\sum_{n=-1}^{k-j-1}\frac{e^{(b)}_m}{v^{m+1}}\frac{\psi^{(a)}_n}{u^{n+1}}\right)\right).
	\end{split}
\end{equation}

By analyzing how the atoms in the molten crystal configuration can be added and removed, we can write down the action of the currents on any crystal state $|\mathfrak{C}\rangle$. Consider an atom $\mathfrak{a}$ of colour $a$ that can be added to (removed from) the molten crystal according to the melting rule. Then we shall use the notation $\mathfrak{a}\in\mathfrak{C}_+$ ($\mathfrak{a}\in\mathfrak{C}_-$) such that $|\mathfrak{C}\rangle$ would become $|\mathfrak{C}+\mathfrak{a}\rangle$ ($|\mathfrak{C}-\mathfrak{a}\rangle$) after the corresponding action. Suppose the initial atom $\mathfrak{o}$ in the crystal has colour $o=1$. We have \cite{Li:2020rij}
\begin{align}
	&\psi^{(a)}(u)|\mathfrak{C}\rangle=\Psi^{(a)}_{\mathfrak{C}}(u)|\mathfrak{C}\rangle,\\
	&e^{(a)}(u)|\mathfrak{C}\rangle=\sum_{\mathfrak{a}\in\mathfrak{C}_+}\frac{\pm\sqrt{-(-1)^{|(a)|}\text{Res}_{\widetilde{\epsilon}(\mathfrak{a})}\Psi^{(a)}_{\mathfrak{C}}(u)}}{u-\widetilde{\epsilon}(\mathfrak{a})}|\mathfrak{C}+\mathfrak{a}\rangle,\\
	&f^{(a)}(u)|\mathfrak{C}\rangle=\sum_{\mathfrak{a}\in\mathfrak{C}_-}\frac{\pm\sqrt{\text{Res}_{\widetilde{\epsilon}(\mathfrak{a})}\Psi^{(a)}_{\mathfrak{C}}(u)}}{u-\widetilde{\epsilon}(\mathfrak{a})}|\mathfrak{C}-\mathfrak{a}\rangle,
\end{align}
where
\begin{align}
	&\Psi^{(a)}_{\mathfrak{C}}(u):=\left(\frac{u+C}{u}\right)^{\delta_{a,1}}\prod_{b\in Q_0}\prod_{\mathfrak{b}\in\mathfrak{C}}\phi^{b\Rightarrow a}(u-\widetilde{\epsilon}(\mathfrak{b})),\\
	&\phi^{b\Rightarrow a}(u)=\frac{\prod\limits_{I\in a\rightarrow b}(u+\widetilde{\epsilon}_I)}{\prod\limits_{I\in b\rightarrow a}(u-\widetilde{\epsilon}_I)},\label{bondfactor}\\
	&\widetilde{\epsilon}(\mathfrak{a})=\sum_{I\in\text{path}[\mathfrak{o}\rightarrow\mathfrak{a}]}\widetilde{\epsilon}_I.
\end{align}
Here, $C$ is some numerical constant known as the vacuum charge\footnote{For toric CY without compact 4-cycles, it can be identified as the central term $\sum\limits_{a\in Q_0}\psi^{(a)}_0$.}. The $\pm$ signs in the actions depend on the statistics of the algebra. Moreover, the charge assignment $\widetilde{\epsilon}_I$ should be compatible with the superpotential\footnote{This means that $\widetilde{\epsilon}_I$ can be viewed as charges under a global symmetry of the quiver quantum mechanics, and this charge constraint is the only role that the superpotential plays in the definition of $\mathtt{Y}$.}. Therefore, the coordinate parameters $\widetilde{\epsilon}_I$ of the arrows should satify the loop constraint
\begin{equation}
	\sum_{I\in L}\widetilde{\epsilon}_I=0,
\end{equation}
for any closed loop $L$ in the periodic quiver. It turns out that the number of coordinate parameters is given by
\begin{equation}
	|Q_1|-|Q_2|-1=|Q_0|+1,
\end{equation}
where $Q_2$ denotes the faces of the periodic quiver, or equivalently, the monomial terms in the superpotential.

There are quite a few properties for the quiver Yangians discussed in \cite{Li:2020rij}. Here, we shall only mention one feature that would be important in the following discussions. As pointed out in \cite{Li:2020rij}, there is a mixing of global and gauge symmetries associated to each node, and this would cause shifts of $\widetilde{\epsilon}_I$. One can then introduce a gauge fixing condition to get rid of this shift. This is known as the vertex constraint:
\begin{equation}
	\sum_{I\in a}\text{sgn}_a(I)\widetilde{\epsilon}_I=0,
\end{equation}
where the sign function $\text{sgn}_a(I)$ is equal to $+1$ ($-1$) when the arrow $I$ starts from (ends at) the node $a$, and $0$ otherwise. As an overall U(1) symmetry decouples, the total number of the vertex constraints is $|Q_0|-1$. Together with the $|Q_0|+1$ loop constraints, we are then left with two independent parameters\footnote{These two coordinate parameters, along with the R-symmetry, give the U(1)$^3$ isometry of the toric CY threefold.} denoted as $\epsilon_{1,2}$. It would also be convenient to introduce a third parameter $\epsilon_3$ such that $\epsilon_1+\epsilon_2+\epsilon_3=0$.

\section{Yang-Baxter Algebras and $\mathcal{R}$-Matrices}\label{YBRmat}
In \cite{Litvinov:2020zeq,Chistyakova:2021yyd}, the MO $\mathcal{R}$-matrices were constructed using the $\mathcal{RTT}$ relation and some current algebras known as the Yang-Baxter algebras for $\widehat{\mathfrak{gl}}_1$ and $\widehat{\mathfrak{gl}}_2$. In this section, we shall first define the YB algebras for general quivers.

\subsection{Yang-Baxter Algebras}\label{YBalg}
Given a quiver $Q$, the YB algebra $\mathtt{YB}_Q$ is defined by the generators $h^{(a)}_i$, $e^{(a)}_i$, $f^{(a)}_i$ and $\psi^{(a)}_j$ ($a\in Q_0$, $i\in\mathbb{N}$, $j\in\mathbb{Z}$) subject to the relations
\begin{align}
	&\left[h^{(a)}_n,h^{(b)}_m\right]=\left[h^{(a)}_n,\psi^{(b)}_m\right]=0,\\
	&\left[h^{(a)}_n,e^{(b)}_m\right]=\delta_{ab}\epsilon_3\sum_{k=0}^nh^{(a)}_{n-k-1}e^{(b)}_{m+k},\\
	&\left[f^{(b)}_m,h^{(a)}_n\right]=\delta_{ab}\epsilon_3\sum_{k=0}^nf^{(b)}_{m+k}h^{(a)}_{n-k-1},\\
	&\left[\psi^{(a)}_n,\psi^{(b)}_m\right]=0,\\
	&\left[e^{(a)}_n,f^{(b)}_m\right\}=-\delta_{ab}\psi^{(a)}_{m+n},\\
	&\sum_{k=0}^{|b\rightarrow a|}(-1)^{|b\rightarrow a|-k}\sigma_{|b\rightarrow a|-k}^{b\rightarrow a}\left[\psi^{(a)}_ne^{(b)}_m\right]_k=\sum_{k=0}^{|a\rightarrow b|}\sigma_{|a\rightarrow b|-k}^{a\rightarrow b}\left[e^{(b)}_m\psi^{(a)}_n\right]^k,\\
	&\sum_{k=0}^{|b\rightarrow a|}(-1)^{|b\rightarrow a|-k}\sigma_{|b\rightarrow a|-k}^{b\rightarrow a}\left[e^{(a)}_ne^{(b)}_m\right]_k=(-1)^{|(a)||(b)|}\sum_{k=0}^{|a\rightarrow b|}\sigma_{|a\rightarrow b|-k}^{a\rightarrow b}\left[e^{(b)}_me^{(a)}_n\right]^k,\\
	&\sum_{k=0}^{|b\rightarrow a|}(-1)^{|b\rightarrow a|-k}\sigma_{|b\rightarrow a|-k}^{b\rightarrow a}\left[f^{(b)}_m\psi^{(a)}_n\right]^k=\sum_{k=0}^{|a\rightarrow b|}\sigma_{|a\rightarrow b|-k}^{a\rightarrow b}\left[\psi^{(a)}_nf^{(b)}_m\right]_k,\\
	&\sum_{k=0}^{|b\rightarrow a|}(-1)^{|b\rightarrow a|-k}\sigma_{|b\rightarrow a|-k}^{b\rightarrow a}\left[f^{(b)}_mf^{(a)}_n\right]^k=(-1)^{|(a)||(b)|}\sum_{k=0}^{|a\rightarrow b|}\sigma_{|a\rightarrow b|-k}^{a\rightarrow b}\left[f^{(a)}_nf^{(b)}_m\right]_k,\\
	&\text{Serre relations}.
\end{align}
As we can see, the relations among $e^{(a)}_i$, $f^{(a)}_i$ and $\psi^{(a)}_j$ are exactly the same as the ones for their namesakes in the quiver Yangian $\mathtt{Y}_Q$ except the extra minus sign in the $ef$ relation. Moreover, similar to $\psi^{(a)}_j$, the modes $h^{(a)}_i$ are Cartan modes and are always bosonic for any node $a$. Denoting the subalgebra of $\mathtt{YB}$ generated by $e^{(a)}_i$, $f^{(a)}_i$ and $\psi^{(a)}_j$ as $\mathtt{YB}_0$, it is then straightforward to see that given a quiver $Q$, the map
\begin{equation}
	\rho:\mathtt{Y}\rightarrow\mathtt{YB}_0,\quad e^{(a)}_i\mapsto e^{(a)}_i,\quad-f^{(a)}_i\mapsto f^{(a)}_i,\quad\psi^{(a)}_i\mapsto\psi^{(a)}_i
\end{equation}
is an isomorphism\footnote{The case for $\mathbb{C}^3$ was proven in \cite{Wang:2022rvj}.}. In general, the YB algebra is strictly larger than the quiver Yangian. For instance, $\mathtt{Y}\left(\widehat{\mathfrak{gl}}_1\right)$ is the factorization of the YB algebra for $\mathbb{C}^3$ over its centre as shown in \cite{Litvinov:2020zeq}. In the remaining of this section (\S\ref{YBRmat}), we shall always refer to $f$ as the generators for the YB algebra.

We may then write the currents
\begin{equation}
	h^{(a)}(u)=1+\sum_{n=0}^\infty\frac{h^{(a)}_n}{u^{n+1}},\quad e^{(a)}(u)=\sum_{n=0}^\infty\frac{e^{(a)}_n}{u^{n+1}},\quad f^{(a)}(u)=\sum_{n=0}^\infty\frac{f^{(a)}_n}{u^{n+1}},\quad\psi^{(a)}(u)=\sum_{n\in\mathbb{Z}}\frac{\psi^{(a)}_n}{u^{n+1}}.
\end{equation}
In particular, we can define $h^{(a)}_{-1}=1$. In terms of currents, the relations read
\begin{align}
	&\left[h^{(a)}(u),h^{(b)}(v)\right]=\left[h^{(a)}(u),\psi^{(b)}(v)\right]=0,\\
	&(u-v-\delta_{ab}\epsilon_3)h^{(a)}(u)e^{(b)}(v)=(u-v)e^{(b)}(v)h^{(a)}(u)-\delta_{ab}\epsilon_3h^{(a)}(u)e^{(b)}(u),\\
	&(u-v-\delta_{ab}\epsilon_3)f^{(b)}(v)h^{(a)}(u)=(u-v)h^{(a)}(u)f^{(b)}(v)-\delta_{ab}\epsilon_3f^{(b)}(u)h^{(a)}(u),
\end{align}
as well as those for $e^{(a)}(u)$, $f^{(a)}(u)$ and $\psi^{(a)}(u)$ being the same as in quiver Yangians (with minus signs correspondingly added due to different conventions of $f$). Again, the terms involving only the parameter $u$ are called local terms.

\begin{remark}
	Instead of introducing an $h^{(a)}(u)$ for each node $a$, we could also consider a single current $h(u)$ such that $h(u):=\prod\limits_{a\in Q_0}h^{(a)}(u)$ with mode expansion $h(u)=\sum\limits_{n=-1}^{\infty}\frac{h_n}{u^{n+1}}$ (where $h_{-1}=1$). This would slightly alter the definition of $\mathtt{YB}$, but the relations would still be very similar. We can simply remove the factors $\delta_{ab}$ (and of course also the superscripts in $h$) to get both the mode and current relations for $h$.
	
	More generally, especially for $\textup{CY}_3$ with compact $4$-cycles, we may also introduce negative modes for $h^{(a)}(u)$ (or $h(u)$) in the definition of YB algebras just like $\psi^{(a)}(u)$. This might be more convenient when discussing the relations between $\mathtt{YB}$ and $\mathtt{Y}$. However, for our purpose here (especially for symmetric quivers without negative $\psi^{(a)}_j$ modes), it suffices to consider $h^{(a)}(u)$ with modes $n\geq-1$.
\end{remark}

As the quiver Yangians have crystal representations, we may also find how $\mathtt{YB}$, or more specifically $h^{(a)}(u)$, would act on the crystals. This can be done with the help of the actions of other generators. Write $h^{(a)}(u)|\mathfrak{C}\rangle=h^{(a)}_{\mathfrak{C}}|\mathfrak{C}\rangle$ for an arbitrary crystal configuration $\mathfrak{C}$. Using the $he$ relation, we have
\begin{equation}
	(u-v-\delta_{ab}\epsilon_3)h^{(a)}(u)e^{(b)}(v)|\mathfrak{C}\rangle=\left((u-v)e^{(b)}(v)h^{(a)}(u)-\delta_{ab}\epsilon_3h^{(a)}(u)e^{(b)}(u)\right)|\mathfrak{C}\rangle.
\end{equation}
Then
\begin{equation}
	\begin{split}
		&(u-v-\delta_{ab}\epsilon_3)h^{(a)}(u)\sum_{\mathfrak{b}\in\mathfrak{C}_+}\frac{\text{Num}(\mathfrak{b})}{v-\widetilde{\epsilon}(\mathfrak{b})}|\mathfrak{C}+\mathfrak{b}\rangle\\
		=&(u-v)e^{(b)}(v)h^{(a)}_{\mathfrak{C}}|\mathfrak{C}\rangle-\delta_{ab}\epsilon_3h^{(a)}(u)\sum_{\mathfrak{b}\in\mathfrak{C}_+}\frac{\text{Num}(\mathfrak{b})}{u-\widetilde{\epsilon}(\mathfrak{b})}|\mathfrak{C}+\mathfrak{b}\rangle,
	\end{split}
\end{equation}
where the numerator in the action of $e^{(b)}$ is denoted as $\text{Num}(\mathfrak{b})$. The explicit expression can be found in \S\ref{QY}, but it is not important here. This yields
\begin{equation}
	\begin{split}
		&(u-v-\delta_{ab}\epsilon_3)\sum_{\mathfrak{b}\in\mathfrak{C}_+}\frac{\text{Num}(\mathfrak{b})}{v-\widetilde{\epsilon}(\mathfrak{b})}h^{(a)}_{\mathfrak{C}+\mathfrak{b}}|\mathfrak{C}+\mathfrak{b}\rangle\\
		=&(u-v)\sum_{\mathfrak{b}\in\mathfrak{C}_+}\frac{\text{Num}(\mathfrak{b})}{v-\widetilde{\epsilon}(\mathfrak{b})}h^{(a)}_{\mathfrak{C}}|\mathfrak{C}+\mathfrak{b}\rangle-\delta_{ab}\epsilon_3\sum_{\mathfrak{b}\in\mathfrak{C}_+}\frac{\text{Num}(\mathfrak{b})}{u-\widetilde{\epsilon}(\mathfrak{b})}h^{(a)}_{\mathfrak{C}+\mathfrak{b}}|\mathfrak{C}+\mathfrak{b}\rangle.
	\end{split}
\end{equation}
In other words,
\begin{equation}
	(u-v-\delta_{ab}\epsilon_3)\frac{\text{Num}(\mathfrak{b})}{v-\widetilde{\epsilon}(\mathfrak{b})}h^{(a)}_{\mathfrak{C}+\mathfrak{b}}=(u-v)\frac{\text{Num}(\mathfrak{b})}{v-\widetilde{\epsilon}(\mathfrak{b})}h^{(a)}_{\mathfrak{C}}-\delta_{ab}\epsilon_3\frac{\text{Num}(\mathfrak{b})}{u-\widetilde{\epsilon}(\mathfrak{b})}h^{(a)}_{\mathfrak{C}+\mathfrak{b}}.
\end{equation}
By taking the contour integral $\oint_{v=\infty}$ (or equivalently, the large $v$ expansion), we have
\begin{equation}
	\frac{h^{(a)}_{\mathfrak{C}+\mathfrak{b}}}{h^{(a)}_{\mathfrak{C}}}=\frac{u-\widetilde{\epsilon}(\mathfrak{b})}{u-\widetilde{\epsilon}(\mathfrak{b})-\delta_{ab}\epsilon_3}.
\end{equation}
Let us choose the normalization $h^{(a)}(u)|\varnothing\rangle=|\varnothing\rangle$. Then we get
\begin{equation}
	h^{(a)}(u)|\mathfrak{C}\rangle=\prod_{\mathfrak{a}\in\mathfrak{C}}\frac{u-\widetilde{\epsilon}(\mathfrak{a})}{u-\widetilde{\epsilon}(\mathfrak{a})-\epsilon_3}|\mathfrak{C}\rangle
\end{equation}
for any crystal configuration $\mathfrak{C}$. Thus, $h^{(a)}(u)$ only sees the atoms of colour $a$ in the crystal\footnote{Hence, if we consider the action of $h(u)=\prod\limits_{a\in Q_0}h^{(a)}(u)$, then we would get
\begin{equation}
	h(u)|\mathfrak{C}\rangle=\prod_{a\in Q_0}\prod_{\mathfrak{a}\in\mathfrak{C}}\frac{u-\widetilde{\epsilon}(\mathfrak{a})}{u-\widetilde{\epsilon}(\mathfrak{a})-\epsilon_3}|\mathfrak{C}\rangle.\nonumber
\end{equation}}.

By comparing the actions of $h^{(a)}(u)$ and $\psi^{(a)}(u)$ (with vertex constraints taken into account), we can write $\psi^{(a)}(u)$ in terms of $h^{(a)}(u)$. For instance, for generalized conifolds, we have the relation
\begin{equation}
	\begin{split}
		\psi^{(a)}(u)=&\left(\frac{u+\psi_0}{u}\right)^{\delta_{a1}}h^{(a-1)}(u+\sigma_a\epsilon_1)h^{(a-1)}(u+\sigma_a\epsilon_2)\\
		&h^{(a+1)}(u+\sigma_{a+1}\epsilon_1)h^{(a+1)}(u+\sigma_{a+1}\epsilon_2)\left(h^{(a)}(u)h^{(a)}(u+\epsilon_3)\right)^{\frac{\sigma_a+\sigma_{a+1}}{2}},
	\end{split}
\end{equation}
where we have used $C=\sum\limits_{a\in Q_0}\psi^{(a)}_0=:\psi_0$ for the vacuum charge as shown in \cite{Li:2020rij} for generalized conifolds. The detailed description of the quivers and the definition of $\sigma_a$ can be found in Appendix \ref{quivergencon}.

\subsection{Crystal Melting and the $\mathcal{RTT}$ Relation}\label{crystalRTT}
Given a quiver and its quiver Yangian, we shall construct the $\mathcal{R}$-matrices by acting the $\mathcal{RTT}$ relation on the Fock modules of the algebra. For any quiver, we propose that we can consider a particular representation whose states are labelled by molten crystal configurations at depth 0 in the crystal melting model. In other words, such representation is a 2d crystal which is a surface of the 3d crystal constructed from the periodic quiver. Indeed, the Fock representation would arise when one considers the D4-brane framing for the quiver. On the other hand, it was shown in \cite{Nishinaka:2013mba} that the torus fixed points of the D4 moduli space are in one-to-one correspondence with the 2d molten crystal configurations. Moreover, the 2d crystal structure, that is, the specific surface in the 3d crystal, is determined by the correpsonding (non-compact) divisor in the toric diagram\footnote{As studied in \cite{Li:2020rij}, the representation of $\mathtt{Y}$ constructed from cyrstal configurations would become reducible for some special values of $\widetilde{\epsilon}_I$. In terms of crystals, truncations would appear to stop the molten crystal growing at certain atoms. Therefore, some $\text{Res}\Psi^{(a)}_{\mathfrak{C}}(u)$ would vanish in the actions of $e^{(a)}(u)$ and $f^{(a)}(u)$. The representation would then become irreducible in the truncated algebra. As the 2d crystal is essentially a surface of the 3d crystal, it could be possible to study this from the perspective of truncations. It would be interesting to see if there could be any new insights for the truncations by considering the relations between $\mathtt{Y}$ and $\mathtt{YB}$.}.

In fact, this agrees with the modules used in \cite{Litvinov:2020zeq,Chistyakova:2021yyd}, where the states are labelled by partitions and bi-coloured partitions for $\widehat{\mathfrak{gl}}_1$ and $\widehat{\mathfrak{gl}}_2$ respectively (see also \cite{Galakhov:2020vyb}). Now, if we know how the currents of $\mathtt{YB}$ are connected to $\mathcal{T}$, the actions of the $\mathcal{R}$-matrix can then be found using the relations among these currents.

The strategy is to consider the matrix element obtained by sandwiching $\mathcal{T}$ between two states $|\mu_{1,2}\rangle\in\bigoplus\limits_a\mathcal{F}_{(a),0}(u)$, viz, $\mathcal{T}_{\mu_1,\mu_2}(u):=\langle\mu_1|\mathcal{T}(u)|\mu_2\rangle$. Here, we have further labelled the auxiliary spaces $\mathcal{F}_{(a),0}(u)$ with the colours $a$ as the 2d crystals can have different initial atoms of different colours. As the name of YB algebras suggests, we propose that the first matrix elements are related to our currents of $\mathtt{YB}$ by
\begin{equation}
	\begin{split}
		&h^{(a)}(u)=\mathcal{T}_{\varnothing_{(a)},\varnothing_{(a)}}(u),\quad h^{(a)}(u)e^{(a)}(u)=\mathcal{T}_{\varnothing_{(a)},\square_{(a)}}(u),\quad f^{(a)}(u)h^{(a)}(u)=\mathcal{T}_{\square_{(a)},\varnothing_{(a)}}(u),\\
		&\psi^{(a)}(u-\epsilon_3)=\left(\mathcal{T}_{\square_{(a)},\square_{(a)}}(u)-\mathcal{T}_{\varnothing_{(a)},\square_{(a)}}(u)h^{(a)}(u)^{-1}\mathcal{T}_{\square_{(a)},\varnothing_{(a)}}(u)\right)h^{(a)}(u)^{-1},
	\end{split}
\end{equation}
where $\varnothing_{(a)}$ and $\square_{(a)}$ denote the empty 2d crystal and one single atom of colour $a$ respectively. Intuitively, starting with the ``empty'' $h^{(a)}(u)$, we can create an atom by acting $f^{(a)}(u)$ ($e^{(a)}(u)$) on the empty bra (ket) vector. Nevertheless, the actual situation is more complicated (although we would have a conjectural expression for higher levels with a similar intuition involving integrals). Indeed, the expression for $\psi^{(a)}(u)$ in terms of the matrix elements already looks somewhat intricate.

Now we can try to find the actions of the $\mathcal{R}$-matrix on these states via the $\mathcal{RTT}$ relation. Let us take the normalization $\mathcal{R}_{12}(u-v)|\varnothing_{(a)},\varnothing_{(b)}\rangle=|\varnothing_{(a)},\varnothing_{(b)}\rangle$. Then
\begin{equation}
	\langle\varnothing_{(a)},\varnothing_{(b)}|\mathcal{R}_{12}(u-v)\mathcal{T}_1(u)\mathcal{T}_2(v)|\varnothing_{(a)},\varnothing_{(b)}\rangle=\langle\varnothing_{(a)},\varnothing_{(b)}|\mathcal{T}_2(v)\mathcal{T}_1(u)\mathcal{R}_{12}(u-v)|\varnothing_{(a)},\varnothing_{(b)}\rangle
\end{equation}
simply yields the $hh$ relation
\begin{equation}
	h^{(a)}(u)h^{(b)}(v)=h^{(b)}(v)h^{(a)}(u).
\end{equation}

Next, we can consider
\begin{equation}
	\langle\square_{(a)},\varnothing_{(b)}|\mathcal{R}_{12}(u-v)\mathcal{T}_1(u)\mathcal{T}_2(v)|\varnothing_{(a)},\varnothing_{(b)}\rangle=\langle\square_{(a)},\varnothing_{(b)}|\mathcal{T}_2(v)\mathcal{T}_1(u)\mathcal{R}_{12}(u-v)|\varnothing_{(a)},\varnothing_{(b)}\rangle.
\end{equation}
The right hand side is actually
\begin{equation}
	\mathcal{T}_{\varnothing_{(b)},\varnothing_{(b)}}(v)\mathcal{T}_{\square_{(a)},\varnothing_{(a)}}(u)=h^{(b)}(v)f^{(a)}(u)h^{(a)}(u).
\end{equation}
By applying the $hf$ and $hh$ relations, we get
\begin{equation}
	\begin{split}
		&\langle\square_{(a)},\varnothing_{(b)}|\mathcal{R}_{12}(u-v)\mathcal{T}_1(u)\mathcal{T}_2(v)|\varnothing_{(a)},\varnothing_{(b)}\rangle\\
		=&\frac{1}{v-u}\left((v-u-\delta_{ab}\epsilon_3)f^{(a)}(u)h^{(b)}(v)h^{(a)}(u)+\delta_{ab}\epsilon_3f^{(a)}(v)h^{(b)}(v)h^{(a)}(u)\right)\\
		=&\frac{1}{v-u}\left((v-u-\delta_{ab}\epsilon_3)f^{(a)}(u)h^{(a)}(u)h^{(b)}(v)+\delta_{ab}\epsilon_3f^{(a)}(v)h^{(b)}(v)h^{(a)}(u)\right)\\
		=&\frac{1}{v-u}(v-u-\delta_{ab}\epsilon_3)\mathcal{T}_{\square_{(a)},\varnothing_{(a)}}(u)\mathcal{T}_{\varnothing_{(b)},\varnothing_{(b)}}(v)+\frac{1}{v-u}\delta_{ab}\epsilon_3\mathcal{T}_{\square_{(b)},\varnothing_{(b)}}(v)\mathcal{T}_{\varnothing_{(a)},\varnothing_{(a)}}(u).
	\end{split}
\end{equation}
Therefore, we find that
\begin{equation}
	\langle\square_{(a)},\varnothing_{(b)}|\mathcal{R}_{12}(u-v)\mathcal=\langle\square_{(a)},\varnothing_{(b)}|\frac{v-u-\delta_{ab}\epsilon_3}{v-u}+\langle\varnothing_{(a)},\square_{(b)}|\frac{\delta_{ab}\epsilon_3}{v-u}.
\end{equation}
Likewise, $\mathcal{R}_{12}(u-v)|\square_{(a)},\varnothing_{(b)}\rangle$ can be obtained by using the $he$ and $ee$ relations.

One can then proceed to higher levels with more atoms. However, we do not know how general $\mathcal{T}_{\mu_1,\mu_2}$ correspond to the currents. A possible way is to look for currents at higher levels that appear in the local terms from $ee$ and $ff$ relations. These higher currents would then give rise to matrix elements of $\mathcal{T}$ at higher levels. However, the computations would get rather involved even at the levels with 2 atoms for a general quiver. In \cite{Litvinov:2020zeq,Chistyakova:2021yyd}, for $\widehat{\mathfrak{gl}}_1$ and $\widehat{\mathfrak{gl}}_2$, it was found that any such matrix element can be expressed as some contour integral in terms of the currents. Here, we conjecture that this remains true for any general quiver. Explicitly, we have
\begin{equation}
	\mathcal{T}_{\mu_1,\mu_2}(u)=\frac{1}{(2\pi i)^n}\oint_{\mathcal{C}_1}\text{d}z_1\dots\oint_{\mathcal{C}_n}\text{d}z_nF(\bm{z})\left(\prod_{j=k+1}^nf^{(a_j)}(z_j)\right)h^{(a_0)}(u)\left(\prod_{j=1}^ke^{(a_j)}(z_j)\right),
\end{equation}
where the rational function $F(\bm{z})$ has poles at $z_j=u$. The clockwise contour $\mathcal{C}_j$ goes around $z_j=u,\infty$ and can be deformed in a way such that the contributions from local terms would be cancelled when applying current relations to swap $e^{(a)}(z_j)$ or $f^{(a)}(z_j)$ with other currents. Moreover, the indices $a_j$ (including $a_0$) should correspond to the colours of the atoms in $\mu_1$ and $\mu_2$.

This conjecture does not tell us how to compute $F(\bm{z})$, which is the key to get the exact results. Nevertheless, we may still verify this with the expression at level 1. Indeed,
\begin{equation}
	\begin{split}
		\frac{1}{2\pi i}\oint_\mathcal{C}\frac{1}{u-z}f^{(a)}(z)h^{(a)}(u)\text{d}z=&-\text{Res}_u\left(\frac{f^{(a)}(z)h^{(a)}(u)}{u-z}\right)-\text{Res}_{\infty}\left(\frac{f^{(a)}(z)h^{(a)}(u)}{u-z}\right)\\
		=&f^{(a)}(u)h^{(a)}(u)+\text{Res}_0\left(\frac{1}{z^2}\frac{f^{(a)}(1/z)h^{(a)}(u)}{u-1/z}\right)\\
		=&f^{(a)}(u)h^{(a)}(u)
	\end{split}
\end{equation}
recovers $\mathcal{T}_{\square_{(a)},\varnothing_{(a)}}(u)$ with $F(z)=1/(u-z)$. We also give some examples for states at higher levels in Appendix \ref{exhigherlv}.

The motivation of this conjecture stems from the $\mathcal{R}$-matrix being the intertwiner between certain free field representations. For the $\mathbb{C}^3$ case, it was found in \cite{Prochazka:2015deb,Litvinov:2020zeq} that we have the relations
\begin{equation}
	\begin{split}
		&[a_{-n},\mathcal{T}_{\mu_1,\mu_2}]=\mathcal{T}_{\mu_1,\mu_2'},\quad[\mathcal{T}_{\mu_1,\mu_2},a_n]=\mathcal{T}_{\mu_1',\mu_2},\\
		&a_{-n}=\frac{1}{\epsilon_3^k(n-1)!}\text{ad}_{e_1}^{n-1}e_0,\quad a_n=\frac{1}{\epsilon_3^k(n-1)!}\text{ad}_{f_1}^{n-1}f_0,
	\end{split}
\end{equation}
where $a_{-n}|\mu\rangle=|\mu'\rangle$ ($n>0$) creates boxes/atoms in the Young tableau. Therefore,
\begin{equation}
	\begin{split}
		&a_{-n}=\frac{1}{(-\epsilon_3)^k(n-1)!}\frac{1}{(2\pi i)^n}\oint\text{d}\bm{z}\left(\prod_{j=1}^nz_j\right)\left(\sum_{j=1}^n\frac{(-1)^{j-1}\binom{n-1}{j-1}}{z_j}\right)\prod_{j=1}^ne(z_j),\\
		&a_n=\frac{1}{(-\epsilon_3)^k(n-1)!}\frac{1}{(2\pi i)^n}\oint\text{d}\bm{z}\left(\prod_{j=1}^nz_j\right)\left(\sum_{j=1}^n\frac{(-1)^{j-1}\binom{n-1}{j-1}}{z_j}\right)\prod_{j=1}^nf(z_j).
	\end{split}
\end{equation}
The integral expression for matrix elements of $\mathcal{T}$ would then follow from their commutation relations with the modes. In general, such process is still not clear, and the explicit expression for the rational function $F(\bm{z})$ is desired. Nevertheless, we can still apply this to certain problems without the knowledge of its precise form. We will also further expound the contour integral conjecture in \S\ref{generators} for a certain class of quivers.

\subsection{Bethe Ansatz}\label{Bethe}
As an application of our previous results, let us now try to generalize the results in \cite{Litvinov:2020zeq,Chistyakova:2021yyd} and obtain the Bethe ansatz equation for any quiver $Q$. Consider the quantum space which is the tensor product of $n$ Fock spaces, $\mathcal{F}(u_1)\otimes\dots\otimes\mathcal{F}(u_n)$. We can define the Knizhinik-Zamolodchikov (KZ) operator
\begin{equation}
	T_1:=t_1^{L_1}\dots t_{G}^{L_G}\mathcal{R}_{1,n}(u_1-u_n)\dots\mathcal{R}_{1,2}(u_1-u_2),
\end{equation}
where $t_a\in[0,1)$ are the twist parameters and $G=|Q_0|$. The quantum space is graded under each level operator $L_a$ via
\begin{equation}
	L_a:=\sum_{j=1}^nL_{a,j}\quad\text{such that}\quad L_{a,j}|\bm{\mu}\rangle_u=N_{a,j}|\bm{\mu}\rangle_u
\end{equation}
gives the number $N_{a,j}$ of atoms with colour $a$ in the $j^\text{th}$ 2-dimensional crystal, where the subscript $u$ indicates that the state belongs to $\bigotimes\limits\mathcal{F}(u_j)$. By considering\footnote{Note added in version 3: It was very recently pointed out in \cite{Galakhov:2022uyu} that due to the non-trivial coproduct of the algebra from soliton contributions, in general $|\chi\rangle_x$ should be a mixed state of the chains of crystals rather than simply being identified as a chain of single-atom states. See \cite{Galakhov:2022uyu} for the modification of this subtlety. In the following discussions, this would change the eigenvalue of the KZ operator. Nevertheless, we expect that the effects of $|\chi\rangle_x$ (namely actions of $h^{(a)}$) would eventually cancel out.}
\begin{equation}
	|\chi\rangle_x:=|\square_1,\dots,\square_1,\dots,\square_G,\dots,\square_G\rangle_x\in\mathcal{F}_1(x_{1,1})\otimes\dots\otimes\mathcal{F}_1(x_{1,N_1})\otimes\dots\otimes\mathcal{F}_G(x_{G,1})\otimes\dots\otimes\mathcal{F}_G(x_{G,N_G}),
\end{equation}
let us further introduce the off-shell Bethe vector
\begin{equation}
	\begin{split}
		|B(x)\rangle_u:={}_x\langle\varnothing|&\mathcal{R}_{x_{1,1},~u_1}\dots\mathcal{R}_{x_{1,1},~u_n}\dots\mathcal{R}_{x_{1,N_1},~u_1}\dots\mathcal{R}_{x_{1,N_1},~u_n}\\
		&\dots\mathcal{R}_{x_{G,1},~u_1}\dots\mathcal{R}_{x_{G,1},~u_n}\dots\mathcal{R}_{x_{G,N_G},~u_1}\dots\mathcal{R}_{x_{G,N_G},~u_n}|\varnothing\rangle_u|\chi\rangle_x
	\end{split}
\end{equation}
in the quantum space, where ${}_x\langle\varnothing,\dots,\varnothing|$ is abbreviated as ${}_x\langle\varnothing|$ for brevity. We would like to find the condition such that the off-shell Bethe vector is an eigenvector of the KZ operator, that is, $T_1|B(x)\rangle_u=\mathtt{t}|B(x)\rangle_u$. Pictorially, we have
\begin{equation}
	\includegraphics[width=7cm]{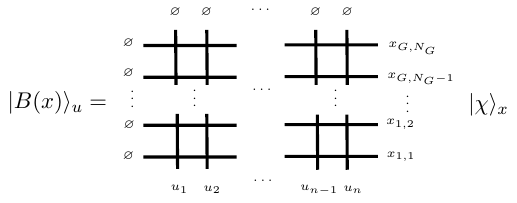}
\end{equation}
and
\begin{equation}
	\includegraphics[width=14cm]{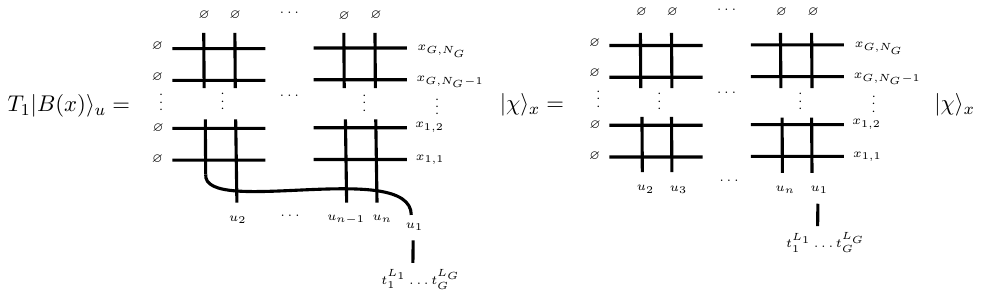}
\end{equation}
following the $\mathcal{RTT}$ relations along with $\mathcal{R}_{1,j}|\varnothing,\varnothing\rangle=|\varnothing,\varnothing\rangle$.

If we project the eigenvalue equation onto some state ${}_u\langle\bm{\mu}|$ satisfying $\sum\limits_{j=1}^nN_{a,j}=N_a$ for all $a$, then
\begin{equation}
	\begin{split}
		{}_u\langle\bm{\mu}|T_1|B(x)\rangle_u&=t_1^{N_{1,1}}\dots t_n^{N_{G,1}}{}_x\langle\varnothing|\mathcal{T}_{\mu_2,\emptyset}(u_2)\dots\mathcal{T}_{\mu_n,\varnothing}(u_n)\mathcal{T}_{\mu_1,\varnothing}(u_1)|\chi\rangle_x\\
		&=\mathtt{t}~{}_x\langle\varnothing|\mathcal{T}_{\mu_1,\varnothing}(u_1)\mathcal{T}_{\mu_2,\varnothing}(u_2)\dots\mathcal{T}_{\mu_n,\varnothing}(u_n)|\chi\rangle_x.
	\end{split}
\end{equation}
By setting ${}_x\langle\bm{\mu}|={}_x\langle\varnothing,\mu_2,\dots|$, i.e., $\mu_1=\varnothing$, this equation becomes
\begin{equation}
	t_1^0\dots t_G^0~{}_x\langle\varnothing|\mathcal{T}_{\mu_2,\varnothing}(u_2)\dots\mathcal{T}_{\mu_n,\varnothing}(u_n)h^{(a)}(u_1)|\chi\rangle_x=\mathtt{t}~{}_x\langle\varnothing|h^{(a)}(u_1)\mathcal{T}_{\mu_2,\varnothing}(u_2)\dots\mathcal{T}_{\mu_n,\varnothing}(u_n)|\chi\rangle_x.
\end{equation}
The actions of the currents/modes in $\mathtt{YB}$ on the 2d crystal are completely analogous to the actions on the 3d crystal discussed above. Therefore\footnote{Notice that the coordinates are now given by $y+\widetilde{\epsilon}(\mathfrak{a})=y+\sum\limits_IN_I\widetilde{\epsilon}_I$ for an atom $\mathfrak{a}$ in $|\mu\rangle_y$ \cite{Litvinov:2020zeq,Chistyakova:2021yyd}.},
\begin{equation}
	h^{(a)}(u)|\chi\rangle_x=\prod_{j=1}^{N_a}\frac{u-x_{a,j}}{u-x_{a,j}+\epsilon}|\chi\rangle_x,
\end{equation}
where we have used $\epsilon:=\epsilon_1+\epsilon_2=-\epsilon_3$. As a result,
\begin{equation}
	\mathtt{t}=\prod_{j=1}^{N_1}\frac{u_1-x_{1,j}}{u_1-x_{1,j}+\epsilon}.
\end{equation}

Now let us consider the state ${}_x\langle\bm{\mu}|$ with $N_{a,1}=\delta_{aa'}$ for some colour $a'$. Using the contour integral form of $\mathcal{T}_{\mu_j,\varnothing}$, the eigenvalue equation reads
\begin{equation}
	\begin{split}
		&t_{1}\bigg._x\bigg\langle\varnothing\bigg|\oint F(\bm{z})\left(f^{(1)}\left(z_{1,1}^{(n)}\right)\dots f^{(1)}\left(z_{1,N_{1,n}}^{(n)}\right)\dots f^{(G)}\left(z_{G,1}^{(n)}\right)\dots f^{(G)}\left(z_{G,N_{G,n}}^{(n)}\right)h^{(1)}(u_n)\right)\\
		&\dots\left(f^{(1)}\left(z_{1,1}^{(2)}\right)\dots f^{(G)}\left(z_{G,N_{G,2}}^{(2)}\right)h^{(1)}(u_2)\right)f^{(1)}\left(z_{1,1}^{(1)}\right)h^{(1)}(u_1)\text{d}\bm{z}\bigg|\chi\bigg\rangle_x\\
		=&\mathtt{t}\bigg._x\bigg\langle\varnothing\bigg|\oint F(\bm{z})f^{(1)}\left(z_{1,1}^{(1)}\right)h^{(1)}(u_1)\left(f^{(1)}\left(z_{1,1}^{(n)}\right)\dots h^{(1)}(u_n)\right)\dots\left(f^{(1)}\left(z_{1,1}^{(2)}\right)\dots h^{(1)}(u_2)\right)\text{d}\bm{z}\bigg|\chi\bigg\rangle_x.
	\end{split}
\end{equation}
Given the different variables with three indices, it would be better to clarify the notation here. For $z_{\ast,\star}^{(\divideontimes)}$ indicating the $\divideontimes^\text{th}$ state/2d crystal, $\ast$ denotes the colour of an atom (as in $f^{(\ast)}$), and $\star$ enumerates the number of the atoms of such colour. Recall that each Fock space in the quantum space is $\mathcal{F}_1$ whose initial atom is of colour 1 (as in $h^{(1)}$).

Using the $hf$ and $ff$ relations, we can get
\begin{equation}
	t_1\left(\prod_{i=2}^n\frac{u_i-x_{1,1}+\epsilon}{u_i-x_{1,1}}\right)(-1)^{|(1)|\sum\limits_{a=1}^G|(a)|N_a}\left(\prod_{a=1}^G\prod_{\substack{j=1\\(a,j)\neq(1,1)}}^{N_a}\frac{g_{1a}(x_{1,1}-x_{a,j})}{\overline{g}_{a1}(x_{1,1}-x_{a.j})}\right)\left(\prod_{l=1}^{N_1}\frac{u_1-x_{1,l}}{u_1-x_{1,l}+\epsilon}\right)=\mathtt{t}.
\end{equation}
Notice that the parameters $\widetilde{\epsilon}_{1a,i}$ and $\widetilde{\epsilon}_{a1,i}$ in $g_{1a}$ and $\overline{g}_{a1}$ should be correspondingly changed to $\epsilon_j$ in terms of the loop and vertex constraints from the quiver Yangian. Plugging in the value of $\mathtt{t}$ yields the Bethe equation
\begin{equation}
	\left(\prod_{i=2}^n\frac{u_i-x_{1,1}}{u_i-x_{1,1}+\epsilon}\right)\left(\prod_{a=1}^G\prod_{\substack{j=1\\(a,j)\neq(1,1)}}^{N_a}\frac{\overline{g}_{a1}(x_{1,1}-x_{a.j})}{g_{1a}(x_{1,1}-x_{a,j})}\right)=(-1)^{|(1)|\sum\limits_{a=1}^G|(a)|N_a}t_1.
\end{equation}
One may then consider other states ${}_x\langle\bm{\mu}|$ with different 2d crystal configurations (whose initial atoms are still labelled by 1) so that the other twist parameters $t_j$ would also appear in the Bethe equations. There should be a set of $G$ independent such equations as the sufficient and necessary condition for the off-shell Bethe vector $|B(x)\rangle_u$ to be an eigenstate of the KZ operator $T_1$. These equations can then be labelled by (eqn)$_{1,\dots,G}$ so that they are chosen by considering the state where the atom of colour $a$ first appears in the 2d crystal for (eqn)$_a$.

\paragraph{Examples} Consider the Jordan quiver, that is, one node with one loop. Taking the quantum (auxiliary) space to be a tensor product of $L$ ($M$) Fock space $\mathcal{F}$, we simply have
\begin{equation}
	\prod_{l=1}^{L}\frac{x_j-u_l}{x_j-u_l+\epsilon}=t\prod_{k\neq j}^M\frac{x_j-x_k-\epsilon}{x_j-x_k+\epsilon}.
\end{equation}
This reduces to the familiar Bethe equation
\begin{equation}
	\left(\frac{x_j+i}{x_j-i}\right)^L=t\prod_{k\neq j}^M\frac{x_j-x_k+2i}{x_j-x_k-2i}
\end{equation}
for the XXX spin chain under $u_l=-i$, $\epsilon=-2i$.

As the simplest toric CY example, consider $\mathbb{C}^3$ whose quiver Yangian is the affine Yangian $\mathtt{Y}\left(\widehat{\mathfrak{gl}}_1\right)$. Taking the quantum (auxiliary) space to be a tensor product of $n$ ($N$) Fock space $\mathcal{F}$ (notice that we only have one node in the quiver), we get the equation
\begin{equation}
	\prod_{l=1}^{n}\frac{x_j-u_l}{x_j-u_l-\epsilon_3}=t\prod_{k\neq j}^N\prod_{\alpha=1}^3\frac{x_j-x_k+\epsilon_\alpha}{x_j-x_k-\epsilon_\alpha}
\end{equation}
for any $j=1,\dots,N$, as obtained in \cite{Litvinov:2020zeq}.

The connection to Bethe ansatz equation would be of particular interest in the context of Bethe/gauge correspondence \cite{Nekrasov:2009uh,Nekrasov:2009ui,Nekrasov:2009rc} (see \cite{Galakhov:2022uyu} for a more recent discussion on this). For instance, the rapidities (denoted as $x_j$ in the above examples) in the Bethe equations correspond to the supersymmetric vacua of the associated 2d $\mathcal{N}=(2,2)$ theory. In terms of the $S$-matrix\footnote{We would like to thank the referee for pointing out this very interesting question.}, due to its factorized scattering property, we expect that each (2-magnon) $S$-matrix would correspond to a bond factor $\phi^{b\Rightarrow a}(x_j-x_k)$ as in \eqref{bondfactor} on the quiver side.

\section{Yangians and Coproducts}\label{coproduct}
Let us now have a brief discussion on the connections of quiver Yangians to some other Yangian algebras. As their coproducts have been explicitly constructed, this might shed light on the study of coproducts and $\mathcal{R}$-matrices for quiver Yangians. In this section, $f^{(a)}_n$ will be used to denote the generators in $\mathtt{Y}$ (instead of $\mathtt{YB}$).

\subsection{Cartan doubled Yangians}\label{CartandoubledY}
For $\mathbb{C}\times\mathbb{C}^2/\mathbb{Z}_N$, the quiver Yangian (with vertex constraints) is exactly Guay's affine Yangian \cite{guay2005cherednik,guay2007affine} as pointed out in \cite{Li:2020rij}. As a warm-up, we shall only consider a one-parameter ``degeneration'' here. In this subsection, we will mainly focus on the affine Lie algebra $A_{N-1}^{(1)}$ with $N>2$. As introduced in \cite{Braverman:2016pwk,finkelberg2018comultiplication} (see also \cite{guay2018coproduct}), given a symmetrizable Kac-Moody algebra with simple roots $\{\alpha_a\}_{a\in G}$, the Cartan doubled Yangian $\mathcal{Y}_{\infty}$ is the $\mathbb{C}$-algebra with generators $E^{(a)}_i$, $F^{(a)}_i$, $H^{(a)}_j$ ($i\in\mathbb{N}^*$, $j\in\mathbb{Z}$ and $a\in\mathcal{I}$)\footnote{Notice that we have a different convention to put the indices from the one in literature so as to be more consistent with the notation of the generators for $\mathtt{Y}$ above.} satisfying the relations
\begin{align}
	&\left[H^{(a)}_n,H^{(b)}_m\right]=0,\\
	&\left[E^{(a)}_n,F^{(b)}_m\right]=\delta_{ab}H^{(a)}_{m+n},\\
	&\left[H^{(a)}_{n+1},E^{(b)}_m\right]-\left[H^{(a)}_n,E^{(b)}_{m+1}\right]=\frac{(\alpha_a,\alpha_b)}{2}\left\{H^{(a)}_n,E^{(b)}_m\right\},\\
	&\left[H^{(a)}_n,F^{(b)}_{m+1}\right]-\left[H^{(a)}_{n+1},F^{(b)}_m\right]=\frac{(\alpha_a,\alpha_b)}{2}\left\{H^{(a)}_n,F^{(b)}_m\right\},\\
	&\left[E^{(a)}_{n+1},E^{(b)}_m\right]-\left[E^{(a)}_n,E^{(b)}_{m+1}\right]=\frac{(\alpha_a,\alpha_b)}{2}\left\{E^{(a)}_n,E^{(b)}_m\right\},\\
	&\left[F^{(a)}_n,F^{(b)}_{m+1}\right]-\left[F^{(a)}_{n+1},F^{(b)}_m\right]=\frac{(\alpha_a,\alpha_b)}{2}\left\{F^{(a)}_n,F^{(b)}_m\right\},\\
	&\text{Sym}_{\mathfrak{S}_k}\left[E^{(a)}_{n_1},\left[E^{(a)}_{n_2},\dots,\left[E^{(a)}_{n_k},E^{(b)}_m\right]\dots\right]\right]=0~(a\neq b,~k=1-\alpha_a\cdot\alpha_b),\\
	&\text{Sym}_{\mathfrak{S}_k}\left[F^{(a)}_{n_1},\left[F^{(a)}_{n_2},\dots,\left[F^{(a)}_{n_k},F^{(b)}_m\right]\dots\right]\right]=0~(a\neq b,~k=1-\alpha_a\cdot\alpha_b).
\end{align}

In particular, for $A_{N-1}^{(1)}$, its Dynkin diagram has tripled quiver being the quiver for $\mathbb{C}\times\mathbb{C}^2/\mathbb{Z}_N$. In fact, under the special choice $\epsilon_1=\epsilon_2$, the map
\begin{equation}
	\iota:\mathtt{Y}_{\mathbb{C}\times\mathbb{C}^2/\mathbb{Z}_N}\rightarrow\mathcal{Y}_{\infty}\left(A_{N-1}^{(1)}\right),\quad e^{(a)}_m\mapsto\frac{1}{\sqrt{\epsilon_3}}E^{(a)}_{m+1},\quad f^{(a)}_m\mapsto\frac{1}{\sqrt{\epsilon_3}}F^{(a)}_{m+1},\quad\psi^{(a)}_m\mapsto\frac{1}{\sqrt{\epsilon_3}}H^{(a)}_{m+1}
\end{equation}
is an isomorphism. The proof is also straightforward by checking the commutation relations of the generators on both sides. Since $\mathcal{Y}_{\infty}$ is a one-parameter Yangian algebra\footnote{One can introduce an extra parameter for $\mathcal{Y}_\infty$ as in \cite{guay2018coproduct}, but it is straightforward to see that the algebra remains the same for any value of the parameter.} while $\mathtt{Y}$ (with the vertex constraints) has two parameters, $\epsilon_{1,2}$ have to take special values for this isomorphism to hold.

As proven in \cite{guay2018coproduct}, the Cartan doubled Yangian $\mathcal{Y}_{\infty}(\mathfrak{g})$ has a coproduct uniquely determined by
\begin{equation}
	\begin{split}
		&\Delta\left(H^{(a)}_0\right)=H^{(a)}_0\otimes1+1\otimes H^{(a)}_0,\quad\Delta\left(E^{(a)}_0\right)=E^{(a)}_0\otimes1+1\otimes E^{(a)}_0,\quad\Delta\left(F^{(a)}_0\right)=F^{(a)}_0\otimes1+1\otimes F^{(a)}_0,\\
		&\Delta\left(H^{(a)}_1\right)=H^{(a)}_1\otimes1+1\otimes H^{(a)}_1+H^{(a)}_0\otimes H^{(a)}_0-\sum_{\alpha\in\Phi_+}\sum_{l=1}^{\dim\mathfrak{g}_{\alpha}}(\alpha_a,\alpha)x^{(-\alpha)}_l\otimes x^{(\alpha)}_l.
	\end{split}\label{coproduct1}
\end{equation}
The notation here requires some explanation. Here, $\mathfrak{g}_\alpha$ denotes the positive root $\alpha$ space with basis $\{x^{(\alpha)}_l\}$ and dual basis $\{x^{(-\alpha)}_l\}$ for $\mathfrak{g}_{-\alpha}$ such that $\left(x^{(\alpha)}_k,x^{(-\alpha)}_l\right)=\delta_{kl}$. Moreover, there is a homomorphism $\tau$ from $\mathfrak{g}$ to $\mathcal{Y}_{\infty}(\mathfrak{g})$. Then in the above expression, we simply denote $\tau\left(x^{(\alpha)}_l\right)$ as $x^{(\alpha)}_l$ for brevity. In particular, $x^{(\alpha_a)}=e^{(a)}_0$ and $x^{(-\alpha_a)}=f^{(a)}_0$ for all $a\in G$.

With the isomorphism $\iota$, we now have the coproduct $\Delta$ for the quiver Yangian $\mathtt{Y}\left(\widehat{\mathfrak{gl}}_N\right)$ (at least under some special condition). By considering the permutation map $\pi:x\otimes y\mapsto y\otimes x$, we can write another coproduct $\Delta':=\pi\circ\Delta$. More generally, for any quiver Yangian, we would have $\pi:x\otimes y\mapsto(-1)^{|x||y|}y\otimes x$. Once we get the complete coproduct for (any) quiver Yangian, we can obtain the universal $\mathcal{R}$-matrix satisfying $\Delta'(x)\mathcal{R}=\mathcal{R}\Delta(x)$, $(\text{id}\otimes\Delta)\mathcal{R}=\mathcal{R}_{13}\mathcal{R}_{12}$ and $(\Delta\otimes\text{id})\mathcal{R}=\mathcal{R}_{13}\mathcal{R}_{23}$. By considering the action on the crystals, this would lead to the $\mathcal{R}$-matrices in the specific representation discussed above.

\subsection{Affine super Yangians}\label{affinesuperY}
Now, let us consider the quiver Yangians for generalized conifolds and compare them with Ueda's affine super Yangians introduced in \cite{ueda2019affine}. Let $M,N\geq2$ and $M\neq N$. Ueda's affine super Yangian $Y_{\hbar_1,\hbar_2}\left(\widehat{\mathfrak{sl}}_{M|N}\right)$ is the $\mathbb{C}$-algebra with two parameters $\hbar_{1,2}\in\mathbb{C}$ and generators $\mathtt{x}^{\pm,(a)}_i$, $\mathtt{h}^{(a)}_i$ ($i\in\mathbb{N}$, $a\in\mathbb{Z}_{M+N}$) satisfying the relations
\begin{align}
	&\left[\mathtt{h}^{(a)}_n,\mathtt{h}^{(b)}_m\right]=0,\\
	&\left[\mathtt{x}^{+,(a)}_n,\mathtt{x}^{-,(b)}_m\right\}=\delta_{ab}\mathtt{h}^{(a)}_{m+n},\\
	&\left[\mathtt{h}^{(a)}_0,\mathtt{x}^{\pm,(b)}_m\right]=\pm\mathtt{a}_{ab}\mathtt{x}^{\pm,(b)}_m,\\
	&\left[\mathtt{h}^{(a)}_{n+1},\mathtt{x}^{\pm,(b)}_m\right]-\left[\mathtt{h}^{(a)}_n,\mathtt{x}^{\pm,(b)}_{m+1}\right]=\pm\mathtt{a}_{ab}\frac{\hbar_1+\hbar_2}{2}\left\{\mathtt{h}^{(a)}_n,\mathtt{x}^{\pm,(b)}_m\right\}-\mathtt{b}_{ab}\frac{\hbar_1-\hbar_2}{2}\left[\mathtt{h}^{(a)}_n,\mathtt{x}^{\pm,(b)}_m\right],\label{checkillustration}\\
	&\left[\mathtt{x}^{\pm,(a)}_{n+1},\mathtt{x}^{\pm,(b)}_m\right]-\left[\mathtt{x}^{\pm,(a)}_n,\mathtt{x}^{\pm,(b)}_{m+1}\right]=\pm\mathtt{a}_{ab}\frac{\hbar_1+\hbar_2}{2}\left\{\mathtt{x}^{\pm,(a)}_n,\mathtt{x}^{\pm,(b)}_m\right\}-\mathtt{b}_{ab}\frac{\hbar_1-\hbar_2}{2}\left[\mathtt{x}^{\pm,(a)}_n,\mathtt{x}^{\pm,(b)}_m\right],\\
	&\text{Sym}_{\mathfrak{S}_{1+|\mathtt{a}_{ab}|}}\left[\mathtt{x}^{\pm,(a)}_{n_1},\left[\mathtt{x}^{\pm,(a)}_{n_2},\dots,\left[\mathtt{x}^{\pm,(a)}_{n_{1+|\mathtt{a}_{ab}|}},\mathtt{x}^{\pm,(b)}_m\right\}\dots\right\}\right\}=0~(a\neq b),\\
	&\left\{\mathtt{x}^{\pm,(a)}_m,\mathtt{x}^{\pm,(a)}_n\right\}=0~(a=0,M),\\
	&\left\{\left[\mathtt{x}^{\pm,(a-1)}_m,\mathtt{x}^{\pm,(a)}_0\right],\left[\mathtt{x}^{\pm,(a)}_0,\mathtt{x}^{\pm,(a+1)}_n\right]\right\}=0~(a=0,M),\label{Serre2affinesuperY}
\end{align}
where
\begin{equation}
	\begin{split}
		&\mathtt{a}_{ab}=
		\begin{cases}
			(-1)^{p(a)}+(-1)^{p(a+1)},&a=b\\
			-(-1)^{p(a+1)},&b=a+1\\
			-(-1)^{p(a)},&b=a-1\\
			0,&\text{otherwise},
		\end{cases}
		\qquad\mathtt{b}_{ab}=
		\begin{cases}
			(-1)^{p(a)},&b=a+1\\
			-(-1)^{p(a+1)},&b=a-1\\
			0,&\text{otherwise},
		\end{cases}\\
		&p(a)=
		\begin{cases}
			0,&1\leq a\leq M\\
			1,&M+1\leq a\leq M+N,
		\end{cases}
	\end{split}
\end{equation}
and the generators $\mathtt{x}^{\pm,{(a)}}_i$ are fermionic for $a=0,M$ while all the other generators are bosonic.

It is worth noting that the defining relations bear resemblance to those for the quiver Yangian $\mathtt{Y}\left(\widehat{\mathfrak{gl}}_{M|N}\right)$. However, as we are now going to discuss, it seems that only a very special case would make the two Yangians (almost) coincide.

First of all, we need to identify the quivers associated to the two algebras. As described in Appendix \ref{quivergencon}, the $\mathbb{Z}_2$ grading of the $M+N$ nodes can be determined by the triangulation of the toric diagram. On the other hand, since $Y_{\hbar_1,\hbar_2}\left(\widehat{\mathfrak{sl}}_{M|N}\right)$ only has fermionic generators when $a=0,M$, we would expect the associated quiver (if there is one) should be related to the Dynkin diagram of the affine super algebra which has two fermionic nodes. Fortunately, given a generalized conifold, there always exists a quiver with precisely two fermionic nodes. This can be shown from the toric diagrams:
\begin{equation}
	\includegraphics[width=12cm]{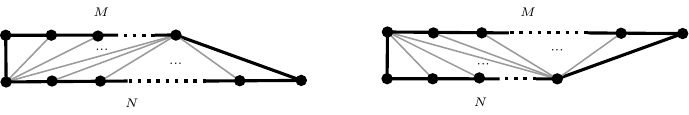}.
\end{equation}
In particular, we shall choose $\sigma_{1,\dots,M}=1$ and $\sigma_{M+1,\dots,M+N}=-1$. Moreover, different choices of $\sigma$/quivers for a given generalized conifold should give isomorphic quiver Yangians as they are the rational limit of the same quantum toroidal $\mathfrak{gl}_{M|N}$ algebra \cite{Li:2020rij,bezerra2021quantum,bezerra2021braid} \footnote{More generally, for any toric CY$_3$, as the corresponding quivers in different toric phases for are related by Seiberg duality (namely, crossing the wall of second kind \cite{Aganagic:2010qr}), it is conjectured that these quiver Yangians are isomorphic.}. Therefore, we have found the presentation of the quiver Yangian for a generalized conifold with fermionic generators $e^{(a)}_n$ and $f^{(a)}_n$ for $a=M,M+N$, and the quiver is actually the ``tripled'' quiver (in the sense of Appendix \ref{quivergencon}) of the Dynkin diagram with exactly two fermionic nodes.

With the choice of $\sigma_a$ as above, it is not hard to see that $(-1)^{p(a)}=\sigma_a$, and therefore
\begin{equation}
	\mathtt{a}_{ab}=
	\begin{cases}
		\sigma_a+\sigma_{a+1},&a=b\\
		-\sigma_{a+1},&b=a+1\\
		-\sigma_{a},&b=a-1\\
		0,&\text{otherwise},
	\end{cases}
	\qquad\mathtt{b}_{ab}=
	\begin{cases}
		\sigma_{a},&b=a+1\\
		-\sigma_{a+1},&b=a-1\\
		0,&\text{otherwise}.
	\end{cases}
\end{equation}
For convenience, let us further introduce $\widetilde{\mathtt{Y}}$ and $\widetilde{Y}_{\hbar_1,\hbar_2}$ such that we have the quotients $\mathtt{Y}=\widetilde{\mathtt{Y}}/\eqref{Serre2gencon}$ and $Y_{\hbar_1,\hbar_2}=\widetilde{Y}_{\hbar_1,\hbar_2}/\eqref{Serre2affinesuperY}$.

Now, consider the special case $\epsilon_1=\epsilon_2$ and $\hbar_1=\hbar_2$. Then the map
\begin{equation}
	\xi:\widetilde{\mathtt{Y}}\rightarrow\widetilde{Y}_{\hbar_1,\hbar_2},\quad\psi^{(a)}_n\mapsto\epsilon_3\mathtt{h}^{(a)}_n,\quad e^{(a)}_n\mapsto\epsilon_3^{1/2}\mathtt{x}^{+,(a)}_n,\quad f^{(a)}_n\mapsto\epsilon_3^{1/2}\mathtt{x}^{-,(a)}_n
\end{equation}
with $\hbar_1=\hbar_2=-\epsilon_1=-\epsilon_2=\epsilon_3/2$ is an isomorphism. In other words, when the parameters of the two Yangians take certain special values, $\mathtt{Y}\left(\widehat{\mathfrak{gl}}_{M|N}\right)$ and $Y_{\hbar_1,\hbar_2}\left(\widehat{\mathfrak{sl}}_{M|N}\right)$ are isomorphic up to one Serre relation. The proof is rather straightforward by checking the defining relations for the generators of the two algebras. Nevertheless, let us give an explicit check for one of the relations as an illustration. For instance, consider the relation \eqref{checkillustration}. When $b=a$, we have
\begin{equation}
	\left[\mathtt{h}^{(a)}_{n+1},\mathtt{x}^{\pm,(a)}_m\right]-\left[\mathtt{h}^{(a)}_n,\mathtt{x}^{\pm,(a)}_{m+1}\right]=\pm\frac{\sigma_a+\sigma_{a+1}}{2}(\hbar_1+\hbar_2)\left\{\mathtt{h}^{(a)}_n,\mathtt{x}^{\pm,(a)}_m\right\}.
\end{equation}
This recovers the relations
\begin{equation}
	\begin{split}
		&\left[\psi^{(a)}_{n+1},e^{(a)}_m\right]-\left[\psi{(a)}_n,e^{(a)}_{m+1}\right]=\frac{\sigma_a+\sigma_{a+1}}{2}\epsilon_3\left\{\psi^{(a)}_n,e^{(a)}_m\right\},\\
		&\left[\psi^{(a)}_{n+1},f^{(a)}_m\right]-\left[\psi{(a)}_n,f^{(a)}_{m+1}\right]=-\frac{\sigma_a+\sigma_{a+1}}{2}\epsilon_3\left\{\psi^{(a)}_n,f^{(a)}_m\right\}
	\end{split}
\end{equation}
in the quiver Yangian. When $b=a+1$, we have
\begin{equation}
	\left[\mathtt{h}^{(a)}_{n+1},\mathtt{x}^{\pm,(a+1)}_m\right]-\left[\mathtt{h}^{(a)}_n,\mathtt{x}^{\pm,(a+1)}_{m+1}\right]=\mp\sigma_{a+1}\frac{\hbar_1+\hbar_2}{2}\left\{\mathtt{h}^{(a)}_n,\mathtt{x}^{\pm,(a+1)}_m\right\}-\sigma_a\frac{\hbar_1-\hbar_2}{2}\left[\mathtt{h}^{(a)}_n,\mathtt{x}^{\pm,(a+1)}_m\right].\label{b=a+1}
\end{equation}
This has two cases. When $\sigma_a=\sigma_{a+1}$, the right hand side of \eqref{b=a+1} equals
\begin{equation}
	\begin{cases}
		-\sigma_{a+1}\left(\hbar_1\mathtt{h}^{(a)}_n\mathtt{x}^{+,(a+1)}_m+\hbar_2\mathtt{x}^{+,(a+1)}_m\mathtt{h}^{(a)}_n\right)&\\
		\sigma_{a+1}\left(\hbar_1\mathtt{x}^{-,(a+1)}_m\mathtt{h}^{(a)}_n+\hbar_2\mathtt{h}^{(a)}_n\mathtt{x}^{-,(a+1)}_m\right).&
	\end{cases}
\end{equation}
This recovers the relations
\begin{equation}
	\begin{cases}
		\left[\psi^{(a)}_{n+1},e^{(a+1)}_m\right]-\left[\psi{(a)}_n,e^{(a+1)}_{m+1}\right]=\sigma_{a+1}\left(\epsilon_1\psi^{(a)}_ne^{(a+1)}_m+\epsilon_2e^{(a+1)}_m\psi^{(a)}_n\right),&0<a<M\\
		\left[\psi^{(a)}_{n+1},e^{(a+1)}_m\right]-\left[\psi{(a)}_n,e^{(a+1)}_{m+1}\right]=\sigma_{a+1}\left(\epsilon_2\psi^{(a)}_ne^{(a+1)}_m+\epsilon_1e^{(a+1)}_m\psi^{(a)}_n\right),&M<a<M+N
	\end{cases}\label{swap1}
\end{equation}
in the quiver Yangian and likewise for the $\psi f$ relations. When $\sigma_a=-\sigma_{a+1}$, the right hand side of \eqref{b=a+1} equals
\begin{equation}
	\begin{cases}
		-\sigma_{a+1}\left(\hbar_1\mathtt{x}^{+,(a+1)}_m\mathtt{h}^{(a)}_n+\hbar_2\mathtt{h}^{(a)}_n\mathtt{x}^{+,(a+1)}_m\right)&\\
		\sigma_{a+1}\left(\hbar_1\mathtt{h}^{(a)}_n\mathtt{x}^{-,(a+1)}_m+\hbar_2\mathtt{x}^{-,(a+1)}_m\mathtt{h}^{(a)}_n\right).&
	\end{cases}
\end{equation}
This recovers the relations
\begin{equation}
	\begin{cases}
		\left[\psi^{(a)}_{n+1},e^{(a+1)}_m\right]-\left[\psi{(a)}_n,e^{(a+1)}_{m+1}\right]=\sigma_{a+1}\left(\epsilon_2\psi^{(a)}_ne^{(a+1)}_m+\epsilon_1e^{(a+1)}_m\psi^{(a)}_n\right),&a=M\\
		\left[\psi^{(a)}_{n+1},e^{(a+1)}_m\right]-\left[\psi{(a)}_n,e^{(a+1)}_{m+1}\right]=\sigma_{a+1}\left(\epsilon_1\psi^{(a)}_ne^{(a+1)}_m+\epsilon_2e^{(a+1)}_m\psi^{(a)}_n\right),&a=M+N
	\end{cases}\label{swap2}
\end{equation}
in the quiver Yangian and likewise for the $\psi f$ relations. When $b=a-1$, it is completely analogous to the discussions for $b=a+1$. When $b\neq a, a\pm1$, we have
\begin{equation}
	\left[\mathtt{h}^{(a)}_{n+1},\mathtt{x}^{\pm,(b)}_m\right]-\left[\mathtt{h}^{(a)}_n,\mathtt{x}^{\pm,(b)}_{m+1}\right]=0.
\end{equation}
This recovers the relations
\begin{equation}
		\left[\psi^{(a)}_{n+1},e^{(b)}_m\right]-\left[\psi{(a)}_n,e^{(b)}_{m+1}\right]=0,\quad\left[\psi^{(a)}_{n+1},f^{(b)}_m\right]-\left[\psi{(a)}_n,f^{(b)}_{m+1}\right]=0
\end{equation}
in the quiver Yangian. One may check that the other relations would also match correspondingly\footnote{When checking some relations, it is useful to notice that there are no consecutive fermionic nodes in the quiver.}.

With this explicit check, we can also see the reason of this isomorphism holding only at $\hbar_1=\hbar_2$, or equivalently, $\epsilon_1=\epsilon_2$. Due to the existence of the two fermionic nodes, the weights of the two quiver arrows connecting two consecutive nodes would have $\epsilon_1$ and $\epsilon_2$ swapped in order to satisfy the loop and vertex constraints. This then leads to \eqref{swap1} and \eqref{swap2} while the parameters $\hbar_{1,2}$ do not get swapped in the corresponding relations for $Y_{\hbar_1,\hbar_2}$.

Another subtlety that prevents us to get a perfect isomorphism between $\mathtt{Y}$ and $Y_{\hbar_1,\hbar_2}$ is the discrepancy between one of their Serre relations, that is, \eqref{Serre2gencon} and \eqref{Serre2affinesuperY}. Nevertheless, it would be natural to expect that the properties (such as the coproduct and the relation with certain $\mathcal{W}$-algebras mentioned below) would still hold if we replace \eqref{Serre2affinesuperY} with \eqref{Serre2gencon}. The proofs (if they are true) should be similar to the ones in \cite{ueda2019affine,ueda2022affine}.

As before, we can write the coproduct for Ueda's affine super Yangian, which is uniquely determined by \cite{ueda2019affine}
\begin{equation}
	\begin{split}
		&\Delta\left(H^{(a)}_0\right)=H^{(a)}_0\otimes1+1\otimes H^{(a)}_0,\quad\Delta\left(E^{(a)}_0\right)=E^{(a)}_0\otimes1+1\otimes E^{(a)}_0,\quad\Delta\left(F^{(a)}_0\right)=F^{(a)}_0\otimes1+1\otimes F^{(a)}_0,\\
		&\Delta\left(H^{(a)}_1\right)=H^{(a)}_1\otimes1+1\otimes H^{(a)}_1+(\hbar_1+\hbar_2)H^{(a)}_0\otimes H^{(a)}_0-(\hbar_1+\hbar_2)\sum_{\alpha\in\Phi_+}\sum_{l=1}^{\dim\mathfrak{g}_{\alpha}}(\alpha_a,\alpha)x^{(-\alpha)}_l\otimes x^{(\alpha)}_l.
	\end{split}\label{coproduct2}
\end{equation}
Notice that now $\mathfrak{g}$ is $\widehat{\mathfrak{sl}}_{M|N}$.

Moreover, in \cite{ueda2022affine}, it was proven that there exists a surjective homomorphism from $Y_{\hbar_1,\hbar_2}$ to the universal enveloping algebra of rectangular $\mathcal{W}$-superalgebras \cite{kac2004quantum,Kac2005CorrigendumT,arakawa2005representation} when the parameters are set to be $\hbar_1=\alpha/(M-N)$ and $\hbar_2=1-\alpha/(M-N)$ for some complex $\alpha$. In fact, this is closely related to the $\mathcal{W}_{M|N\times\infty}$ algebras that will be mentioned in \S\ref{Wmninfty}. The generators with spin 1 and 2 of the $\mathcal{W}_{M|N\times\infty}$ algebra, $U_{(1),AB}$ and $U_{(2),AB}$, are also part of the generators of the rectangular $\mathcal{W}$-algebras (and for certain values of $\alpha$, they fully generate the rectangular $\mathcal{W}$-algebras). As pointed out in \cite{Eberhardt:2019xmf}, $U_{(1),AB}$ and $U_{(2),AB}$ are sufficient to generate the whole $\mathcal{W}_{M|N\times\infty}$ algebra.

It could be possible that the above discussions can be properly extended to arbitrary $\epsilon_i$ and arbitrary $M,N$, or even more generally, to any quiver Yangians $\mathtt{Y}_Q$. We would expect that they would also have a similar coproduct as in \eqref{coproduct1} and \eqref{coproduct2} although we still need to figure out what the last term for $\Delta\left(\psi^{(a)}_1\right)$ would be.

\subsection{Generators of Quiver Yangians}\label{generators}
Analogous to the Yangian algebras discussed above, for toric CYs without compact divisors whose quivers have more than two nodes\footnote{Notice that in this subsection, we do not have any further restrictions on the numbers of bosonic and fermionic nodes. More generally, the discussions here should work for any symmetric quiver with at most one pair of arrows between any two nodes.}, the quiver Yangians are actually generated by finitely many generators.

Recall that the generators are $e^{(a)}_i$, $f^{(a)}_i$ and $\psi^{(a)}_j$ with $a\in Q_0$, $i\in\mathbb{N}$ and $j\in\mathbb{Z}_{\geq-1}$. In particular, $\psi^{(a)}_{-1}=1$. As $|a\rightarrow b|\leq1$, we have the relations
\begin{equation}
	\begin{split}
		&\left[\psi^{(a)}_{n+1},e^{(b)}_m\right]=\sigma_1e^{(b)}_m\psi^{(a)}_n+\sigma_1'\psi^{(a)}_ne^{(b)}_m+\left[\psi^{(a)}_n,e^{(b)}_{m+1}\right],\\
		&\left[\psi^{(a)}_{n+1},f^{(b)}_m\right]=-\sigma_1'f^{(b)}_m\psi^{(a)}_n-\sigma_1\psi^{(a)}_nf^{(b)}_m+\left[\psi^{(a)}_n,f^{(b)}_{m+1}\right],
	\end{split}
\end{equation}
where $\sigma_1:=\sigma^{a\rightarrow b}_1$ and $\sigma_1':=\sigma^{b\rightarrow a}_1$. Then for $n=-1$, we have
\begin{equation}
	\left[\psi^{(a)}_0,e^{(b)}_m\right]=\widetilde{\sigma}_1e^{(b)}_m,\quad\left[\psi^{(a)}_0,f^{(b)}_m\right]=-\widetilde{\sigma}_1f^{(b)}_m,
\end{equation}
where $\widetilde{\sigma}_1:=\sigma_1+\sigma_1'$. Notice that these were also used when discussing the relation of quiver Yangians and Ueda's affine super Yangians. Therefore, for $n=0$, we get
\begin{equation}
	\begin{split}
		&\left[\psi^{(a)}_1,e^{(b)}_m\right]=\frac{\sigma_1}{\widetilde{\sigma}_1}\psi^{(a)}_0\left[\psi^{(a)}_0,e^{(b)}_m\right]+\frac{\sigma_1'}{\widetilde{\sigma}_1}\left[\psi^{(a)}_0,e^{(b)}_m\right]\psi^{(a)}_0+\widetilde{\sigma}_1e^{(b)}_{m+1},\\
		&\left[\psi^{(a)}_1,f^{(b)}_m\right]=\frac{\sigma_1'}{\widetilde{\sigma}_1}\psi^{(a)}_0\left[\psi^{(a)}_0,f^{(b)}_m\right]+\frac{\sigma_1}{\widetilde{\sigma}_1}\left[\psi^{(a)}_0,f^{(b)}_m\right]\psi^{(a)}_0-\widetilde{\sigma}_1f^{(b)}_{m+1}.
	\end{split}
\end{equation}
Notice that we have chosen $a$ and $b$ with arrows connecting them so that $\sigma_1$ and $\sigma_1'$ are non-zero.

For $|(a)|=0$, choosing $b=a$, we have
\begin{equation}
	e^{(a)}_{m+1}=\frac{1}{2\sigma_1}\left[\psi^{(a)}_1-\frac{1}{2}\left(\psi^{(a)}_0\right)^2,e^{(a)}_m\right],\quad f^{(a)}_{m+1}=-\frac{1}{2\sigma_1}\left[\psi^{(a)}_1-\frac{1}{2}\left(\psi^{(a)}_0\right)^2,f^{(a)}_m\right].
\end{equation}
For $|(a)|=1$, choosing $a=b+1$, we have
\begin{equation}
	\begin{split}
		&e^{(b)}_{m+1}=\frac{1}{\widetilde{\sigma}_1}\left[\psi^{(b+1)}_1-\frac{\sigma_1}{\widetilde{\sigma}_1}\left(\psi^{(b+1)}_0\right)^2,e^{(b)}_m\right]-\frac{\sigma_1-\sigma_1'}{\widetilde{\sigma}_1}e^{(b)}_m\psi^{(b+1)}_0,\\
		&f^{(b)}_{m+1}=-\frac{1}{\widetilde{\sigma}_1}\left[\psi^{(b+1)}_1-\frac{\sigma_1}{\widetilde{\sigma}_1}\left(\psi^{(b+1)}_0\right)^2,f^{(b)}_m\right]+\frac{\sigma_1-\sigma_1'}{\widetilde{\sigma}_1}\psi^{(b+1)}_0f^{(b)}_m.
	\end{split}
\end{equation}
Notice that we can always write $\sigma_1$ and $\sigma_1'$ in terms of $\epsilon_{1,2,3}$ due to vertex constraints. For both bosonic and fermionic nodes, define
\begin{equation}
	\widetilde{\psi}^{(a)}_1:=\psi^{(a)}_1-\frac{\sigma^{a\rightarrow b}_1}{\sigma^{a\rightarrow b}_1+\sigma^{b\rightarrow a}_1}\left(\psi^{(a)}_0\right)^2.
\end{equation}
We can then compactly write the relations as
\begin{equation}
	\begin{split}
		&e^{(a)}_{m+1}=\frac{1}{\sigma^{b\rightarrow a}_1+\sigma^{a\rightarrow b}_1}\left[\widetilde{\psi}^{(b)}_1,e^{(a)}_m\right]-\frac{\sigma^{b\rightarrow a}_1-\sigma^{a\rightarrow b}_1}{\sigma^{b\rightarrow a}_1+\sigma^{a\rightarrow b}_1}e^{(a)}_m\psi^{(b)}_0,\\
		&f^{(a)}_{m+1}=-\frac{1}{\sigma^{b\rightarrow a}_1+\sigma^{a\rightarrow b}_1}\left[\widetilde{\psi}^{(b)}_1,f^{(a)}_m\right]+\frac{\sigma^{b\rightarrow a}_1-\sigma^{a\rightarrow b}_1}{\sigma^{b\rightarrow a}_1+\sigma^{a\rightarrow b}_1}\psi^{(b)}_0f^{(a)}_m,\label{ind1}
	\end{split}
\end{equation}
where $b=a$ for $|(a)|=0$ and $b=a+1$ for $|(a)|=1$. Moreover, we have
\begin{equation}
	\psi^{(a)}_{m+1}=\left[e^{(a)}_{m+1},f^{(a)}_0\right\}.\label{ind2}
\end{equation}
As a result, we have shown that the quiver Yangian in this case is generated by $e^{(a)}_0$, $f^{(a)}_0$ and $\psi^{(a)}_{0,1}$. The other modes can actually be inductively obtained using \eqref{ind1} and \eqref{ind2}.

It could then be possible that this would enable us to write down a minimalistic presentation of the quiver Yangian similar to those in \cite{guay2018coproduct,ueda2019affine}. This would help us study the coproduct of the quiver Yangian and its relation to $\mathcal{W}$-algebras.

Moreover, we can now also get the matrix elements of $\mathcal{T}(u)$ at higher levels inductively. For instance,
\begin{equation}
	\mathcal{T}_{\varnothing,\mu'}(u)=\langle\varnothing|\mathcal{T}(u)|\mu'\rangle=\langle\varnothing|\mathcal{T}(u)e^{(a)}_m|\mu\rangle.
\end{equation}
Then using \eqref{ind1} and \eqref{ind2}, we can express all $\mathcal{T}_{\mu_1,\mu_2}(u)$ in terms of $e^{(a)}_0$, $f^{(a)}_0$ and $\psi^{(a)}_{0,1}$. If we can write the contour integral expressions for all the states generated only by $e^{(a)}_0$ (and $f^{(a)}_0$), then we can write the contour integral expression for any $\mathcal{T}_{\mu_1,\mu_2}(u)$ and hence obtain the action of the $\mathcal{R}$-matrix (recall that we know the actions of $\psi$ from \S\ref{QY} and \S\ref{Bethe}).

In fact, analogous to the known cases, we conjecture that for $|\mu'\rangle=e^{(a)}_0|\mu\rangle$, where $|\mu\rangle$ is a state generated only from $e^{(a_i)}_0$, we have
\begin{equation}
	\begin{split}
		\mathcal{T}_{\varnothing,\mu'}(u)=&\frac{1}{2\pi i}\oint_{\infty+u}\text{d}z\frac{1}{\epsilon_3}\left(1-\frac{u-z-\epsilon_3}{u-z}\prod_{i=1}^k\frac{\overline{g}_{aa_i}(u-z)}{g_{a_ia}(u-z)}\right)\mathcal{T}_{\varnothing,\mu}(u)e^{(a)}(z)\\
		=&\frac{1}{2\pi i}\oint_{\infty+u}\text{d}z\frac{1}{\epsilon_3}\left(1-\frac{u-z-\epsilon_3}{u-z}\prod_{i=1}^k\frac{1}{\phi^{a\Rightarrow a_i}(u-z)}\right)\mathcal{T}_{\varnothing,\mu}(u)e^{(a)}(z),
	\end{split}
\end{equation}
where $\mathcal{T}_{\varnothing,\mu}(u)=\frac{1}{(2\pi i)^k}\oint\text{d}\bm{z}F(\bm{z})h^{(a_0)}(u)e^{(a_1)}(z_1)\dots e^{(a_k)}(z_k)$. Similarly,\footnote{Recall that the convention of $f$ is the one for $\mathtt{Y}$ instead of $\mathtt{YB}$ in this section.}
\begin{equation}
	\begin{split}
		\mathcal{T}_{\mu',\varnothing}(u)=&\frac{1}{2\pi i}\oint_{\infty+u}\text{d}z\frac{1}{\epsilon_3}\left(1-\frac{u-z-\epsilon_3}{u-z}\prod_{i=1}^k\frac{\overline{g}_{aa_i}(u-z)}{g_{a_ia}(u-z)}\right)\left(-f^{(a)}(z)\right)\mathcal{T}_{\mu,\varnothing}(u)\\
		=&\frac{1}{2\pi i}\oint_{\infty+u}\text{d}z\frac{1}{\epsilon_3}\left(1-\frac{u-z-\epsilon_3}{u-z}\prod_{i=1}^k\frac{1}{\phi^{a\Rightarrow a_i}(u-z)}\right)\left(-f^{(a)}(z)\right)\mathcal{T}_{\mu,\varnothing}(u).
	\end{split}
\end{equation}
We can then get any $\mathcal{T}_{\mu_1,\mu_2}(u)$ inductively using the contour integral expressions. For instance, at level 1, we simply have
\begin{equation}
	\mathcal{T}_{\varnothing,\square_{(a)}}(u)=\frac{1}{2\pi i}\oint_{\infty+u}\text{d}z\frac{1}{\epsilon_3}\left(1-\frac{u-z-\epsilon_3}{u-z}\right)\mathcal{T}_{\varnothing,\varnothing}(u)e^{(a)}(z)=\frac{1}{2\pi i}\oint_{\infty+u}\text{d}z\frac{1}{u-z}h^{(a)}(u)e^{(a)}(z),
\end{equation}
which agrees with our discussions in \S\ref{crystalRTT}. More examples at higher levels can be found in Appendix \ref{exhigherlv}.

\section{Generalized Conifolds and $\mathcal{W}_{m|n\times\infty}$ Algebras}\label{Wmninfty}
Now, let us slightly digress from the previous discussions and consider the matrix extensions of $\mathcal{W}_{1+\infty}$ for generalized conifolds. As the quiver Yangian for the generalized conifold defined by $xy=z^Mw^N$ is the affine Yangian $\mathtt{Y}\left(\widehat{\mathfrak{gl}}_{M|N}\right)$, it is expected to be intimately related to the corresponding $\mathcal{W}_{M|N\times\infty}$ algebra as mentioned in \S\ref{intro}.

\subsection{Miura Transformations}\label{Miura}
Let us consider the fields $J_{AB}$ that generate the $\widehat{\mathfrak{gl}}(M|N)_\kappa$ Kac-Moody (super)algebra with OPE \cite{Rapcak:2019wzw,Eberhardt:2019xmf}
\begin{equation}
	\begin{split}
		J_{AB}(z)J_{CD}(w)\sim&\frac{(-1)^{|B||C|}\kappa\delta_{AD}\delta_{CB}+\delta_{AB}\delta_{CD}}{(z-w)^2}\\
		&+\frac{(-1)^{|A||B|+|C||D|+|C||B|}\delta_{AD}J_{CB}(w)-(-1)^{|B||C|}\delta_{CB}J_{AD}(w)}{z-w}.
	\end{split}
\end{equation}
As $J$ is an $m|n$ supermatrix, we have $|A|=0$ for $1\leq A\leq M$ and $|A|=1$ for $M+1\leq A\leq M+N$. We can then write the mode expansion of $J_{AB}$,
\begin{equation}
	J_{AB}(z)=\sum_{k\in\mathbb{Z}}\frac{a_{AB,k}}{z^{k+1}},
\end{equation}
and find the commutation relation
\begin{equation}
	\begin{split}
		[a_{AB,n},a_{CD,m}\}=&\delta_{n,-m}n\left((-1)^{|B||C|}\kappa\delta_{AD}\delta_{CB}+\delta_{AB}\delta_{CD}\right)\\
		&+(-1)^{|A||B|+|C||D|+|C||B|}\delta_{AD}a_{CB,m+n}-(-1)^{|B||C|}\delta_{CB}a_{AD,m+n}.
	\end{split}
\end{equation}

Now, we need to introduce a matrix-valued differential operator known as the Miura operator, $\mathcal{L}_i:=\kappa\mathbbm{1}\partial+J_i$. One can then consider the product of $k$ Miura operators as
\begin{equation}
	\mathcal{L}_1\mathcal{L}_2\dots\mathcal{L}_k=\sum_{i=0}^kU_{(i)}(\kappa\partial)^{k-i}.
\end{equation}
The $U_{(i),AB}$ operators at each spin $i$ generate the $\mathcal{W}_{M|N\times\infty}$ algebra \cite{Rapcak:2019wzw,Eberhardt:2019xmf}, which is a matrix extension of the $\mathcal{W}_{1+\infty}$ algebra. As mentioned before, $U_{(1),AB}$ and $U_{(2),AB}$ should fully generate the whole algebra (while one further need $U_{(3)}$ in the case of $\mathcal{W}_{1+\infty}$) \cite{Eberhardt:2019xmf}. The $\mathcal{W}_{M|N\times\infty}$ algebras are closely related to the rectangular $\mathcal{W}$-algebras. In particular, the construction using Miura basis reveals some remarkable features by considering the OPEs of $U_{(i),AB}$. See \cite{Eberhardt:2019xmf} for more details.

Here, we shall consider the following intertwiner\footnote{Although we are using the same letter $\mathcal{R}$ here, it does not mean that this intertwiner should coincide with the $\mathcal{R}$-matrices discussed in the previous sections. In fact, it could also be possible for one to define a different operator that intertwines between two matrix-valued Miura operators, i.e., $\mathcal{R}_{12}\mathcal{L}_1\mathcal{L}_2=\mathcal{L}_2\mathcal{L}_1\mathcal{R}_{12}$. Nevertheless, the precise connection between quiver Yangians and $\mathcal{W}_{M|N\times\infty}$ still requires further study.}:
\begin{equation}
	\mathcal{R}_{12}(\mathcal{L}_1(z)\mathcal{L}_2(z))_{AB}=(\mathcal{L}_2(z)\mathcal{L}_1(z))_{AB}\mathcal{R}_{12}.
\end{equation}
It is not hard to see that $\mathcal{R}_{ij}$ satisfies the Yang-Baxter equation. Consider
\begin{equation}
	(\mathcal{L}_3\mathcal{L}_2\mathcal{L}_1)_{AB}=\sum_{k}(\mathcal{L}_3\mathcal{L}_2)_{Ak}\mathcal{L}_{1,kB}=\mathcal{R}_{23}\sum_{k}(\mathcal{L}_2\mathcal{L}_3)_{Ak}\mathcal{L}_{1,kB}\mathcal{R}_{23}^{-1}=\mathcal{R}_{23}(\mathcal{L}_2\mathcal{L}_3\mathcal{L}_1)_{AB}\mathcal{R}_{23}^{-1}.
\end{equation}
Further conjugated by $\mathcal{R}_{13}$ and then by $\mathcal{R}_{12}$, this becomes
\begin{equation}
	(\mathcal{L}_3\mathcal{L}_2\mathcal{L}_1)_{AB}=\mathcal{R}_{23}\mathcal{R}_{13}\mathcal{R}_{12}(\mathcal{L}_1\mathcal{L}_2\mathcal{L}_3)_{AB}\mathcal{R}_{12}^{-1}\mathcal{R}_{13}^{-1}\mathcal{R}_{23}^{-1}.
\end{equation}
On the other hand, conjugating $(\mathcal{L}_3\mathcal{L}_2\mathcal{L}_1)_{AB}$ by $\mathcal{R}_{12}$, $\mathcal{R}_{13}$ and $\mathcal{R}_{23}$ successively, we get
\begin{equation}
	(\mathcal{L}_3\mathcal{L}_2\mathcal{L}_1)_{AB}=\mathcal{R}_{12}\mathcal{R}_{13}\mathcal{R}_{23}(\mathcal{L}_1\mathcal{L}_2\mathcal{L}_3)_{AB}\mathcal{R}_{23}^{-1}\mathcal{R}_{13}^{-1}\mathcal{R}_{12}^{-1}.
\end{equation}
Thus, we have $\mathcal{R}_{23}\mathcal{R}_{13}\mathcal{R}_{12}=\mathcal{R}_{12}\mathcal{R}_{13}\mathcal{R}_{23}$. By writing $\mathcal{T}_0\equiv\mathcal{R}_{0,1}\dots\mathcal{R}_{0,k}$, we can also get the $\mathcal{RTT}$ relation.

By definition of the Miura operator, we can write
\begin{equation}
	\mathcal{R}_{12}((\kappa\partial+J_1)(\kappa\partial+J_2))_{AB}=((\kappa\partial+J_2)(\kappa\partial+J_1))_{AB}\mathcal{R}_{12}.
\end{equation}
Let us expand this and compare the terms at different orders of $\partial$ on both sides. At order $\partial^1$, we have
\begin{equation}
	\mathcal{R}_{12}(J_1+J_2)_{AB}=(J_1+J_2)_{AB}\mathcal{R}_{12}.
\end{equation}
If we define $J_+:=J_1+J_2$ and $J_-:=J_1-J_2$, we can see that $\mathcal{R}_{12}$ commutes with $J_{+,AB}$. At order $\partial^0$, we get
\begin{equation}
	\mathcal{R}_{12}((J_1J_2)(z)+\kappa\partial J_2(z))_{AB}=((J_2J_1)(z)+\kappa\partial J_1(z))_{AB}\mathcal{R}_{12},
\end{equation}
where we have used brackets to indicate normal ordering for simplicity. In \cite{Prochazka:2019dvu}, for the $\widehat{\mathfrak{gl}}(1)$ case where every matrix is just $1\times1$, we can use the above two equations to write
\begin{equation}
	\mathcal{R}_{12}((J_-J_-)(z)+2\kappa\partial J_-(z))=((J_-J_-)(z)-2\kappa\partial J_-(z))\mathcal{R}_{12}.
\end{equation}
This shows the connection to Liouville reflection operators \cite{maulik2012quantum,Zhu:2015nha}. Here, we would like to write such equation for $J_-$ as well. However, we find that this would give rise to some extra terms:
\begin{equation}
	\begin{split}
		&\mathcal{R}_{12}((J_-J_-)(z)+2\kappa\partial J_-(z)+2(J_2J_1)(z)-2(J_1J_2)(z))_{AB}\\
		=&((J_-J_-)(z)-2\kappa\partial J_-(z)-2(J_2J_1)(z)+2(J_1J_2)(z))_{AB}\mathcal{R}_{12}.
	\end{split}
\end{equation}
Nevertheless, let us still consider its mode expansion. Henceforth, we shall write $J\equiv J_-$ and $a_{AB}\equiv a_{-,AB}$ for brevity. At level $n$, we have
\begin{equation}
	\begin{split}
		&\mathcal{R}_{12}\left(\sum_{k=1}^{M+N}\sum_{l\in\mathbb{Z}}(a_{Ak,l}a_{kB,n-l})-2\kappa(n+1)a_{AB,n}\right.\\
		&\left.-2\sum_{k=1}^{M+N}\sum_{l\in\mathbb{Z}}(a_{1,Ak,l}a_{2,kB,n-l})+2\sum_{k=1}^{M+N}\sum_{l\in\mathbb{Z}}(a_{2,Ak,l}a_{1,kB,n-l})\right)\\
		=&\left(\sum_{k=1}^{M+N}\sum_{l\in\mathbb{Z}}(a_{Ak,l}a_{kB,n-l})+2\kappa(n+1)a_{AB,n}\right.\\
		&\left.-2\sum_{k=1}^{M+N}\sum_{l\in\mathbb{Z}}(a_{1,Ak,l}a_{2,kB,n-l})+2\sum_{k=1}^{M+N}\sum_{l\in\mathbb{Z}}(a_{2,Ak,l}a_{1,kB,n-l})\right)\mathcal{R}_{12}.
	\end{split}
\end{equation}

Similar to \cite{Prochazka:2019dvu,Zhu:2015nha}, it would be more useful to consider the mode expansion on the cylinder whose coordinate map relating the one on the complex plane is given by $z=\text{e}^{\widetilde{z}}$. We can then make a conformal transformation to get the expression on the cylinder. As the stress tensor is \cite{Prochazka:2018tlo}
\begin{equation}
	T(z)=\frac{1}{2}\sum_{A,B}(J_{AB}J_{AB})(z)+\frac{\kappa}{2}\sum_A\partial J_{AA}(z),
\end{equation}
we find that
\begin{equation}
	T(z)J_{AB}(w)\sim-\frac{2\delta_{AB}\kappa}{(z-w)^3}+\frac{J_{AB}(w)}{(z-w)^2}+\frac{\partial J_{AB}(w)}{z-w}
\end{equation}
(with a proper normalization of $T(z)$). Therefore, the off-diagonal fields transform as primaries while the diagonal ones have anomalous transformations. More explicitly,
\begin{equation}
	J_{AB}(z)\rightarrow\widetilde{J}_{AB}(\widetilde{z})=\left(\frac{\text{d}\widetilde{z}}{\text{d}z}\right)^{-1}J_{AB}(z)+\delta_{AB}\kappa\left(\frac{\text{d}\widetilde{z}}{\text{d}z}\right)^{-2}\left(\frac{\text{d}^2\widetilde{z}}{\text{d}z^2}\right).
\end{equation}
On the cylinder, we then have
\begin{equation}
	\widetilde{J}_{AB}=\sum_{k\in\mathbb{Z}}a_{AB,k}\text{e}^{-k\widetilde{z}}-\delta_{AB}\kappa.
\end{equation}
As we can see, this is the usual mode expansion on the cylinder but with the zero mode shifted by a constant. As we will mainly work with $\widetilde{a}_{AB,n}$ on the cylinder in the followings, we shall simply denote $\widetilde{a}_{AB,n}$ as $a_{AB,n}$. By virtue of the factor $\delta_{AB}$, the equation for the intertwiner becomes\footnote{As pointed out in \cite{Prochazka:2019dvu}, $\mathcal{R}$ should intertwine between two opposite conformal transformations on $a_{AB,n}$ on the left and right hands.}
\begin{equation}
	\begin{split}
		&\mathcal{R}_{12}\left(\sum_{k=1}^{M+N}\sum_{l\in\mathbb{Z}}(a_{Ak,l}a_{kB,n-l})-2\kappa na_{AB,n}\right.\\
		&\left.-2\sum_{k=1}^{M+N}\sum_{l\in\mathbb{Z}}(a_{1,Ak,l}a_{2,kB,n-l})+2\sum_{k=1}^{M+N}\sum_{l\in\mathbb{Z}}(a_{2,Ak,l}a_{1,kB,n-l})\right)\\
		=&\left(\sum_{k=1}^{M+N}\sum_{l\in\mathbb{Z}}(a_{Ak,l}a_{kB,n-l})+2\kappa na_{AB,n}\right.\\
		&\left.-2\sum_{k=1}^{M+N}\sum_{l\in\mathbb{Z}}(a_{1,Ak,l}a_{2,kB,n-l})+2\sum_{k=1}^{M+N}\sum_{l\in\mathbb{Z}}(a_{2,Ak,l}a_{1,kB,n-l})\right)\mathcal{R}_{12}
	\end{split}\label{Ra-}
\end{equation}
on the cylinder. This should be equivalent to the result from taking the mode expansions of $J_{AB}$ directly on the cylinder.

Incidentally, if we add all such equations for $A=B$, then we find that the extra pieces with $a_{1,2}$ get cancelled:
\begin{equation}
	\begin{split}
		&\mathcal{R}_{12}\sum_{A=1}^{M+N}\left(\sum_{k=1}^{M+N}\sum_{l\in\mathbb{Z}}(a_{Ak,l}a_{kA,n-l})-2\kappa na_{AA,n}\right)\\
		=&\sum_{A=1}^{M+N}\left(\sum_{k=1}^{M+N}\sum_{l\in\mathbb{Z}}(a_{Ak,l}a_{kA,n-l})+2\kappa na_{AA,n}\right)\mathcal{R}_{12}.
	\end{split}
\end{equation}
It could be possible that there exist other combinations that would only keep $a$ modes left. However, we shall still focus on \eqref{Ra-} in the following discussions.

In the followings, we shall act \eqref{Ra-} on different states/representations and mainly focus on level 1 with $n=-1$. One may also consider the higher levels for \eqref{Ra-}. For instance, at level 2 (which is necessary for fully analyzing the generators of the algebra), we have two sets of equations, either one factor of $n=-2$ or two factors of $n=-1$. Of course, the calculations would become more tedious when we consider higher levels, as well as larger $M$ and $N$. Therefore, a more systematic study would be quite helpful, especially for the comparison with quiver Yangians.

As $\mathcal{T}_0=\mathcal{R}_{0,1}\dots\mathcal{R}_{0,l}$, we have
\begin{equation}
	(a_{0,AB,n}+a_{1,AB,n}+\dots+a_{l,AB,n})\mathcal{T}_0=\mathcal{T}_0(a_{0,AB,n}+a_{1,AB,n}+\dots+a_{l,AB,n})
\end{equation}
for any $n\in\mathbb{Z}$. Sandwiching this between some states $\langle\lambda|$ and $|\text{vac}\rangle$ for $n>0$,
\begin{equation}
	\left\langle\lambda\left|\sum_{i=0}^la_{i,AB,n}\mathcal{T}_0\right|\text{vac}\right\rangle=\left\langle\lambda\left|\mathcal{T}_0\sum_{i=0}^la_{i,AB,n}\right|\text{vac}\right\rangle,
\end{equation}
we would get
\begin{equation}
	\langle\lambda|a_{0,AB,n}\mathcal{T}_0|\text{vac}\rangle+\sum_{i=1}^la_{i,AB,n}\mathcal{T}_{\lambda,\text{vac}}=0+\mathcal{T}_{\lambda,\text{vac}}\sum_{i=1}^la_{i,AB,n},
\end{equation}
where we have defined the matrix element $\mathcal{T}_{\lambda_1,\lambda_2}:=\langle\lambda_1|\mathcal{T}_0|\lambda_2\rangle$. In other words,
\begin{equation}
	\mathcal{T}_{\lambda',\text{vac}}=\left[\mathcal{T}_{\lambda,\text{vac}},\sum_{i=1}^la_{i,AB,n}\right],
\end{equation}
where $\langle\lambda'|=\langle\lambda|a_{0,AB,n}$. Similarly, for any $\mathcal{T}_{\text{vac},\lambda}$ with $n<0$ we may also write
\begin{equation}
	\mathcal{T}_{\text{vac},\lambda'}=\left[\sum_{i=1}^la_{i,AB,n},\mathcal{T}_{\text{vac},\lambda}\right].
\end{equation}
With these commutation relations, one can in principle get any matrix element $\mathcal{T}_{\lambda_1,\lambda_2}$ from lower levels. It could then be possible that, under some non-trivial change of generators, this would help us verify and find the explicit contour integral form discussed in \S\ref{crystalRTT} although this process is still not clear in general.

\subsection{Harmonic Oscillator States}\label{oscillators}
We would like to find the action of the intertwining operator on some states for certain representations. One natural choice would be the space spanned by the set of $J_{AB}$ oscillators. We choose the normalization as $\mathcal{R}_{12}|\emptyset\rangle=|\emptyset\rangle$. Again, we will omit the labels in the states for brevity. Suppose $a_{AB,n>0}|\emptyset\rangle=|\emptyset\rangle$ and $a_{AB,0}|\emptyset\rangle=u_{AB}|\emptyset\rangle$ for any $A$, $B$ and some coefficient $u_{AB}$ \footnote{Here, we simply write $u_{AB}$ for brevity. A better notation would perhaps be $a_{AB,0}|\emptyset\rangle=(u_{AB}-v_{AB})|\emptyset\rangle$ such that $u_{AB}$ and $v_{AB}$ are for $a_{1,AB,0}$ and $a_{2,AB,0}$ respectively.\label{uvnotation}}.

For $n=-1$, we have\footnote{Notice that for $n\geq0$, \eqref{Ra-} is trivially satisfied.}
\begin{equation}
	\begin{split}
		&\mathcal{R}_{12}\left(\sum_{k=1}^{M+N}u_{kB}a_{Ak,-1}+\sum_{k=1}^{M+N}(-1)^{(|A|+|k|)(|k|+|B|)}u_{Ak}a_{kB,-1}+2\kappa a_{AB,-1}\right.\\
		&\left.-2\sum_{k=1}^{M+N}u_{Ak}a_{2,kB,-1}-2\sum_{k=1}^{M+N}u_{kB}a_{1,Ak,-1}+2\sum_{k=1}^{M+N}u_{Ak}a_{1,kB,-1}+2\sum_{k=1}^{M+N}u_{kB}a_{2,Ak,-1}\right)|\emptyset\rangle\\
		=&\left(\sum_{k=1}^{M+N}u_{kB}a_{Ak,-1}+\sum_{k=1}^{M+N}(-1)^{(|A|+|k|)(|k|+|B|)}u_{Ak}a_{kB,-1}-2\kappa a_{AB,-1}\right.\\
		&\left.+2\sum_{k=1}^{M+N}u_{Ak}a_{2,kB,-1}+2\sum_{k=1}^{M+N}u_{kB}a_{1,Ak,-1}-2\sum_{k=1}^{M+N}u_{Ak}a_{1,kB,-1}-2\sum_{k=1}^{M+N}u_{kB}a_{2,Ak,-1}\right)|\emptyset\rangle.
	\end{split}
\end{equation}
Therefore, we can combine all the sporadic $a_{1,2}$ into $a$ and write
\begin{equation}
	\begin{split}
		&\mathcal{R}_{12}\left(-\sum_{k\neq B}u_{kB}a_{Ak,-1}+\sum_{k\neq A}\left((-1)^{(|A|+|k|)(|k|+|B|)}+2\right)u_{Ak}a_{kB,-1}+(3u_{AA}-u_{BB}+2\kappa)a_{AB,-1}\right)|\emptyset\rangle\\
		=&\left(\sum_{k\neq B}3u_{kB}a_{Ak,-1}+\sum_{k\neq A}\left((-1)^{(|A|+|k|)(|k|+|B|)}-2\right)u_{Ak}a_{kB,-1}+(3u_{BB}-u_{AA}-2\kappa)a_{AB,-1}\right)|\emptyset\rangle.
	\end{split}
\end{equation}
As a result, to get the action of $\mathcal{R}_{12}$ on the states of level 1, we need to solve a set of $(M+N)\times(M+N)$ equations. To simplify this a bit, a natural assumption we can take would be $u_{AB}=u$ for any $A$, $B$.

The simplest example would be $\mathbb{C}^3$ which has been extensively studied in various literature such as \cite{Prochazka:2019dvu} with a dictionary for $\mathcal{W}_{1+\infty}$ and $\mathtt{Y}\left(\widehat{\mathfrak{gl}}_1\right)$. Therefore, let us consider the second simplest examples, that is, $\mathbb{C}\times\mathbb{C}^2/\mathbb{Z}_2$ and the conifold.

\subsubsection{Example 1: $\mathbb{C}\times\mathbb{C}^2/\mathbb{Z}_2$}\label{exCC2Z2}
For the case of $\mathbb{C}\times\mathbb{C}^2/\mathbb{Z}_2$, we have four equations, and all the modes are bosonic. It turns out that
\begin{align}
	&\mathcal{R}_{12}a_{11,-1}|\emptyset\rangle=A_1(u)a_{11,-1}|\emptyset\rangle+A_2(u)a_{12,-1}|\emptyset\rangle+A_3(u)a_{21,-1}|\emptyset\rangle+A_4(u)a_{22,-1}|\emptyset\rangle,\\
	&\mathcal{R}_{12}a_{12,-1}|\emptyset\rangle=A_2(u)a_{11,-1}|\emptyset\rangle+A_1(u)a_{12,-1}|\emptyset\rangle+A_4(u)a_{21,-1}|\emptyset\rangle+A_3(u)a_{22,-1}|\emptyset\rangle,\\
	&\mathcal{R}_{12}a_{21,-1}|\emptyset\rangle=A_3(u)a_{11,-1}|\emptyset\rangle+A_4(u)a_{12,-1}|\emptyset\rangle+A_1(u)a_{21,-1}|\emptyset\rangle+A_2(u)a_{22,-1}|\emptyset\rangle,\\
	&\mathcal{R}_{12}a_{22,-1}|\emptyset\rangle=A_4(u)a_{11,-1}|\emptyset\rangle+A_3(u)a_{12,-1}|\emptyset\rangle+A_2(u)a_{21,-1}|\emptyset\rangle+A_1(u)a_{22,-1}|\emptyset\rangle.
\end{align}
where
\begin{equation}
	\begin{split}
		&A_1(u)=\frac{-(5u^2-\kappa u-\kappa^2)(u+\kappa)}{(3u+\kappa)(2u+\kappa)(u-\kappa)},\quad A_2(u)=\frac{-u(u^2+4\kappa u+\kappa^2)}{(3u+\kappa)(2u+\kappa)(u-\kappa)},\\
		&A_3(u)=\frac{u(7u^2-\kappa^2)}{(3u+\kappa)(2u+\kappa)(u-\kappa)},\quad A_4(u)=\frac{u^2(5u+\kappa)}{(3u+\kappa)(2u+\kappa)(u-\kappa)}.
	\end{split}
\end{equation}

As $a_{AB,n}=a_{1,AB,n}-a_{2,AB,n}$, together with $\mathcal{R}_{12}(a_{1,AB,n}+a_{2,AB,n})=(a_{1,AB,n}+a_{2,AB,n})\mathcal{R}_{12}$, we get
\begin{equation}
	\begin{split}
		\mathcal{R}_{12}a_{1,11,-1}|\emptyset,\emptyset\rangle=&(B_1(u)a_{1,11,-1}-B_1'(u)a_{2,11,-1})|\emptyset,\emptyset\rangle+B_2(u)(a_{1,12,-1}-a_{2,12,-1})|\emptyset,\emptyset\rangle\\
		&+B_3(u)(a_{1,21,-1}-a_{2,21,-1})|\emptyset,\emptyset\rangle+B_4(u)(a_{1,22,-1}-a_{2,22,-1})|\emptyset,\emptyset\rangle,
	\end{split}
\end{equation}
\begin{equation}
	\begin{split}
		\mathcal{R}_{12}a_{1,12,-1}|\emptyset,\emptyset\rangle=&B_2(u)(a_{1,11,-1}-a_{2,11,-1})|\emptyset,\emptyset\rangle+(B_1(u)a_{1,12,-1}-B_1'(u)a_{2,12,-1})|\emptyset,\emptyset\rangle\\
		&+B_4(u)(a_{1,21,-1}-a_{2,21,-1})|\emptyset,\emptyset\rangle+B_3(u)(a_{1,22,-1}-a_{2,22,-1})|\emptyset,\emptyset\rangle,
	\end{split}
\end{equation}
\begin{equation}
	\begin{split}
		\mathcal{R}_{12}a_{1,21,-1}|\emptyset,\emptyset\rangle=&B_3(u)(a_{1,11,-1}-a_{2,11,-1})|\emptyset,\emptyset\rangle+B_4(u)(a_{1,12,-1}-a_{2,12,-1})|\emptyset,\emptyset\rangle\\
		&+(B_1(u)a_{1,21,-1}-B_1'(u)a_{2,21,-1})|\emptyset,\emptyset\rangle+B_2(u)(a_{1,22,-1}-a_{2,22,-1})|\emptyset,\emptyset\rangle,
	\end{split}
\end{equation}
\begin{equation}
	\begin{split}
		\mathcal{R}_{12}a_{1,22,-1}|\emptyset,\emptyset\rangle=&B_4(u)(a_{1,11,-1}-a_{2,11,-1})|\emptyset,\emptyset\rangle+B_3(u)(a_{1,12,-1}-a_{2,12,-1})|\emptyset,\emptyset\rangle\\
		&+B_2(u)(a_{1,21,-1}-a_{2,21,-1})|\emptyset,\emptyset\rangle+(B_1(u)a_{1,22,-1}-B_1'(u)a_{2,22,-1})|\emptyset,\emptyset\rangle,
	\end{split}
\end{equation}
where
\begin{equation}
	B_1(u)=\frac{u(u^2-5\kappa u-2\kappa^2)}{2(3u+\kappa)(2u+\kappa)(u-\kappa)},\quad B_1'(u)=A_1(u)-B_1(u)=-\frac{11u^3+3\kappa u^2-6\kappa^2u-2\kappa^3}{2(3u+\kappa)(2u+\kappa)(u-\kappa)},
\end{equation}
and $B_{k}(u)=A_{k}(u)/2$ ($k=2,3,4$). In particular, we find that $B_1(u)-B_1'(u)=1$. The actions of $\mathcal{R}_{12}a_{2,AB,-1}|\emptyset,\emptyset\rangle$ are completely the same with $a_{1,AB,-1}\leftrightarrow a_{2,AB,-1}$.

Now let us try to construct the generators as matrix elements from the $\mathcal{R}\mathcal{T}\mathcal{T}$ relation. The first generator would be
\begin{equation}
	\mathcal{H}(u)=\langle\emptyset|\mathcal{T}(u)|\emptyset\rangle.
\end{equation}
As $\mathcal{R}_{12}|\emptyset,\emptyset\rangle=|\emptyset,\emptyset\rangle$, following
\begin{equation}
	\langle\emptyset,\emptyset|\mathcal{R}_{12}(u-v)\mathcal{T}_1(u)\mathcal{T}_2(v)|\emptyset,\emptyset\rangle=\langle\emptyset,\emptyset|\mathcal{T}_2(v)\mathcal{T}_1(u)\mathcal{R}_{12}(u-v)|\emptyset,\emptyset\rangle,
\end{equation}
it is straightforward to see that
\begin{equation}
	\mathcal{H}(u)\mathcal{H}(v)=\mathcal{H}(v)\mathcal{H}(u).
\end{equation}
Then at level 1, we would have
\begin{equation}
	\mathcal{E}_{AB}(u)=\langle\emptyset|\mathcal{T}(u)a_{AB,-1}|\emptyset\rangle,\quad\mathcal{F}_{AB}(u)=\langle\emptyset|a_{AB,1}\mathcal{T}(u)|\emptyset\rangle.
\end{equation}
Again, using the $\mathcal{R}\mathcal{T}\mathcal{T}$ relation
\begin{equation}
	\langle\emptyset,\emptyset|\mathcal{R}_{12}\mathcal{T}_1\mathcal{T}_2a_{1,AB,-1}|\emptyset,\emptyset\rangle=\langle\emptyset,\emptyset|\mathcal{T}_2\mathcal{T}_1\mathcal{R}_{12}a_{1,AB,-1}|\emptyset,\emptyset\rangle,
\end{equation}
we get\footnote{Notice that the variable $u$ used in $\mathcal{R}_{12}a_{AB,-1}$ above is actually $u-v$ here. This is due to our notation for $a_{AB,0}|\emptyset\rangle=u|\emptyset\rangle$, where we could have named it $a_{AB,0}|\emptyset\rangle=(u-v)|\emptyset\rangle$. See also Footnote \ref{uvnotation}.}
\begin{equation}
	\begin{split}
		\mathcal{E}_{11}(u)\mathcal{H}(v)=&B_1(u-v)\mathcal{H}(v)\mathcal{E}_{11}(u)-B_1'(u-v)\mathcal{E}_{11}(v)\mathcal{H}(u)+B_2(u-v)(\mathcal{H}(v)\mathcal{E}_{12}(u)-\mathcal{E}_{12}(v)\mathcal{H}(u))\\
		&+B_3(u-v)(\mathcal{H}(v)\mathcal{E}_{21}(u)-\mathcal{E}_{21}(v)\mathcal{H}(u))+B_4(u-v)(\mathcal{H}(v)\mathcal{E}_{22}(u)-\mathcal{E}_{22}(v)\mathcal{H}(u)),
	\end{split}
\end{equation}
\begin{equation}
	\begin{split}
		\mathcal{E}_{12}(u)\mathcal{H}(v)=&B_1(u-v)\mathcal{H}(v)\mathcal{E}_{12}(u)-B_1'(u-v)\mathcal{E}_{12}(v)\mathcal{H}(u)+B_2(u-v)(\mathcal{H}(v)\mathcal{E}_{11}(u)-\mathcal{E}_{11}(v)\mathcal{H}(u))\\
		&+B_3(u-v)(\mathcal{H}(v)\mathcal{E}_{22}(u)-\mathcal{E}_{22}(v)\mathcal{H}(u))+B_4(u-v)(\mathcal{H}(v)\mathcal{E}_{21}(u)-\mathcal{E}_{21}(v)\mathcal{H}(u)),
	\end{split}
\end{equation}
\begin{equation}
	\begin{split}
		\mathcal{E}_{21}(u)\mathcal{H}(v)=&B_1(u-v)\mathcal{H}(v)\mathcal{E}_{21}(u)-B_1'(u-v)\mathcal{E}_{21}(v)\mathcal{H}(u)+B_2(u-v)(\mathcal{H}(v)\mathcal{E}_{22}(u)-\mathcal{E}_{22}(v)\mathcal{H}(u))\\
		&+B_3(u-v)(\mathcal{H}(v)\mathcal{E}_{11}(u)-\mathcal{E}_{11}(v)\mathcal{H}(u))+B_4(u-v)(\mathcal{H}(v)\mathcal{E}_{12}(u)-\mathcal{E}_{12}(v)\mathcal{H}(u)),
	\end{split}
\end{equation}
\begin{equation}
	\begin{split}
		\mathcal{E}_{22}(u)\mathcal{H}(v)=&B_1(u-v)\mathcal{H}(v)\mathcal{E}_{22}(u)-B_1'(u-v)\mathcal{E}_{22}(v)\mathcal{H}(u)+B_2(u-v)(\mathcal{H}(v)\mathcal{E}_{21}(u)-\mathcal{E}_{21}(v)\mathcal{H}(u))\\
		&+B_3(u-v)(\mathcal{H}(v)\mathcal{E}_{12}(u)-\mathcal{E}_{12}(v)\mathcal{H}(u))+B_4(u-v)(\mathcal{H}(v)\mathcal{E}_{11}(u)-\mathcal{E}_{11}(v)\mathcal{H}(u)).
	\end{split}
\end{equation}
Similarly,
\begin{equation}
	\begin{split}
		\mathcal{H}(v)\mathcal{F}_{11}(u)=&B_1(u-v)\mathcal{F}_{11}(u)\mathcal{H}(v)-B_1'(u-v)\mathcal{H}(u)\mathcal{F}_{11}(v)+B_2(u-v)(\mathcal{F}_{12}(u)\mathcal{H}(v)-\mathcal{H}(u)\mathcal{F}_{12}(v))\\
		&+B_3(u-v)(\mathcal{F}_{21}(u)\mathcal{H}(v)-\mathcal{H}(u)\mathcal{F}_{21}(v))+B_4(u-v)(\mathcal{F}_{22}(u)\mathcal{H}(v)-\mathcal{H}(u)\mathcal{F}_{22}(v)),
	\end{split}
\end{equation}
\begin{equation}
	\begin{split}
		\mathcal{H}(v)\mathcal{F}_{12}(u)=&B_1(u-v)\mathcal{F}_{12}(u)\mathcal{H}(v)-B_1'(u-v)\mathcal{H}(u)\mathcal{F}_{12}(v)+B_2(u-v)(\mathcal{F}_{11}(u)\mathcal{H}(v)-\mathcal{H}(u)\mathcal{F}_{11}(v))\\
		&+B_3(u-v)(\mathcal{F}_{22}(u)\mathcal{H}(v)-\mathcal{H}(u)\mathcal{F}_{22}(v))+B_4(u-v)(\mathcal{F}_{21}(u)\mathcal{H}(v)-\mathcal{H}(u)\mathcal{F}_{21}(v)),
	\end{split}
\end{equation}
\begin{equation}
	\begin{split}
		\mathcal{H}(v)\mathcal{F}_{21}(u)=&B_1(u-v)\mathcal{F}_{21}(u)\mathcal{H}(v)-B_1'(u-v)\mathcal{H}(u)\mathcal{F}_{21}(v)+B_2(u-v)(\mathcal{F}_{22}(u)\mathcal{H}(v)-\mathcal{H}(u)\mathcal{F}_{22}(v))\\
		&+B_3(u-v)(\mathcal{F}_{11}(u)\mathcal{H}(v)-\mathcal{H}(u)\mathcal{F}_{11}(v))+B_4(u-v)(\mathcal{F}_{12}(u)\mathcal{H}(v)-\mathcal{H}(u)\mathcal{F}_{12}(v)),
	\end{split}
\end{equation}
\begin{equation}
	\begin{split}
		\mathcal{H}(v)\mathcal{F}_{22}(u)=&B_1(u-v)\mathcal{F}_{22}(u)\mathcal{H}(v)-B_1'(u-v)\mathcal{H}(u)\mathcal{F}_{22}(v)+B_2(u-v)(\mathcal{F}_{21}(u)\mathcal{H}(v)-\mathcal{H}(u)\mathcal{F}_{21}(v))\\
		&+B_3(u-v)(\mathcal{F}_{12}(u)\mathcal{H}(v)-\mathcal{H}(u)\mathcal{F}_{12}(v))+B_4(u-v)(\mathcal{F}_{11}(u)\mathcal{H}(v)-\mathcal{H}(u)\mathcal{F}_{11}(v)).
	\end{split}
\end{equation}
As the relations for $\mathcal{F}_{AB}(u)$ are completely analogous, we shall only discuss those for $\mathcal{E}_{AB}(u)$ explicitly below.

Now, let us take $\kappa=-2\epsilon_3$. In particular, we observe that
\begin{equation}
	\sum_{A,B=1}^2\mathcal{E}_{AB}(u)\mathcal{H}(v)=\frac{u-v}{u-v-\epsilon_3}\mathcal{H}(v)\sum_{A,B=1}^2\mathcal{E}_{AB}(u)-\frac{\epsilon_3}{u-v-\epsilon_3}\sum_{A,B=1}^2\mathcal{E}_{AB}(v)\mathcal{H}(u),\label{obs1}
\end{equation}
and
\begin{equation}
	\sum_{A,B=1}^2(-1)^{A+B}\mathcal{E}_{AB}(u)\mathcal{H}(v)=\sum_{A,B=1}^2(-1)^{A+B}\mathcal{E}_{AB}(v)\mathcal{H}(u).
\end{equation}
Moreover, we have
\begin{equation}
	\begin{split}
		2(\mathcal{E}_{11}(u)+\mathcal{E}_{22}(u))\mathcal{H}(v)=&\frac{u-v}{u-v-\epsilon_3}\mathcal{H}(v)(\mathcal{E}_{11}(u)+\mathcal{E}_{22}(u))-\frac{u-v-2\epsilon_3}{u-v-\epsilon_3}(\mathcal{E}_{11}(v)+\mathcal{E}_{22}(v))\mathcal{H}(u)\\
		&+\frac{u-v}{u-v-\epsilon_3}\left(\mathcal{H}(v)(\mathcal{E}_{12}(u)+\mathcal{E}_{21}(u))-(\mathcal{E}_{12}(v)+\mathcal{E}_{21}(v))\mathcal{H}(u)\right),
	\end{split}
\end{equation}
and
\begin{equation}
	\begin{split}
		2(\mathcal{E}_{12}(u)+\mathcal{E}_{21}(u))\mathcal{H}(v)=&\frac{u-v}{u-v-\epsilon_3}\mathcal{H}(v)(\mathcal{E}_{12}(u)+\mathcal{E}_{21}(u))-\frac{u-v-2\epsilon_3}{u-v-\epsilon_3}(\mathcal{E}_{12}(v)+\mathcal{E}_{21}(v))\mathcal{H}(u)\\
		&+\frac{u-v}{u-v-\epsilon_3}\left(\mathcal{H}(v)(\mathcal{E}_{11}(u)+\mathcal{E}_{22}(u))-(\mathcal{E}_{11}(v)+\mathcal{E}_{22}(v))\mathcal{H}(u)\right).
	\end{split}
\end{equation}
From \eqref{obs1}, it is straightforward to see that this contains the algebra for $\mathbb{C}^3$ as we would expect since\footnote{The derivation of such relations is completely similar to the examples discussed in this section. In terms of $\mathtt{YB}\left(\widehat{\mathfrak{gl}}_1\right)$, we have $h(u)h(v)=h(v)h(u)$ and
	\begin{equation}
		(h(u)e(u))h(v)=\frac{u-v}{u-v-\epsilon_3}h(v)(h(u)e(u))-\frac{\epsilon_3}{u-v-\epsilon_3}(h(v)e(v))h(u).\nonumber
	\end{equation}	
	Strictly speaking, one should also check the $ee$ relations at a higher level.}
\begin{equation}
	\mathcal{E}(u)\mathcal{H}(v)=\frac{u-v}{u-v-\epsilon_3}\mathcal{H}(v)\mathcal{E}(u)-\frac{\epsilon_3}{u-v-\epsilon_3}\mathcal{E}(v)\mathcal{H}(u)
\end{equation}
(and $\mathcal{H}(u)\mathcal{H}(v)=\mathcal{H}(v)\mathcal{H}(u)$) for the case associated to $\widehat{\mathfrak{gl}}_1$.

\subsubsection{Example 2: Conifold}\label{exconifold}
Now let us consider the case of conifold which corresponds to $\widehat{\mathfrak{gl}}_{1|1}$. At level 1, we have
\begin{equation}
	\mathcal{R}_{12}a_{11,-1}|\emptyset\rangle=A_1(u)a_{11,-1}|\emptyset\rangle+A_2(u)a_{12,-1}|\emptyset\rangle+A_3(u)a_{21,-1}|\emptyset\rangle+A_4(u)a_{22,-1}|\emptyset\rangle,
\end{equation}
\begin{equation}
	\mathcal{R}_{12}a_{12,-1}|\emptyset\rangle=A_5(u)a_{11,-1}|\emptyset\rangle+A_6(u)a_{12,-1}|\emptyset\rangle+A_7(u)a_{21,-1}|\emptyset\rangle+A_8(u)a_{22,-1}|\emptyset\rangle,
\end{equation}
\begin{equation}
	\mathcal{R}_{12}a_{21,-1}|\emptyset\rangle=A_8(u)a_{11,-1}|\emptyset\rangle+A_7(u)a_{12,-1}|\emptyset\rangle+A_6(u)a_{21,-1}|\emptyset\rangle+A_5(u)a_{22,-1}|\emptyset\rangle,
\end{equation}
\begin{equation}
	\mathcal{R}_{12}a_{22,-1}|\emptyset\rangle=A_4(u)a_{11,-1}|\emptyset\rangle+A_3(u)a_{12,-1}|\emptyset\rangle+A_2(u)a_{21,-1}|\emptyset\rangle+A_1(u)a_{22,-1}|\emptyset\rangle,
\end{equation}
where
\begin{equation}
	\begin{split}
		&A_1(u)=\frac{-u^3-3\kappa u^2+\kappa^2u+\kappa^3}{(u+\kappa)(u^2-2\kappa u-\kappa^2)},\quad A_2(u)=\frac{-u(2u+\kappa)}{u^2-2\kappa u-\kappa^2},\\
		&A_3(u)=\frac{u(2u+\kappa)}{u^2-2\kappa u-\kappa^2},\quad A_4(u)=\frac{2u^3}{(u+\kappa)(u^2-2\kappa u-\kappa^2)},\\
		&A_5(u)=\frac{-u(2u^3+3\kappa u^2+4\kappa^2u+\kappa^3)}{(u+\kappa)^2(u^2-2\kappa u-\kappa^2)},\quad A_6(u)=\frac{-3u^3-5\kappa u^2+\kappa^2u+\kappa^3}{(u+\kappa)(u^2-2\kappa u-\kappa^2)},\\
		&A_7(u)=\frac{2u^2(2u+\kappa)}{(u+\kappa)(u^2-2\kappa u-\kappa^2)},\quad A_8(u)=\frac{u(2u+\kappa)(u^2+2\kappa u-\kappa^2)}{(u+\kappa)^2(u^2-2\kappa u-\kappa^2)}.
	\end{split}
\end{equation}
Then
\begin{equation}
	\begin{split}
		\mathcal{R}_{12}a_{1,11,-1}|\emptyset,\emptyset\rangle=&(B_1(u)a_{1,11,-1}-B_1'(u)a_{2,11,-1})|\emptyset,\emptyset\rangle+B_2(u)(a_{1,12,-1}-a_{2,12,-1})|\emptyset,\emptyset\rangle\\
		&+B_3(u)(a_{1,21,-1}-a_{2,21,-1})|\emptyset,\emptyset\rangle+B_4(u)(a_{1,22,-1}-a_{2,22,-1})|\emptyset,\emptyset\rangle,
	\end{split}
\end{equation}
\begin{equation}
	\begin{split}
		\mathcal{R}_{12}a_{1,12,-1}|\emptyset,\emptyset\rangle=&B_5(u)(a_{1,11,-1}-a_{2,11,-1})|\emptyset,\emptyset\rangle+(B_6(u)a_{1,12,-1}-B_6'(u)a_{2,12,-1})|\emptyset,\emptyset\rangle\\
		&+B_7(u)(a_{1,21,-1}-a_{2,21,-1})|\emptyset,\emptyset\rangle+B_8(u)(a_{1,22,-1}-a_{2,22,-1})|\emptyset,\emptyset\rangle,
	\end{split}
\end{equation}
\begin{equation}
	\begin{split}
		\mathcal{R}_{12}a_{1,21,-1}|\emptyset,\emptyset\rangle=&B_8(u)(a_{1,11,-1}-a_{2,11,-1})|\emptyset,\emptyset\rangle+B_7(u)(a_{1,12,-1}-a_{2,12,-1})|\emptyset,\emptyset\rangle\\
		&+(B_6(u)a_{1,21,-1}-B_6'(u)a_{2,21,-1})|\emptyset,\emptyset\rangle+B_5(u)(a_{1,22,-1}-a_{2,22,-1})|\emptyset,\emptyset\rangle,
	\end{split}
\end{equation}
\begin{equation}
	\begin{split}
		\mathcal{R}_{12}a_{1,22,-1}|\emptyset,\emptyset\rangle=&B_4(u)(a_{1,11,-1}-a_{2,11,-1})|\emptyset,\emptyset\rangle+B_3(u)(a_{1,12,-1}-a_{2,12,-1})|\emptyset,\emptyset\rangle\\
		&+B_2(u)(a_{1,21,-1}-a_{2,21,-1})|\emptyset,\emptyset\rangle+(B_1(u)a_{1,22,-1}-B_1'(u)a_{2,22,-1})|\emptyset,\emptyset\rangle,
	\end{split}
\end{equation}
where
\begin{equation}
	\begin{split}
		&B_1(u)=\frac{-\kappa u(2u+\kappa)}{(u+\kappa)(u^2-2\kappa u-\kappa^2)},\quad B_1'(u)=A_1(u)-B_1(u)=\frac{-u^3-\kappa u^2+2\kappa^2u+\kappa^3}{(u+\kappa)(u^2-2\kappa u-\kappa^2)},\\
		&B_6(u)=\frac{-u(u^2+3\kappa u+\kappa^2)}{(u+\kappa)(u^2-2\kappa u-\kappa^2)},\quad B_6'(u)=A_6(u)-B_6(u)=\frac{-2u^3-2\kappa u+2\kappa^2u+\kappa^3)}{(u+\kappa)(u^2-2\kappa u-\kappa^2)},
	\end{split}
\end{equation}
and $B_{k}(u)=A_{k}(u)/2$ for the remaining $k$. In particular, we find that $B_1(u)-B_1'(u)=1$ and $B_6(u)-B_6'(u)=1$. The actions of $\mathcal{R}_{12}a_{2,AB,-1}|\emptyset,\emptyset\rangle$ are completely the same with $a_{1,AB,-1}\leftrightarrow a_{2,AB,-1}$.

Again, let us try to construct the generators as matrix elements from the $\mathcal{R}\mathcal{T}\mathcal{T}$ relation. The first generator would be
\begin{equation}
	\mathcal{H}(u)=\langle\emptyset|\mathcal{T}(u)|\emptyset\rangle.
\end{equation}
Following the $\mathcal{R}\mathcal{T}\mathcal{T}$ relation, it is straightforward to see that
\begin{equation}
	\mathcal{H}(u)\mathcal{H}(v)=\mathcal{H}(v)\mathcal{H}(u).
\end{equation}
At level 1, we would still have
\begin{equation}
	\mathcal{E}_{AB}(u)=\langle\emptyset|\mathcal{T}(u)a_{AB,-1}|\emptyset\rangle,\quad\mathcal{F}_{AB}(u)=\langle\emptyset|a_{AB,1}\mathcal{T}(u)|\emptyset\rangle.
\end{equation}
Therefore,
\begin{equation}
	\begin{split}
		\mathcal{E}_{11}(u)\mathcal{H}(v)=&B_1(u-v)\mathcal{H}(v)\mathcal{E}_{11}(u)-B_1'(u-v)\mathcal{E}_{11}(v)\mathcal{H}(u)+B_2(u-v)(\mathcal{H}(v)\mathcal{E}_{12}(u)-\mathcal{E}_{12}(v)\mathcal{H}(u))\\
		&+B_3(u-v)(\mathcal{H}(v)\mathcal{E}_{21}(u)-\mathcal{E}_{21}(v)\mathcal{H}(u))+B_4(u-v)(\mathcal{H}(v)\mathcal{E}_{22}(u)-\mathcal{E}_{22}(v)\mathcal{H}(u)),
	\end{split}
\end{equation}
\begin{equation}
	\begin{split}
		\mathcal{E}_{12}(u)\mathcal{H}(v)=&B_6(u-v)\mathcal{H}(v)\mathcal{E}_{12}(u)-B_6'(u-v)\mathcal{E}_{12}(v)\mathcal{H}(u)+B_5(u-v)(\mathcal{H}(v)\mathcal{E}_{11}(u)-\mathcal{E}_{11}(v)\mathcal{H}(u))\\
		&+B_7(u-v)(\mathcal{H}(v)\mathcal{E}_{22}(u)-\mathcal{E}_{22}(v)\mathcal{H}(u))+B_8(u-v)(\mathcal{H}(v)\mathcal{E}_{21}(u)-\mathcal{E}_{21}(v)\mathcal{H}(u)),
	\end{split}
\end{equation}
\begin{equation}
	\begin{split}
		\mathcal{E}_{21}(u)\mathcal{H}(v)=&B_6(u-v)\mathcal{H}(v)\mathcal{E}_{21}(u)-B_6'(u-v)\mathcal{E}_{21}(v)\mathcal{H}(u)+B_5(u-v)(\mathcal{H}(v)\mathcal{E}_{22}(u)-\mathcal{E}_{22}(v)\mathcal{H}(u))\\
		&+B_7(u-v)(\mathcal{H}(v)\mathcal{E}_{11}(u)-\mathcal{E}_{11}(v)\mathcal{H}(u))+B_8(u-v)(\mathcal{H}(v)\mathcal{E}_{12}(u)-\mathcal{E}_{12}(v)\mathcal{H}(u)),
	\end{split}
\end{equation}
\begin{equation}
	\begin{split}
		\mathcal{E}_{22}(u)\mathcal{H}(v)=&B_1(u-v)\mathcal{H}(v)\mathcal{E}_{22}(u)-B_1'(u-v)\mathcal{E}_{22}(v)\mathcal{H}(u)+B_2(u-v)(\mathcal{H}(v)\mathcal{E}_{21}(u)-\mathcal{E}_{21}(v)\mathcal{H}(u))\\
		&+B_3(u-v)(\mathcal{H}(v)\mathcal{E}_{12}(u)-\mathcal{E}_{12}(v)\mathcal{H}(u))+B_4(u-v)(\mathcal{H}(v)\mathcal{E}_{11}(u)-\mathcal{E}_{11}(v)\mathcal{H}(u)).
	\end{split}
\end{equation}
Similarly,
\begin{equation}
	\begin{split}
		\mathcal{H}(v)\mathcal{F}_{11}(u)=&B_1(u-v)\mathcal{F}_{11}(u)\mathcal{H}(v)-B_1'(u-v)\mathcal{H}(u)\mathcal{F}_{11}(v)+B_2(u-v)(\mathcal{F}_{12}(u)\mathcal{H}(v)-\mathcal{H}(u)\mathcal{F}_{12}(v))\\
		&+B_3(u-v)(\mathcal{F}_{21}(u)\mathcal{H}(v)-\mathcal{H}(u)\mathcal{F}_{21}(v))+B_4(u-v)(\mathcal{F}_{22}(u)\mathcal{H}(v)-\mathcal{H}(u)\mathcal{F}_{22}(v)),
	\end{split}
\end{equation}
\begin{equation}
	\begin{split}
		\mathcal{H}(v)\mathcal{F}_{12}(u)=&B_6(u-v)\mathcal{F}_{12}(u)\mathcal{H}(v)-B_6'(u-v)\mathcal{H}(u)\mathcal{F}_{12}(v)+B_5(u-v)(\mathcal{F}_{11}(u)\mathcal{H}(v)-\mathcal{H}(u)\mathcal{F}_{11}(v))\\
		&+B_7(u-v)(\mathcal{F}_{22}(u)\mathcal{H}(v)-\mathcal{H}(u)\mathcal{F}_{22}(v))+B_8(u-v)(\mathcal{F}_{21}(u)\mathcal{H}(v)-\mathcal{H}(u)\mathcal{F}_{21}(v)),
	\end{split}
\end{equation}
\begin{equation}
	\begin{split}
		\mathcal{H}(v)\mathcal{F}_{21}(u)=&B_6(u-v)\mathcal{F}_{21}(u)\mathcal{H}(v)-B_6'(u-v)\mathcal{H}(u)\mathcal{F}_{21}(v)+B_5(u-v)(\mathcal{F}_{22}(u)\mathcal{H}(v)-\mathcal{H}(u)\mathcal{F}_{22}(v))\\
		&+B_7(u-v)(\mathcal{F}_{11}(u)\mathcal{H}(v)-\mathcal{H}(u)\mathcal{F}_{11}(v))+B_8(u-v)(\mathcal{F}_{12}(u)\mathcal{H}(v)-\mathcal{H}(u)\mathcal{F}_{12}(v)),
	\end{split}
\end{equation}
\begin{equation}
	\begin{split}
		\mathcal{H}(v)\mathcal{F}_{22}(u)=&B_1(u-v)\mathcal{F}_{22}(u)\mathcal{H}(v)-B_1'(u-v)\mathcal{H}(u)\mathcal{F}_{22}(v)+B_2(u-v)(\mathcal{F}_{21}(u)\mathcal{H}(v)-\mathcal{H}(u)\mathcal{F}_{21}(v))\\
		&+B_3(u-v)(\mathcal{F}_{12}(u)\mathcal{H}(v)-\mathcal{H}(u)\mathcal{F}_{12}(v))+B_4(u-v)(\mathcal{F}_{11}(u)\mathcal{H}(v)-\mathcal{H}(u)\mathcal{F}_{11}(v)).
	\end{split}
\end{equation}
As the relations for $\mathcal{F}_{AB}(u)$ are completely analogous, we shall only discuss those for $\mathcal{E}_{AB}(u)$ explicitly below.

Now, let us take $\kappa=-\epsilon_3$. In particular, we observe that
\begin{equation}
	(\mathcal{E}_{11}(u)+\mathcal{E}_{22}(u))\mathcal{H}(v)=\frac{u-v}{u-v-\epsilon_3}\mathcal{H}(v)(\mathcal{E}_{11}(u)+\mathcal{E}_{22}(u))-\frac{\epsilon_3}{u-v-\epsilon_3}(\mathcal{E}_{11}(v)+\mathcal{E}_{22}(v))\mathcal{H}(u),\label{obs2}
\end{equation}
and
\begin{equation}
	\begin{split}
		(\mathcal{E}_{12}(u)+\mathcal{E}_{21}(u))\mathcal{H}(v)=&\frac{u-v}{u-v-\epsilon_3}\mathcal{H}(v)(\mathcal{E}_{12}(u)+\mathcal{E}_{21}(u))-\frac{\epsilon_3}{u-v-\epsilon_3}(\mathcal{E}_{12}(v)+\mathcal{E}_{21}(v))\mathcal{H}(u)\\
		&-\frac{\epsilon_3(u-v)}{(u-v-\epsilon_3)^2}\left(\mathcal{H}(v)(\mathcal{E}_{11}(u)+\mathcal{E}_{22}(u))-(\mathcal{E}_{11}(v)+\mathcal{E}_{22}(v))\mathcal{H}(u)\right).
	\end{split}
\end{equation}
In particular, \eqref{obs2} seems to give a copy of $\mathtt{Y}\left(\widehat{\mathfrak{gl}}_1\right)$.

\subsection{Highest Weight States}\label{Verma}
In general, it would often be more useful to consider the highest weight states when studying the representation theory of the algebra. Recall that the generators are obtained from $\mathcal{L}_1\mathcal{L}_2\dots\mathcal{L}_k=\sum\limits_{i=0}^kU_{(i)}(\kappa\partial)^{k-i}$. The OPEs of $U_{(i),AB}$ (at low levels) have been worked out in \cite{Eberhardt:2019xmf,Rapcak:2019wzw}. If we consider the mode expansion
\begin{equation}
	U_{(s),AB}(z)=\sum_{m\in\mathbb{Z}}\frac{U_{(s),AB,m}}{z^{m+i}},
\end{equation}
then we can get the commutation relations of the modes from
\begin{equation}
	\left[U_{(r),AB,m},U_{(s),CD,n}\right\}=\frac{1}{(2\pi i)^2}\oint_0\text{d}w\oint_w\text{d}zz^{m+r-1}w^{n+s-1}U_{(r),AB,m}(z)U_{(s),CD,n}(w).
\end{equation}
Here, we list the following three relations which will be useful later:
\begin{equation}
	\begin{split}
		\left[U_{(1),AB,m},U_{(1),CD,n}\right\}=&k\delta_{n,-m}n\left((-1)^{|B||C|}\kappa\delta_{AD}\delta_{CB}+\delta_{AB}\delta_{CD}\right)\\
		&+(-1)^{|A||B|+|C||D|+|C||B|}\delta_{AD}U_{(1),CB,m+n}-(-1)^{|B||C|}\delta_{CB}U_{(1),AD,m+n},
	\end{split}
\end{equation}
\begin{equation}
	\begin{split}
		\left[U_{(1),AB,m},U_{(2),CD,n}\right\}=&k(k-1)\frac{m(m-1)}{2}\kappa\delta_{n,-m}n\left((-1)^{|B||C|}\kappa\delta_{AD}\delta_{CB}+\delta_{AB}\delta_{CD}\right)\\
		&+m(k-1)\left((-1)^{|A||B|+|C||D|+|C||B|}\kappa\delta_{AD}U_{(1),CB,m+n}+\delta_{AB}U_{(1),CD,m+n}\right)\\
		&+(-1)^{|A||B|+|C||D|+|C||B|}\delta_{AD}U_{(2),CB,m+n}-(-1)^{|B||C|}\delta_{CB}U_{(2),AD,m+n},
	\end{split}
\end{equation}
\begin{equation}
	\begin{split}
		\left[U_{(1),AB,m},U_{(3),CD,n}\right\}=&k(k-1)(k-2)\frac{m(m-1)(m-2)}{6}\kappa^2\delta_{n,-m}n\left((-1)^{|B||C|}\kappa\delta_{AD}\delta_{CB}+\delta_{AB}\delta_{CD}\right)\\
		&+\frac{m(m-1)}{2}(k-1)(k-2)\kappa\\
		&\left((-1)^{|A||B|+|C||D|+|C||B|}\kappa\delta_{AD}U_{(1),CB,m+n}+\delta_{AB}U_{(1),CD,m+n}\right)\\
		&+m(k-2)\left((-1)^{|A||B|+|C||D|+|C||B|}\kappa\delta_{AD}U_{(2),CB,m+n}+\delta_{AB}U_{(2),CD,m+n}\right)\\
		&+(-1)^{|A||B|+|C||D|+|C||B|}\delta_{AD}U_{(3),CB,m+n}-(-1)^{|B||C|}\delta_{CB}U_{(3),AD,m+n}.
	\end{split}
\end{equation}
Notice that the first relation reduces to the one for $a_{AB,n}$ when $k=1$ as expected.

As studied in \cite{Eberhardt:2019xmf}, a highest weight state of the algebra should then satisfy $U_{(s),AB,m}|z\rangle=0$ for $m>0$ and all possible $s,A,B$, as well as $U_{(1),AB,0}|z\rangle=0$ for $A<B$ such that it is also of highest weight for the global $\mathfrak{gl}(M|N)$ subalgebra. Moreover, $U_{(s),AA,0}|z\rangle=z|z\rangle$ for the Cartan generators. The Verma module can then be obtained by acting the negative modes and $U_{(s),AB,0}$ with $A>B$ on the highest weight state. With the OPEs taken into account, the conditions of a highest weight state are equivalent to\footnote{Strictly speaking, only the $\mathcal{W}_{M|0\times\infty}$ cases were considered in \cite{Eberhardt:2019xmf}, but we expect this to hold for any general $\mathcal{W}_{M|N\times\infty}$.}
\begin{equation}
	\begin{split}
		&U_{(1),AB,0}|z\rangle=0\quad(A=1,\dots,M+N-1\text{ and }B=A+1),\\
		&U_{(1),11,1}|z\rangle=0,\\
		&U_{(s),(M+N)1,1}|z\rangle=0\quad(s=1,2,3).
	\end{split}
\end{equation}

Here, we shall consider a special example, namely $\mathbb{C}\times\mathbb{C}^2/\mathbb{Z}_2$. (The conifold case as the other special example should be completely analogous where the extra signs from the $\mathbb{Z}_2$-grading would not affect the final results). Using the above commutation relations for $U_{(s),AB,m}$, we find that
\begin{equation}
	\begin{split}
		&U_{(1),12,0}U_{(1),21,0}|z\rangle=0+U_{(1),21,0}U_{(1),12,0}|z\rangle=0,\\
		&U_{(1),11,1}U_{(1),21,0}|z\rangle=U_{(1),21,0}U_{(1),11,1}|z\rangle=0,\\
		&U_{(s),21,1}U_{(1),21,0}|z\rangle=U_{(1),21,0}U_{(s),21,1}|z\rangle=0\quad(s=1,2,3).
	\end{split}
\end{equation}
Since we want a unique highest weight state, $U_{(1),21,0}|z\rangle$ should be a null vector, and we would consider the module quotienting out $U_{(1),21,0}|z\rangle=0$ in this case.

As $U_{(1),AB,n}$ is $a_{AB,n}$ when $k=1$, we may also consider the actions of $a_{AB,n}$ on the highest weight state/Verma module. Acting \eqref{Ra-} on $|z\rangle$ with $\mathcal{R}_{12}|z\rangle=|z\rangle$, when $n=-1$, we find that
\begin{equation}
	\mathcal{R}_{12}a_{1,AB,-1}|z\rangle=\frac{z}{z+\kappa}a_{1,AB,-1}|z\rangle-\frac{-\kappa}{z+\kappa}a_{2,AB,-1}|z\rangle
\end{equation}
for any $A$, $B$. Take $|z\rangle=|u,v\rangle$, $z=u-v$ and $\kappa=-\epsilon_3$. We have
\begin{equation}
	\mathcal{R}_{12}a_{1,AB,-1}|u,v\rangle=\frac{u-v}{u-v-\epsilon_3}a_{1,AB,-1}|u,v\rangle-\frac{\epsilon_3}{u-v-\epsilon_3}a_{2,AB,-1}|u,v\rangle.
\end{equation}
Likewise, the result of $\mathcal{R}_{12}a_{2,AB,-1}$ is simply $a_{1,AB,-1}\leftrightarrow a_{2,AB,-1}$. In particular, we may take any linear combination of $a_{i,AB,-1}$ as $\alpha_{(i),n}=\frac{1}{\sum\limits_{A,B}N_{AB}}\left(\sum\limits_{A,B}N_{AB}a_{i,AB,n}\right)$ such that
\begin{equation}
	\mathcal{R}_{12}\alpha_{(1),-1}|u,v\rangle=\frac{u-v}{u-v-\epsilon_3}\alpha_{(1),-1}|u,v\rangle-\frac{\epsilon_3}{u-v-\epsilon_3}\alpha_{(2),-1}|u,v\rangle.
\end{equation}
Let us consider the generators\footnote{Notice that the generators are not the same as the ones discussed in the previous subsection although we are using the same letters. Again, as the relations for $\mathcal{F}_{\alpha}$ would be similar, we shall only explicitly discuss $\mathcal{E}_{\alpha}$ in the followings.}
\begin{equation}
	\mathcal{H}(u)=\langle u|\mathcal{T}(u)|u\rangle,\quad\mathcal{E}_{\alpha}(u)=\langle u|\mathcal{T}(u)\alpha_{-1}|u\rangle,\quad\mathcal{F}_{\alpha}(u)=\langle u|\alpha_{1}\mathcal{T}(u)|u\rangle.
\end{equation}
Using the $\mathcal{RTT}$ relation, we have
\begin{equation}
	\mathcal{H}(u)\mathcal{H}(v)=\mathcal{H}(v)\mathcal{H}(u),
\end{equation}
and
\begin{equation}
	\mathcal{H}(u)\mathcal{E}_{\alpha}(v)=\frac{u-v}{u-v-\epsilon_3}\mathcal{E}_{\alpha}(v)\mathcal{H}(u)-\frac{\epsilon_3}{u-v-\epsilon_3}\mathcal{H}(v)\mathcal{E}_{\alpha}(u)
\end{equation}
from the matrix element between $\langle u,v|$ and $\alpha_{(2),-1}|u,v\rangle$. Recall that in the YB algebra, we can write $h(u)=\prod\limits_{a\in Q_0}h^{(a)}(u)$, and then for any $e^{(a)}(u)$, we have
\begin{equation}
	\begin{split}
		&h(u)h(v)=h(v)h(u),\\
		&h(u)\left(h(v)e^{(a)}(v)\right)=\frac{u-v}{u-v-\epsilon_3}\left(h(v)e^{(a)}(v)\right)h(u)-\frac{\epsilon_3}{u-v-\epsilon_3}h(v)\left(h(u)e^{(a)}(u)\right).
	\end{split}
\end{equation}
Moreover, we can consider
\begin{equation}
	\langle u,v|\beta_{(2),1}\mathcal{R}_{12}\mathcal{T}_1\mathcal{T}_2\alpha_{(1),-1}|u,v\rangle=\langle u,v|\beta_{(2),1}\mathcal{T}_2\mathcal{T}_1\mathcal{R}_{12}\alpha_{(1),-1}|u,v\rangle,
\end{equation}
where $\beta_{(i),n}$ is a different linear combination of $a_{i,AB,n}$, and assume $\mathcal{T}_{\beta_1,\alpha_{-1}}=\langle u|\beta_1\mathcal{T}\alpha_{-1}|u\rangle=0$. Using the actions of $\mathcal{R}_{12}$ as well as the above current relations for $\mathcal{H}$, $\mathcal{E}_{\alpha}$, $\mathcal{F}_{\alpha}$, we get
\begin{equation}
	\left[\mathcal{E}_{\alpha}(u),\mathcal{F}_{\alpha}(v)\right]=\delta_{\alpha\beta}\frac{\varPsi_{\alpha}(u)-\varPsi_{\alpha}(v)}{u-v},
\end{equation}
where
\begin{equation}
	\varPsi_{\alpha}:=\epsilon_3\left(\mathcal{T}_{\alpha_1,\alpha_{-1}}\mathcal{H}(u)^{-1}-\mathcal{E}_{\alpha}(u)\mathcal{H}(u)^{-1}\mathcal{F}_{\alpha}(u)\mathcal{H}(u)^{-1}\right).
\end{equation}
Therefore, it seems that we can give an explicit map from these generators to the ones of the YB algebra. However, the discussions here do not imply that we have found such map. First, we should check higher level relations. Second, we also need to find out what $\alpha$ and $\beta$ should be so that they would correspond to the two colours in $\mathtt{YB}$, or even whether we do have such combinations for constructing the map. More generally, there could be more systematic ways to study the connection between the Yangian algebras and $\mathcal{W}$ algebras (cf. \cite{ueda2022affine}), and we leave this to future work.

\section{Outlook}\label{outlook}
The shallow discussion in this note is just the tip of the iceberg, and there are still many open problems left. We may consider representations different from those in the paper and construct the corresponding $\mathcal{R}$-matrices. For instance, it could be useful to consider the free fermion representations \cite{Jimbo:1983if} in the discussions of $\mathcal{R}$-matrices. As such formalism is intimately related to crystal melting \cite{Okounkov:2003sp,Sulkowski:2009rw}, it could then be possible to give a full description of the contour integral forms in the $\mathcal{RTT}$ relation.

One may also use the Wakimoto representations \cite{Wakimoto:1986gf,Feigin:1990jc}. For the conifold case, this was analyzed very recently in \cite{Kolyaskin:2022tqi}. It was shown that one can correctly recover the corresponding quiver Yangian starting from the $\mathcal{N}=2$ superconformal $\mathcal{W}$ algebra. In general, a notable feature of $\mathcal{R}$-matrices constructed from Wakimoto realizations is that they would depend not only on $u-v$ but also on other more spectral parameters.

In the constructions of $\mathcal{R}$-matrices for various representations to reproduce the quiver Yangian relations, the screening operator is always a useful tool. For instance, a free field realization for the (truncations of) $\mathcal{W}$ algebra was constructed in \cite{Litvinov:2016mgi} as the kernel of some screening fields acting on the tensor product of current algebras. This was shown to be equivalent to the free field realization from Miura operators in \cite{Prochazka:2018tlo}. It would be interesting to investigate this in the context of matrix extended $\mathcal{W}$ algebras.

When starting from certain algebra/theory to construct the $\mathcal{R}$-matrix and reproduce the quiver Yangian relations, one often benefits from the underlying Kac-Moody (super)algebra. Therefore, it would be helpful to see if there is any similar approach for any CY$_3$/quivers that extends the cases of generalized conifolds. Moreover, the study of $\mathcal{R}$-matrices for quiver Yangians might lead to further applications to the Bethe/gauge correspondence \cite{Nekrasov:2009uh,Nekrasov:2009ui,Nekrasov:2009rc}\footnote{Note added in version 3: It was later found in \cite{Galakhov:2022uyu} that a consistent construction of $\mathcal{R}$-matrices is restricted to symmetric quivers (for unshifted quiver Yangians) and hence rules out those associated to CY$_3$ with compact divisors. Therefore, any further generalization would require a more delicate treatment.}.

Apart from the above perspectives, it would be crucial to have a more unified picture for quiver Yangians and $\mathcal{W}_{M|N\times\infty}$. In particular, it is believed that both the quiver Yangians and the $\mathcal{W}_{M|N\times\infty}$ algebras should play the role as the double of the corresponding CoHAs \cite{Li:2020rij,Rapcak:2019wzw}. It is possible that we can find a map between the two types of algebras using the method similar to the one in \cite{ueda2022affine} which connects Ueda's affine super Yangians and rectangular $\mathcal{W}$ algebras. With appropriate presentations of the algebras, it could also help us understand the structures of the algebras better.

Moreover, truncations of both the quiver Yangians and the $\mathcal{W}$ algebras have led to extensive study. For quiver Yangians, the truncations have a nice interpretation in terms of the crystals. On the other hand, truncations of the $\mathcal{W}_{1+\infty}$ algebra give rise to the VOAs at the corner \cite{Gaiotto:2017euk}, and they serve as building blocks for more general VOAs upon gluing $\mathcal{W}_{1+\infty}$ \cite{Prochazka:2017qum,Prochazka:2018tlo}. The truncations of the quiver Yangians should be quite relevant in the context of these $\mathcal{W}$-algebras. Generalizing this to the case of $\mathcal{W}_{M|N\times\infty}$ might give a larger class of VOAs associated to generalized conifolds. As the truncations of both $\mathtt{Y}$ and $\mathcal{W}$ can be realized by D4-branes on the divisors of CY$_3$, it could also be possible to identify these VOAs with the truncations of quiver Yangians similar to some VOAs for the $\mathbb{C}^3$ case \cite{Li:2020rij}. This could be then give new insights in the study of BPS/CFT correspondence \cite{Alday:2009aq,Nekrasov:2015wsu,Feigin:2018bkf}.

As both the quiver Yangians and MO Yangians are constructed from quivers, it is natural to expect some connections of the two Yangian algebras. However, the precise relation between them is still not known in general. A possible direction could be the notion of tripled quivers. Further explorations of these quantum algebras might give us a deeper understanding of various physical and mathematical problems.

\section*{Acknowledgement}
I am grateful to Alexey Litvinov, Tom\'a\v{s} Proch\'azka, Mamoru Ueda for enlightening explanations on various problems. I would also like to thank Zezhuang Hao, Yang-Hui He, Suvajit Majumder, Dmitrii Riabchenko and Ali Zahabi for enjoyable discussions. The research is supported by a CSC scholarship.

\appendix

\section{Quivers for Generalized Conifolds}\label{quivergencon}
Almost all the toric CY$_3$ without compact 4-cycles are generalized conifolds. Their quiver Yangians have salient features and have been systematically studied in \cite{Li:2020rij}. Given a generalized conifold defined by $xy=z^Mw^N$ ($M,N\in\mathbb{N}$), its toric diagram is
\begin{equation}
	\includegraphics[width=5cm]{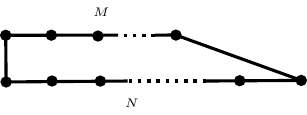},
\end{equation}
and its quiver Yangian is essentially the affine Yangian $\mathtt{Y}\left(\widehat{\mathfrak{gl}}_{M|N}\right)$.

\begin{figure}[h]
	\centering
	\includegraphics[width=6cm]{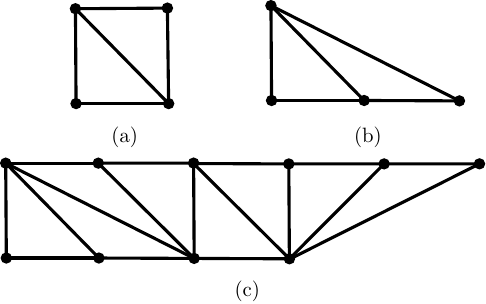}
	\caption{Figure taken from \cite{Bao:2022oyn}. In these examples, we have (a) $\sigma=\{+1,-1\}$, (b) $\sigma=\{+1,+1\}$ and (c) $\sigma=\{+1,+1,-1,-1,+1,-1,-1,-1\}$.}\label{sigmaex}
\end{figure}

Its quivers can be conveniently obtained from the triangulations of the lattice polygon \cite{nagao2008derived,Nagao:2009rq}, corresponding to different toric phases. The triangulations can in turn be concisely encoded by a sequence of signs $\sigma=\{\sigma_a|a\in\mathbb{Z}_{M+N}\}$, one for each node in the quiver/simplex in the toric diagram. There are $M$ plus ones and $N$ minus ones. When two simplices are glued side by side, they have the same sign. When they are glued in the alternative way, they have opposite signs. An illustration can be found in Figure \ref{sigmaex}.

The quiver is then constructed as follows. First, there is always a pair of opposite arrows connecting node $a$ and node $a+1$ ($a\in\mathbb{Z}_{M+N}$). Then the node $a$ is bosonic and has a self-loop if $\sigma_a=\sigma_{a+1}$. If $\sigma_a=-\sigma_{a+1}$, then it is fermionic and has no self-loops.

\begin{figure}[h]
	\centering
	\includegraphics[width=12cm]{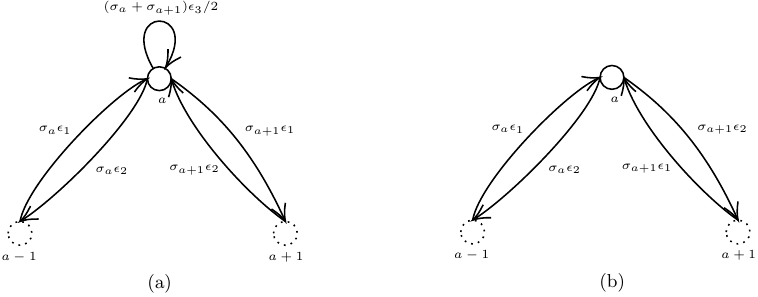}
	\caption{The charge assignment to bifundamentals and adjoints with the loop and vertex constraints for generalized conifolds. We have (a) $\sigma_a=\sigma_{a+1}$ and (b) $\sigma_a=-\sigma_{a+1}$.}\label{epsilongencon}
\end{figure}

The superpotential can be read off from $\sigma$, which leads to the loop constraints. Recall that we can also have vertex constraints, which would especially be useful when comparing $\mathtt{Y}$ with other algebras. Here, we shall just report the resulting coordinate parameter assignment to the arrows with the loop and vertex constraints. The general rule is given in Figure \ref{epsilongencon}.

\begin{figure}[h]
	\centering
	\includegraphics[width=10cm]{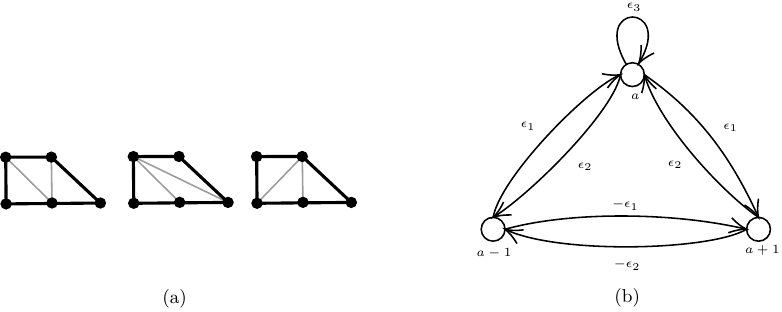}
	\caption{(a) The triangulations of the toric diagram for SPP. (b) The corresponding quiver of SPP. In the three triangulations, the signs are $\{+1,-1,+1\}$, $\{+1,+1,-1\}$, $\{-1,+1,+1\}$ respectively from left to right. They correspond to the same quiver but with the initial node $a+1$, $a$ and $a-1$ respectively.}\label{exSPP}
\end{figure}

Let us give an explicit example to make this more concrete. The possible triangulations of the toric diagram for the suspended pinch point (SPP) with the defining relation $xy=zw^2$ are given in Figure \ref{exSPP}(a). Using the inverse algorithm in \cite{Feng:2004uq,Gulotta:2008ef}, we can obtain the associated toric quiver. It turns out that in this case they all correspond to the same quiver as shown in \ref{exSPP}(b).
Suppose the only bosonic node is labelled by $a$ with the two fermionic nodes labelled by $a\pm1$ respectively. The charge assignment to the arrows is given in Figure \ref{exSPP}(b) where we take $\sigma_a=+1$. The relations for the quiver Yangian as in \S\ref{QY} in this example reads
\begin{align}
	&\left[\psi^{(a)}_n,\psi^{(b)}_m\right]=0,\\
	&\left[e^{(a)}_n,f^{(b)}_m\right\}=\delta_{ab}\psi^{(a)}_{m+n},\\
	&\left[\psi^{(a)}_{n+1},e^{(b)}_m\right]-\left[\psi^{(a)}_n,e^{(b)}_{m+1}\right]=\sigma^{ba}_1\psi^{(a)}_ne^{(b)}_m+\sigma^{ab}_1e^{(b)}_m\psi^{(a)}_n,\\
	&\left[\psi^{(a)}_{n+1},f^{(b)}_m\right]-\left[\psi^{(a)}_n,f^{(b)}_{m+1}\right]=-\sigma^{ab}_1\psi^{(a)}_nf^{(b)}_m-\sigma^{ba}_1f^{(b)}_m\psi^{(a)}_n,\\
	&\left[e^{(a)}_{n+1},e^{(b)}_m\right\}-\left[e^{(a)}_n,e^{(b)}_{m+1}\right\}=\sigma^{ba}_1e^{(a)}_ne^{(b)}_m+(-1)^{|(a)||(b)|}\sigma^{ab}_1e^{(b)}_me^{(a)}_n,\\
	&\left[f^{(a)}_{n+1},f^{(b)}_m\right\}-\left[f^{(a)}_n,f^{(b)}_{m+1}\right\}=-\sigma^{ab}_1f^{(a)}_nf^{(b)}_m-(-1)^{|(a)||(b)|}\sigma^{ba}_1f^{(b)}_mf^{(a)}_n,\\
	&\textup{Sym}_{n_1,n_2}\textup{Sym}_{m_1,m_2}\left[e^{(0)}_{n_1},\left[e^{(2)}_{m_1},\left[e^{(0)}_{n_2},\left[e^{(2)}_{m_2},e^{(1)}_{k}\right]\right\}\right]\right\}=\left(e^{(0)}_{n_1}\leftrightarrow e^{(2)}_{m_1},e^{(0)}_{n_2}\leftrightarrow e^{(2)}_{m_2}\right),\\
	&\textup{Sym}_{n_1,n_2}\textup{Sym}_{m_1,m_2}\left[f^{(0)}_{n_1},\left[f^{(2)}_{m_1},\left[f^{(0)}_{n_2},\left[f^{(2)}_{m_2},f^{(1)}_{k}\right]\right\}\right]\right\}=\left(f^{(0)}_{n_1}\leftrightarrow f^{(2)}_{m_1},f^{(0)}_{n_2}\leftrightarrow f^{(2)}_{m_2}\right),
\end{align}
where we have set the bosonic node to have label $a=1$ in the Serre relations\footnote{Notice that the expected Serre relations are slightly different from the generalized conifolds for $MN\neq2$. This stems from the corresponding quantum toroidal algebra in \cite{bezerra2021quantum}.}. Here, $\sigma^{ab}_1:=\sigma^{a\rightarrow b}_1$ is simply equal to the charge $\pm\epsilon_i$ associated to the single arrow $a\rightarrow b$ as in Figure \ref{exSPP}(b).

Before ending this section, let us give a brief remark on the quivers for generalized conifolds. For $\mathbb{C}\times\mathbb{C}^2/\mathbb{Z}_N$ whose quiver Yangian is $\mathtt{Y}\left(\widehat{\mathfrak{gl}}_N\right)$, its quiver is the tripled quiver of the (oriented) Dynkin diagram for the affine Lie algebra $A_{N-1}^{(1)}$. In particular, this could be closely related to the connection between quiver Yangians and MO Yangians. It is then tempting to wonder if there could be a similar notion for any generalized conifold with $\mathtt{Y}\left(\widehat{\mathfrak{gl}}_{M|N}\right)$. Suppose the quiver has $B$ bosonic nodes and $F$ fermionic ones (satisfying $B+F=M+N$). Here, we propose that we can think of the quiver as a ``tripled'' quiver of the corresponding Dynkin diagram of $\widehat{\mathfrak{sl}}_{M|N}$ (with $B$ bosonic nodes and $F$ fermionic ones at the same positions) in the following sense. Every arrow (i.e., line endowed with an orientation) in the Dynkin diagram is still doubled. Then the loops are only added to the bosonic nodes. Alternatively, we can think of the loops added to the fermionic nodes having weight 0. In other words, the self-loops on the fermionic nodes would cancel themselves. This is also consistent with the assignment in Figure \ref{epsilongencon}.

\section{Examples at Higher Levels}\label{exhigherlv}
Here, we shall consider some examples at higher levels using the contour integral expressions for the matrix elements of $\mathcal{T}$.

\subsection{Example 1: Conifold}\label{exhigherlv1}
Let us first consider the conifold whose 2d crystal description can be found in \cite[Figure 29 and 30]{Nishinaka:2013mba}. In particular, there is one atom of colour $a$ (or $b\neq a$) at the first level and only one atom of colour $b$ (or $a$) can be added at the second level. Following the $\mathcal{RTT}$ relation, we can write
\begin{equation}
	\langle\mathfrak{C}_{(a)},\varnothing_{(c)}|\mathcal{R}_{12}(u-v)\mathcal{T}_1(u)\mathcal{T}_2(v)|\varnothing_{(a)},\varnothing_{(c)}\rangle=\langle\mathfrak{C}_{(a)},\varnothing_{(c)}|\mathcal{T}_2(v)\mathcal{T}_1(u)\mathcal{R}_{12}(u-v)|\varnothing_{(a)},\varnothing_{(c)}\rangle,
\end{equation}
where $c$ is either $a$ or $b$ and $\mathfrak{C}_{(a)}$ here stands for the 2d crystal with two atoms whose initial atom is of colour $a$. Based on the first part of contour integral conjecture (\S\ref{crystalRTT}), the right hand side is then
\begin{equation}
	\mathcal{T}_{\varnothing_{(c)},\varnothing_{(c)}}(v)\mathcal{T}_{\mathfrak{C}_{(a)},\varnothing_{(a)}}=\frac{1}{2\pi i}\oint_{\infty+u}\text{d}zF(z)h^{(c)}(v)f^{(b)}(z)f^{(a)}(u)h^{(a)}(u),\label{TTconifoldex}
\end{equation}
where we have used the fact that $\mathcal{T}_{\square_{(a)},\varnothing_{(a)}}(u)=f^{(a)}(u)h^{(a)}(u)$. Suppose $a=c\neq b$. After applying the current relations, we have
\begin{equation}
	\begin{split}
		\langle\mathfrak{C}_{(a)},\varnothing_{(a)}|\mathcal{R}_{12}\mathcal{T}_1\mathcal{T}_2|\varnothing_{(a)},\varnothing_{(a)}\rangle=&\left(\frac{v-u-\epsilon_3}{v-u}-\epsilon_3^2\right)\frac{1}{2\pi i}\oint_{\infty+u}\text{d}zF(z)f^{(b)}(z)f^{(a)}(u)h^{(a)}(u)h^{(a)}(v)\\
		&+\frac{(u-v)\epsilon_3}{u-v-\epsilon_3}\frac{1}{2\pi i}\oint_{\infty+u}\text{d}zF(z)f^{(b)}(z)h^{(a)}(u)f^{(a)}(v)h^{(a)}(v).
	\end{split}\label{intconifoldex}
\end{equation}
The first term clearly leads to a bra vector $\langle\mathfrak{C}_{(a)},\varnothing_{(a)}|$. For the second term, suppose the contour integral gives
\begin{equation}
	\oint_{\infty+u}\text{d}zF(z)f^{(b)}(z)=P(u)f^{(b)}(u)+\sum_{j}Q_j(u)f^{(b)}_j,
\end{equation}
where $P(u)$ comes from $-\text{Res}_u(F(z)f^{(b)}(z))=P(u)f^{(b)}(u)+\dots$ with the ellipsis denoting terms only with modes of $f^{(b)}$ (if $F(z)$ has a higher order pole at $z=u$). The terms with $Q_j(u)$ then include both such terms and those from the residue at infinity. Thus, using the $hf$ relation and writing the modes as contour integrals of the current, the second term in \eqref{intconifoldex} becomes
\begin{equation}
	\frac{(u-v)\epsilon_3}{u-v-\epsilon_3}h^{(a)}(u)P(u)\left(f^{(b)}(v)+\frac{1}{P(u)}\sum_{j}Q_j(u)f^{(b)}_j\right)f^{(a)}(v)h^{(a)}(v).
\end{equation}
However, since this must become some matrix element(s) composed of allowed states/2d molten crystal configurations (with levels no greater than 2), we propose that $Q_j(u)$ must vanish or equal $P(u)$. Therefore,
\begin{equation}
	\frac{(u-v)\epsilon_3}{u-v-\epsilon_3}\frac{1}{2\pi i}\oint_{\infty+v}\text{d}zF(z)\frac{P(u)}{P(v)}h^{(a)}(u)f^{(b)}(z)f^{(a)}(v)h^{(a)}(v).
\end{equation}
Hence, we get
\begin{equation}
	\langle\mathfrak{C}_{(a)},\varnothing_{(a)}|\mathcal{R}_{12}=\langle\mathfrak{C}_{(a)},\varnothing_{(a)}|\left(\frac{v-u-\epsilon_3}{v-u}-\epsilon_3^2\right)+\langle\varnothing_{(a)},\mathfrak{C}_{(a)}|\frac{(u-v)\epsilon_3P(u)}{(u-v-\epsilon_3)P(v)}.
\end{equation}

Now let us consider the case $b=c\neq a$. Then \eqref{TTconifoldex} becomes
\begin{equation}
	\frac{1}{2\pi i}\oint_{\infty+u}\text{d}zF(z)\frac{v-z-\epsilon_3}{v-z}f^{(b)}(z)f^{(a)}(u)h^{(a)}(u)h^{(b)}(v).
\end{equation}
The residue of the contour integral would be
\begin{equation}
	P(u)\frac{v-u-\epsilon_3}{v-u}f^{(b)}(u)+\sum_{j}Q_j'(u)f^{(b)}_j.
\end{equation}
Therefore, $Q_j'(u)$ should be equal to either $P(u)(v-u-\epsilon_3)/(v-u)$ or 0. As a result,
\begin{equation}
	\langle\mathfrak{C}_{(a)},\varnothing_{(b)}|\mathcal{R}_{12}=\langle\mathfrak{C}_{(a)},\varnothing_{(b)}|\frac{v-u-\epsilon_3}{v-u}P(u).
\end{equation}

As the two-atom configuration is $e^{(b)}_0e^{(a)}_0|\varnothing_{(a)}\rangle$ (and $\langle\varnothing_{(a)}|f^{(a)}_0f^{(b)}_0$) for $a\neq b$, we can use the second part of the contour integral conjecture (\S\ref{generators}) to write\footnote{Notice that here the convention of $f$ is the one for $\mathtt{YB}$ instead of $\mathtt{Y}$.}
\begin{equation}
	\begin{split}
		\langle\varnothing_{(a)}|f^{(a)}_0f^{(b)}_0\mathcal{T}|\varnothing_{(a)}\rangle=&\frac{1}{2\pi i}\oint_{\infty+u}\text{d}z\frac{1}{\epsilon_3}\left(1-\frac{u-z-\epsilon_3}{u-z}\frac{(u-z-\epsilon_2)(u-z+\epsilon_2)}{(u-z-\epsilon_1)(u-z+\epsilon_1)}\right)f^{(b)}(z)f^{(a)}(u)h^{(a)}(u)\\
		=&-\frac{\epsilon_2^2}{\epsilon_1^2}f^{(b)}(u)f^{(a)}(u)h^{(a)}(u).
	\end{split}
\end{equation}
Therefore, $P(u)=\epsilon_2^2/\epsilon_1^2$ (and indeed $Q_j$, $Q_j'$ vanish). Hence,
\begin{equation}
	\langle\mathfrak{C}_{(a)},\varnothing_{(b)}|\mathcal{R}_{12}=
	\begin{cases}
		\langle\mathfrak{C}_{(a)},\varnothing_{(a)}|\left(\frac{u-v+\epsilon_3}{u-v}-\epsilon_3^2\right)+\langle\varnothing_{(a)},\mathfrak{C}_{(a)}|\frac{(u-v)\epsilon_3}{(u-v-\epsilon_3)},&a=b\\
		\langle\mathfrak{C}_{(a)},\varnothing_{(b)}|\frac{\epsilon_2^2}{\epsilon_1^2}\frac{u-v+\epsilon_3}{u-v},&a\neq b.
	\end{cases}
\end{equation}

\subsection{Example 2: $\mathbb{C}\times\mathbb{C}^2/\mathbb{Z}_3$}\label{exhigherlv2}
Now, let us discuss $\mathbb{C}\times\mathbb{C}^2/\mathbb{Z}_3$ with the specific state, say, $e^{(2)}_1e^{(3)}_0e^{(1)}_0|\varnothing_{(1)}\rangle$. At level 1, we simply have $\mathcal{T}_{\varnothing_{(1)},\square_{(1)}}=h^{(1)}(u)e^{(1)}(u)$. At level 2, we have
\begin{equation}
	\begin{split}
		\langle\varnothing_{(1)}|\mathcal{T}e^{(3)}_0e^{(1)}_0|\varnothing_{(1)}\rangle=&\frac{1}{2\pi i}\oint_{\infty+u}\text{d}z\frac{1}{\epsilon_3}\left(1-\frac{u-z-\epsilon_3}{u-z}\frac{u-z-\epsilon_1}{u-z+\epsilon_2}\right)h^{(1)}(u)e^{(1)}(u)e^{(3)}(z)\\
		=&-\frac{\epsilon_1}{\epsilon_2}h^{(1)}(u)e^{(1)}(u)e^{(3)}(u).
	\end{split}
\end{equation}
Then at level 3, recall that
\begin{equation}
	e^{(2)}_1=\frac{1}{2\epsilon_3}\left[\psi^{(2)}_1-\frac{1}{2}\left(\psi^{(2)}_0\right)^2,e^{(2)}_0\right]
\end{equation}
following the results in \S\ref{generators}. Therefore,
\begin{equation}
	e^{(2)}_1e^{(3)}_0e^{(1)}_0|\varnothing_{(1)}\rangle=\frac{1}{2\epsilon_3}\left(\psi^{(2)}_1e^{(2)}_0-\frac{1}{2}\psi^{(2)}_0\psi^{(2)}_0e^{(2)}_0-e^{(2)}_0\psi^{(2)}_1+\frac{1}{2}e^{(2)}_0\psi^{(2)}_0\psi^{(2)}_0\right)e^{(3)}_0e^{(1)}_0|\varnothing_{(1)}\rangle.
\end{equation}
By considering the action of the current $\psi^{(2)}(z)$ and taking the contour integral around $\infty$, we get
\begin{equation}
	\begin{split}
		&\psi^{(2)}_0e^{(3)}_0e^{(1)}_0|\varnothing_{(1)}\rangle=\left(4u-2\epsilon_3\right)e^{(3)}_0e^{(1)}_0|\varnothing_{(1)}\rangle,\quad\psi^{(2)}_0e^{(2)}_0e^{(3)}_0e^{(1)}_0|\varnothing_{(1)}\rangle=6ue^{(2)}_0e^{(3)}_0e^{(1)}_0|\varnothing_{(1)}\rangle,\\
		&\psi^{(2)}_1e^{(3)}_0e^{(1)}_0|\varnothing_{(1)}\rangle=\left(8u^2-8\epsilon_3u+\epsilon_3^2\right)e^{(3)}_0e^{(1)}_0|\varnothing_{(1)}\rangle,\quad\psi^{(2)}_1e^{(2)}_0e^{(3)}_0e^{(1)}_0|\varnothing_{(1)}\rangle=18u^2e^{(2)}_0e^{(3)}_0e^{(1)}_0|\varnothing_{(1)}\rangle.
	\end{split}
\end{equation}
Moreover,
\begin{equation}
	\begin{split}
		&\langle\varnothing_{(1)}|\mathcal{T}e^{(2)}_0e^{(3)}_0e^{(1)}_0|\varnothing_{(1)}\rangle\\
		=&\frac{1}{2\pi i}\oint_{\infty+u}\text{d}z\frac{-\epsilon_1}{\epsilon_2\epsilon_3}\left(1-\frac{u-z-\epsilon_3}{u-z}\frac{u-z-\epsilon_1}{u-z+\epsilon_2}\frac{u-z-\epsilon_2}{u-z+\epsilon_1}\right)h^{(1)}(u)e^{(1)}(u)e^{(3)}(u)e^{(2)}(z)\\
		=&-\frac{\epsilon_1}{\epsilon_2}h^{(1)}(u)e^{(1)}(u)e^{(3)}(u)e^{(2)}(u).
	\end{split}
\end{equation}
Hence,
\begin{equation}
	\langle\varnothing_{(1)}|\mathcal{T}(u)e^{(2)}_1e^{(3)}_0e^{(1)}_0|\varnothing_{(1)}\rangle=-\frac{\epsilon_1\epsilon_3}{2\epsilon_2}h^{(1)}(u)e^{(1)}(u)e^{(3)}(u)e^{(2)}(u).
\end{equation}
One can then obtain, for example, $\mathcal{R}_{12}(u-v)\left(e^{(2)}_1e^{(3)}_0e^{(1)}_0|\varnothing_{(1)}\rangle\right)\otimes|\varnothing_{(a)}\rangle$ using the $\mathcal{RTT}$ relation and the relations among the currents.

\addcontentsline{toc}{section}{References}
\bibliographystyle{utphys}
\bibliography{references}

\providecommand{\href}[2]{#2}\begingroup\raggedright\begin{thebibliography}{10}

\bibitem{Harvey:1996gc}
J.~A. Harvey and G.~W. Moore, ``{On the algebras of BPS states},''
  \href{http://dx.doi.org/10.1007/s002200050461}{{\em Commun. Math. Phys.}
  {\bfseries 197} (1998) 489--519},
  \href{http://arxiv.org/abs/hep-th/9609017}{{\ttfamily arXiv:hep-th/9609017}}.

\bibitem{Nakajima:1994nid}
H.~Nakajima, ``{Instantons on ALE spaces, quiver varieties, and Kac-Moody
  algebras},'' \href{http://dx.doi.org/10.1215/S0012-7094-94-07613-8}{{\em Duke
  Math. J.} {\bfseries 76} no.~2, (1994) 365--416}.

\bibitem{Douglas:1996sw}
M.~R. Douglas and G.~W. Moore, ``{D-branes, quivers, and ALE instantons},''
  \href{http://arxiv.org/abs/hep-th/9603167}{{\ttfamily arXiv:hep-th/9603167}}.

\bibitem{Hanany:2005ve}
A.~Hanany and K.~D. Kennaway, ``{Dimer models and toric diagrams},''
  \href{http://arxiv.org/abs/hep-th/0503149}{{\ttfamily arXiv:hep-th/0503149}}.

\bibitem{Franco:2005rj}
S.~Franco, A.~Hanany, K.~D. Kennaway, D.~Vegh, and B.~Wecht, ``{Brane dimers
  and quiver gauge theories},''
  \href{http://dx.doi.org/10.1088/1126-6708/2006/01/096}{{\em JHEP} {\bfseries
  01} (2006) 096}, \href{http://arxiv.org/abs/hep-th/0504110}{{\ttfamily
  arXiv:hep-th/0504110}}.

\bibitem{Franco:2005sm}
S.~Franco, A.~Hanany, D.~Martelli, J.~Sparks, D.~Vegh, and B.~Wecht, ``{Gauge
  theories from toric geometry and brane tilings},''
  \href{http://dx.doi.org/10.1088/1126-6708/2006/01/128}{{\em JHEP} {\bfseries
  01} (2006) 128}, \href{http://arxiv.org/abs/hep-th/0505211}{{\ttfamily
  arXiv:hep-th/0505211}}.

\bibitem{Feng:2005gw}
B.~Feng, Y.-H. He, K.~D. Kennaway, and C.~Vafa, ``{Dimer models from mirror
  symmetry and quivering amoebae},''
  \href{http://dx.doi.org/10.4310/ATMP.2008.v12.n3.a2}{{\em Adv. Theor. Math.
  Phys.} {\bfseries 12} no.~3, (2008) 489--545},
  \href{http://arxiv.org/abs/hep-th/0511287}{{\ttfamily arXiv:hep-th/0511287}}.

\bibitem{Okounkov:2003sp}
A.~Okounkov, N.~Reshetikhin, and C.~Vafa, ``{Quantum Calabi-Yau and classical
  crystals},'' \href{http://dx.doi.org/10.1007/0-8176-4467-9_16}{{\em Prog.
  Math.} {\bfseries 244} (2006) 597},
  \href{http://arxiv.org/abs/hep-th/0309208}{{\ttfamily arXiv:hep-th/0309208}}.

\bibitem{Iqbal:2003ds}
A.~Iqbal, N.~Nekrasov, A.~Okounkov, and C.~Vafa, ``{Quantum foam and
  topological strings},''
  \href{http://dx.doi.org/10.1088/1126-6708/2008/04/011}{{\em JHEP} {\bfseries
  04} (2008) 011}, \href{http://arxiv.org/abs/hep-th/0312022}{{\ttfamily
  arXiv:hep-th/0312022}}.

\bibitem{Ooguri:2009ijd}
H.~Ooguri and M.~Yamazaki, ``{Crystal Melting and Toric Calabi-Yau
  Manifolds},'' \href{http://dx.doi.org/10.1007/s00220-009-0836-y}{{\em Commun.
  Math. Phys.} {\bfseries 292} (2009) 179--199},
  \href{http://arxiv.org/abs/0811.2801}{{\ttfamily arXiv:0811.2801 [hep-th]}}.

\bibitem{Li:2020rij}
W.~Li and M.~Yamazaki, ``{Quiver Yangian from Crystal Melting},''
  \href{http://dx.doi.org/10.1007/JHEP11(2020)035}{{\em JHEP} {\bfseries 11}
  (2020) 035}, \href{http://arxiv.org/abs/2003.08909}{{\ttfamily
  arXiv:2003.08909 [hep-th]}}.

\bibitem{Galakhov:2020vyb}
D.~Galakhov and M.~Yamazaki, ``{Quiver Yangian and Supersymmetric Quantum
  Mechanics},'' \href{http://arxiv.org/abs/2008.07006}{{\ttfamily
  arXiv:2008.07006 [hep-th]}}.

\bibitem{Nagao:2010kx}
K.~Nagao and H.~Nakajima, ``{Counting invariant of perverse coherent sheaves
  and its wall-crossing},'' \href{http://arxiv.org/abs/0809.2992}{{\ttfamily
  arXiv:0809.2992 [math.AG]}}.

\bibitem{Jafferis:2008uf}
D.~L. Jafferis and G.~W. Moore, ``{Wall crossing in local Calabi Yau
  manifolds},'' \href{http://arxiv.org/abs/0810.4909}{{\ttfamily
  arXiv:0810.4909 [hep-th]}}.

\bibitem{Chuang:2009crq}
W.-y. Chuang and D.~L. Jafferis, ``{Wall Crossing of BPS States on the Conifold
  from Seiberg Duality and Pyramid Partitions},''
  \href{http://dx.doi.org/10.1007/s00220-009-0832-2}{{\em Commun. Math. Phys.}
  {\bfseries 292} (2009) 285--301},
  \href{http://arxiv.org/abs/0810.5072}{{\ttfamily arXiv:0810.5072 [hep-th]}}.

\bibitem{Aganagic:2010qr}
M.~Aganagic and K.~Schaeffer, ``{Wall Crossing, Quivers and Crystals},''
  \href{http://dx.doi.org/10.1007/JHEP10(2012)153}{{\em JHEP} {\bfseries 10}
  (2012) 153}, \href{http://arxiv.org/abs/1006.2113}{{\ttfamily arXiv:1006.2113
  [hep-th]}}.

\bibitem{Bao:2022oyn}
J.~Bao, Y.-H. He, and A.~Zahabi, ``{Crystal melting, BPS quivers and
  plethystics},'' \href{http://dx.doi.org/10.1007/JHEP06(2022)016}{{\em JHEP}
  {\bfseries 06} (2022) 016}, \href{http://arxiv.org/abs/2202.12850}{{\ttfamily
  arXiv:2202.12850 [hep-th]}}.

\bibitem{Galakhov:2021xum}
D.~Galakhov, W.~Li, and M.~Yamazaki, ``{Shifted quiver Yangians and
  representations from BPS crystals},''
  \href{http://dx.doi.org/10.1007/JHEP08(2021)146}{{\em JHEP} {\bfseries 08}
  (2021) 146}, \href{http://arxiv.org/abs/2106.01230}{{\ttfamily
  arXiv:2106.01230 [hep-th]}}.

\bibitem{Kontsevich:2010px}
M.~Kontsevich and Y.~Soibelman, ``{Cohomological Hall algebra, exponential
  Hodge structures and motivic Donaldson-Thomas invariants},''
  \href{http://dx.doi.org/10.4310/CNTP.2011.v5.n2.a1}{{\em Commun. Num. Theor.
  Phys.} {\bfseries 5} (2011) 231--352},
  \href{http://arxiv.org/abs/1006.2706}{{\ttfamily arXiv:1006.2706 [math.AG]}}.

\bibitem{Prochazka:2017qum}
T.~Proch\'azka and M.~Rap\v{c}\'ak, ``{Webs of W-algebras},''
  \href{http://dx.doi.org/10.1007/JHEP11(2018)109}{{\em JHEP} {\bfseries 11}
  (2018) 109}, \href{http://arxiv.org/abs/1711.06888}{{\ttfamily
  arXiv:1711.06888 [hep-th]}}.

\bibitem{Prochazka:2018tlo}
T.~Proch\'azka and M.~Rap\v{c}\'ak, ``{$ \mathcal{W} $ -algebra modules, free
  fields, and Gukov-Witten defects},''
  \href{http://dx.doi.org/10.1007/JHEP05(2019)159}{{\em JHEP} {\bfseries 05}
  (2019) 159}, \href{http://arxiv.org/abs/1808.08837}{{\ttfamily
  arXiv:1808.08837 [hep-th]}}.

\bibitem{Eberhardt:2019xmf}
L.~Eberhardt and T.~Proch\'azka, ``{The matrix-extended $W_{1+\infty}$
  algebra},'' \href{http://dx.doi.org/10.1007/JHEP12(2019)175}{{\em JHEP}
  {\bfseries 12} (2019) 175}, \href{http://arxiv.org/abs/1910.00041}{{\ttfamily
  arXiv:1910.00041 [hep-th]}}.

\bibitem{Rapcak:2019wzw}
M.~Rap\v{c}\'ak, ``{On extensions of $
  \mathfrak{gl}\widehat{\left(\left.m\right|n\right)} $ Kac-Moody algebras and
  Calabi-Yau singularities},''
  \href{http://dx.doi.org/10.1007/JHEP01(2020)042}{{\em JHEP} {\bfseries 01}
  (2020) 042}, \href{http://arxiv.org/abs/1910.00031}{{\ttfamily
  arXiv:1910.00031 [hep-th]}}.

\bibitem{Gaberdiel:2017dbk}
M.~R. Gaberdiel, R.~Gopakumar, W.~Li, and C.~Peng, ``{Higher Spins and Yangian
  Symmetries},'' \href{http://dx.doi.org/10.1007/JHEP04(2017)152}{{\em JHEP}
  {\bfseries 04} (2017) 152}, \href{http://arxiv.org/abs/1702.05100}{{\ttfamily
  arXiv:1702.05100 [hep-th]}}.

\bibitem{Prochazka:2019dvu}
T.~Proch\'azka, ``{Instanton R-matrix and $ \mathcal{W} $-symmetry},''
  \href{http://dx.doi.org/10.1007/JHEP12(2019)099}{{\em JHEP} {\bfseries 12}
  (2019) 099}, \href{http://arxiv.org/abs/1903.10372}{{\ttfamily
  arXiv:1903.10372 [hep-th]}}.

\bibitem{Gaberdiel:2014cha}
M.~R. Gaberdiel and R.~Gopakumar, ``{Higher Spins \& Strings},''
  \href{http://dx.doi.org/10.1007/JHEP11(2014)044}{{\em JHEP} {\bfseries 11}
  (2014) 044}, \href{http://arxiv.org/abs/1406.6103}{{\ttfamily arXiv:1406.6103
  [hep-th]}}.

\bibitem{Henneaux:2010xg}
M.~Henneaux and S.-J. Rey, ``{Nonlinear $W_{infinity}$ as Asymptotic Symmetry
  of Three-Dimensional Higher Spin Anti-de Sitter Gravity},''
  \href{http://dx.doi.org/10.1007/JHEP12(2010)007}{{\em JHEP} {\bfseries 12}
  (2010) 007}, \href{http://arxiv.org/abs/1008.4579}{{\ttfamily arXiv:1008.4579
  [hep-th]}}.

\bibitem{Campoleoni:2010zq}
A.~Campoleoni, S.~Fredenhagen, S.~Pfenninger, and S.~Theisen, ``{Asymptotic
  symmetries of three-dimensional gravity coupled to higher-spin fields},''
  \href{http://dx.doi.org/10.1007/JHEP11(2010)007}{{\em JHEP} {\bfseries 11}
  (2010) 007}, \href{http://arxiv.org/abs/1008.4744}{{\ttfamily arXiv:1008.4744
  [hep-th]}}.

\bibitem{Gaberdiel:2010pz}
M.~R. Gaberdiel and R.~Gopakumar, ``{An AdS$_{3}$ Dual for Minimal Model
  CFTs},'' \href{http://dx.doi.org/10.1103/PhysRevD.83.066007}{{\em Phys. Rev.
  D} {\bfseries 83} (2011) 066007},
  \href{http://arxiv.org/abs/1011.2986}{{\ttfamily arXiv:1011.2986 [hep-th]}}.

\bibitem{Gaberdiel:2017hcn}
M.~R. Gaberdiel, W.~Li, C.~Peng, and H.~Zhang, ``{The supersymmetric affine
  Yangian},'' \href{http://dx.doi.org/10.1007/JHEP05(2018)200}{{\em JHEP}
  {\bfseries 05} (2018) 200}, \href{http://arxiv.org/abs/1711.07449}{{\ttfamily
  arXiv:1711.07449 [hep-th]}}.

\bibitem{Gaberdiel:2018nbs}
M.~R. Gaberdiel, W.~Li, and C.~Peng, ``{Twin-plane-partitions and
  $\mathcal{N}=2$ affine Yangian},''
  \href{http://dx.doi.org/10.1007/JHEP11(2018)192}{{\em JHEP} {\bfseries 11}
  (2018) 192}, \href{http://arxiv.org/abs/1807.11304}{{\ttfamily
  arXiv:1807.11304 [hep-th]}}.

\bibitem{Li:2019nna}
W.~Li and P.~Longhi, ``{Gluing two affine Yangians of $\mathfrak{gl}_1$},''
  \href{http://dx.doi.org/10.1007/JHEP10(2019)131}{{\em JHEP} {\bfseries 10}
  (2019) 131}, \href{http://arxiv.org/abs/1905.03076}{{\ttfamily
  arXiv:1905.03076 [hep-th]}}.

\bibitem{Li:2019lgd}
W.~Li, ``{Gluing affine Yangians with bi-fundamentals},''
  \href{http://dx.doi.org/10.1007/JHEP06(2020)182}{{\em JHEP} {\bfseries 06}
  (2020) 182}, \href{http://arxiv.org/abs/1910.10129}{{\ttfamily
  arXiv:1910.10129 [hep-th]}}.

\bibitem{Rapcak:2021hdh}
M.~Rapcak, ``{Branes, Quivers and BPS Algebras},''
  \href{http://arxiv.org/abs/2112.13878}{{\ttfamily arXiv:2112.13878
  [hep-th]}}.

\bibitem{Yamazaki:2022cdg}
M.~Yamazaki, ``{Quiver Yangians and Crystal Melting: A Concise Summary},'' in
  {\em {International Congress on Mathematical Physics}}.
\newblock 3, 2022.
\newblock \href{http://arxiv.org/abs/2203.14314}{{\ttfamily arXiv:2203.14314
  [hep-th]}}.

\bibitem{Drinfeld1985HopfAA}
V.~Drinfeld, ``Hopf algebras and the quantum yang-baxter equation,'' {\em
  Proceedings of the USSR Academy of Sciences} {\bfseries 32} (1985) 254--258.

\bibitem{maulik2012quantum}
D.~Maulik and A.~Okounkov, ``Quantum groups and quantum cohomology,''
  \href{http://arxiv.org/abs/1211.1287}{{\ttfamily arXiv:1211.1287 [math.AG]}}.

\bibitem{Davison:2013nza}
B.~Davison, ``{The critical CoHA of a quiver with potential},''
  \href{http://dx.doi.org/10.1093/qmath/haw053}{{\em Quart. J. Math. Oxford
  Ser.} {\bfseries 68} no.~2, (2017) 635--703},
  \href{http://arxiv.org/abs/1311.7172}{{\ttfamily arXiv:1311.7172 [math.AG]}}.

\bibitem{Litvinov:2020zeq}
A.~Litvinov and I.~Vilkoviskiy, ``{Liouville reflection operator, affine
  Yangian and Bethe ansatz},''
  \href{http://dx.doi.org/10.1007/JHEP12(2020)100}{{\em JHEP} {\bfseries 12}
  (2020) 100}, \href{http://arxiv.org/abs/2007.00535}{{\ttfamily
  arXiv:2007.00535 [hep-th]}}.

\bibitem{Chistyakova:2021yyd}
E.~Chistyakova, A.~Litvinov, and P.~Orlov, ``{Affine Yangian of $ \mathfrak{gl}
  $(2) and integrable structures of superconformal field theory},''
  \href{http://dx.doi.org/10.1007/JHEP03(2022)102}{{\em JHEP} {\bfseries 03}
  (2022) 102}, \href{http://arxiv.org/abs/2110.05870}{{\ttfamily
  arXiv:2110.05870 [hep-th]}}.

\bibitem{Wang:2022rvj}
N.~Wang and K.~Wu, ``{Yang\textendash{}Baxter algebra and MacMahon
  representation},'' \href{http://dx.doi.org/10.1063/5.0064593}{{\em J. Math.
  Phys.} {\bfseries 63} no.~2, (2022) 021702}.

\bibitem{Nishinaka:2013mba}
T.~Nishinaka, S.~Yamaguchi, and Y.~Yoshida, ``{Two-dimensional crystal melting
  and D4-D2-D0 on toric Calabi-Yau singularities},''
  \href{http://dx.doi.org/10.1007/JHEP05(2014)139}{{\em JHEP} {\bfseries 05}
  (2014) 139}, \href{http://arxiv.org/abs/1304.6724}{{\ttfamily arXiv:1304.6724
  [hep-th]}}.

\bibitem{Prochazka:2015deb}
T.~Proch\'azka, ``{$ \mathcal{W} $ -symmetry, topological vertex and affine
  Yangian},'' \href{http://dx.doi.org/10.1007/JHEP10(2016)077}{{\em JHEP}
  {\bfseries 10} (2016) 077}, \href{http://arxiv.org/abs/1512.07178}{{\ttfamily
  arXiv:1512.07178 [hep-th]}}.

\bibitem{Galakhov:2022uyu}
D.~Galakhov, W.~Li, and M.~Yamazaki, ``{Gauge/Bethe correspondence from quiver
  BPS algebras},'' \href{http://arxiv.org/abs/2206.13340}{{\ttfamily
  arXiv:2206.13340 [hep-th]}}.

\bibitem{Nekrasov:2009uh}
N.~A. Nekrasov and S.~L. Shatashvili, ``{Supersymmetric vacua and Bethe
  ansatz},'' \href{http://dx.doi.org/10.1016/j.nuclphysbps.2009.07.047}{{\em
  Nucl. Phys. B Proc. Suppl.} {\bfseries 192-193} (2009) 91--112},
  \href{http://arxiv.org/abs/0901.4744}{{\ttfamily arXiv:0901.4744 [hep-th]}}.

\bibitem{Nekrasov:2009ui}
N.~A. Nekrasov and S.~L. Shatashvili, ``{Quantum integrability and
  supersymmetric vacua},'' \href{http://dx.doi.org/10.1143/PTPS.177.105}{{\em
  Prog. Theor. Phys. Suppl.} {\bfseries 177} (2009) 105--119},
  \href{http://arxiv.org/abs/0901.4748}{{\ttfamily arXiv:0901.4748 [hep-th]}}.

\bibitem{Nekrasov:2009rc}
N.~A. Nekrasov and S.~L. Shatashvili,
  \href{http://dx.doi.org/10.1142/9789814304634_0015}{``{Quantization of
  Integrable Systems and Four Dimensional Gauge Theories},''} in {\em {16th
  International Congress on Mathematical Physics}}, pp.~265--289.
\newblock 8, 2009.
\newblock \href{http://arxiv.org/abs/0908.4052}{{\ttfamily arXiv:0908.4052
  [hep-th]}}.

\bibitem{guay2005cherednik}
N.~Guay, ``{Cherednik algebras and Yangians},'' {\em International Mathematics
  Research Notices} {\bfseries 2005} no.~57, (2005) 3551--3593.

\bibitem{guay2007affine}
N.~Guay, ``{Affine Yangians and deformed double current algebras in type A},''
  {\em Advances in Mathematics} {\bfseries 211} no.~2, (2007) 436--484.

\bibitem{Braverman:2016pwk}
A.~Braverman, M.~Finkelberg, and H.~Nakajima, ``{Coulomb branches of $3d$
  $\mathcal{N}=4$ quiver gauge theories and slices in the affine
  Grassmannian},'' \href{http://dx.doi.org/10.4310/ATMP.2019.v23.n1.a3}{{\em
  Adv. Theor. Math. Phys.} {\bfseries 23} (2019) 75--166},
  \href{http://arxiv.org/abs/1604.03625}{{\ttfamily arXiv:1604.03625
  [math.RT]}}.

\bibitem{finkelberg2018comultiplication}
M.~Finkelberg, J.~Kamnitzer, K.~Pham, L.~Rybnikov, and A.~Weekes,
  ``{Comultiplication for shifted Yangians and quantum open Toda lattice},''
  {\em Advances in Mathematics} {\bfseries 327} (2018) 349--389,
  \href{http://arxiv.org/abs/1608.03331}{{\ttfamily arXiv:1608.03331
  [math.RT]}}.

\bibitem{guay2018coproduct}
N.~Guay, H.~Nakajima, and C.~Wendlandt, ``Coproduct for yangians of affine
  kac--moody algebras,'' {\em Advances in Mathematics} {\bfseries 338} (2018)
  865--911, \href{http://arxiv.org/abs/1701.05288}{{\ttfamily arXiv:1701.05288
  [math.QA]}}.

\bibitem{ueda2019affine}
M.~Ueda, ``{Affine Super Yangian},''
  \href{http://arxiv.org/abs/1911.06666}{{\ttfamily arXiv:1911.06666
  [math.RT]}}.

\bibitem{bezerra2021quantum}
L.~Bezerra and E.~Mukhin, ``{Quantum Toroidal Algebra Associated with
  $\mathfrak{gl}_{m|n}$},'' {\em Algebras and Representation Theory} {\bfseries
  24} no.~2, (2021) 541--564, \href{http://arxiv.org/abs/1904.07297}{{\ttfamily
  arXiv:1904.07297 [math.QA]}}.

\bibitem{bezerra2021braid}
L.~Bezerra and E.~Mukhin, ``{Braid actions on quantum toroidal
  superalgebras},'' {\em Journal of Algebra} {\bfseries 585} (2021) 338--369,
  \href{http://arxiv.org/abs/1912.08729}{{\ttfamily arXiv:1912.08729
  [math.QA]}}.

\bibitem{ueda2022affine}
M.~Ueda, ``{Affine super Yangians and rectangular W-superalgebras},'' {\em
  Journal of Mathematical Physics} {\bfseries 63} no.~5, (2022) 051701,
  \href{http://arxiv.org/abs/2002.03479}{{\ttfamily arXiv:2002.03479
  [math.RT]}}.

\bibitem{kac2004quantum}
V.~G. Kac and M.~Wakimoto, ``Quantum reduction and representation theory of
  superconformal algebras,'' {\em Advances in Mathematics} {\bfseries 185}
  no.~2, (2004) 400--458,
  \href{http://arxiv.org/abs/math-ph/0304011}{{\ttfamily
  arXiv:math-ph/0304011}}.

\bibitem{Kac2005CorrigendumT}
V.~G. Kac and M.~Wakimoto, ``{Corrigendum to ``Quantum reduction and
  representation theory of superconformal algebras'': [Adv. Math. 185 (2004)
  400-458]},'' {\em Advances in Mathematics} {\bfseries 193} (2005) 453--455.

\bibitem{arakawa2005representation}
T.~Arakawa, ``{Representation theory of superconformal algebras and the
  Kac-Roan-Wakimoto conjecture},'' {\em Duke Mathematical Journal} {\bfseries
  130} no.~3, (2005) 435--478,
  \href{http://arxiv.org/abs/math-ph/0405015}{{\ttfamily
  arXiv:math-ph/0405015}}.

\bibitem{Zhu:2015nha}
R.-D. Zhu and Y.~Matsuo, ``{Yangian associated with 2D \ensuremath{\mathscr{N}}
  = 1 SCFT},'' \href{http://dx.doi.org/10.1093/ptep/ptv116}{{\em PTEP}
  {\bfseries 2015} no.~9, (2015) 093A01},
  \href{http://arxiv.org/abs/1504.04150}{{\ttfamily arXiv:1504.04150
  [hep-th]}}.

\bibitem{Jimbo:1983if}
M.~Jimbo and T.~Miwa, ``{Solitons and Infinite Dimensional Lie Algebras},''
  \href{http://dx.doi.org/10.2977/prims/1195182017}{{\em Publ. Res. Inst. Math.
  Sci. Kyoto} {\bfseries 19} (1983) 943}.

\bibitem{Sulkowski:2009rw}
P.~Sulkowski, ``{Wall-crossing, free fermions and crystal melting},''
  \href{http://dx.doi.org/10.1007/s00220-010-1153-1}{{\em Commun. Math. Phys.}
  {\bfseries 301} (2011) 517--562},
  \href{http://arxiv.org/abs/0910.5485}{{\ttfamily arXiv:0910.5485 [hep-th]}}.

\bibitem{Wakimoto:1986gf}
M.~Wakimoto, ``{Fock representations of the affine lie algebra A1(1)},''
  \href{http://dx.doi.org/10.1007/BF01211068}{{\em Commun. Math. Phys.}
  {\bfseries 104} (1986) 605--609}.

\bibitem{Feigin:1990jc}
B.~L. Feigin and E.~V. Frenkel, ``{Affine Kac-Moody algebras and semiinfinite
  flag manifolds},'' \href{http://dx.doi.org/10.1007/BF02097051}{{\em Commun.
  Math. Phys.} {\bfseries 128} (1990) 161--189}.

\bibitem{Kolyaskin:2022tqi}
D.~Kolyaskin, A.~Litvinov, and A.~Zhukov, ``{R-matrix formulation of affine
  Yangian of $\widehat{\mathfrak{gl}}(1|1)$},''
  \href{http://arxiv.org/abs/2206.01636}{{\ttfamily arXiv:2206.01636
  [hep-th]}}.

\bibitem{Litvinov:2016mgi}
A.~Litvinov and L.~Spodyneiko, ``{On W algebras commuting with a set of
  screenings},'' \href{http://dx.doi.org/10.1007/JHEP11(2016)138}{{\em JHEP}
  {\bfseries 11} (2016) 138}, \href{http://arxiv.org/abs/1609.06271}{{\ttfamily
  arXiv:1609.06271 [hep-th]}}.

\bibitem{Gaiotto:2017euk}
D.~Gaiotto and M.~Rap\v{c}\'ak, ``{Vertex Algebras at the Corner},''
  \href{http://dx.doi.org/10.1007/JHEP01(2019)160}{{\em JHEP} {\bfseries 01}
  (2019) 160}, \href{http://arxiv.org/abs/1703.00982}{{\ttfamily
  arXiv:1703.00982 [hep-th]}}.

\bibitem{Alday:2009aq}
L.~F. Alday, D.~Gaiotto, and Y.~Tachikawa, ``{Liouville Correlation Functions
  from Four-dimensional Gauge Theories},''
  \href{http://dx.doi.org/10.1007/s11005-010-0369-5}{{\em Lett. Math. Phys.}
  {\bfseries 91} (2010) 167--197},
  \href{http://arxiv.org/abs/0906.3219}{{\ttfamily arXiv:0906.3219 [hep-th]}}.

\bibitem{Nekrasov:2015wsu}
N.~Nekrasov, ``{BPS/CFT correspondence: non-perturbative Dyson-Schwinger
  equations and qq-characters},''
  \href{http://dx.doi.org/10.1007/JHEP03(2016)181}{{\em JHEP} {\bfseries 03}
  (2016) 181}, \href{http://arxiv.org/abs/1512.05388}{{\ttfamily
  arXiv:1512.05388 [hep-th]}}.

\bibitem{Feigin:2018bkf}
B.~Feigin and S.~Gukov, ``{VOA[$M_4$]},''
  \href{http://dx.doi.org/10.1063/1.5100059}{{\em J. Math. Phys.} {\bfseries
  61} no.~1, (2020) 012302}, \href{http://arxiv.org/abs/1806.02470}{{\ttfamily
  arXiv:1806.02470 [hep-th]}}.

\bibitem{nagao2008derived}
K.~Nagao, ``Derived categories of small toric calabi-yau 3-folds and counting
  invariants,'' \href{http://arxiv.org/abs/0809.2994}{{\ttfamily
  arXiv:0809.2994 [math.AG]}}.

\bibitem{Nagao:2009rq}
K.~Nagao and M.~Yamazaki, ``{The Non-commutative Topological Vertex and Wall
  Crossing Phenomena},''
  \href{http://dx.doi.org/10.4310/ATMP.2010.v14.n4.a3}{{\em Adv. Theor. Math.
  Phys.} {\bfseries 14} no.~4, (2010) 1147--1181},
  \href{http://arxiv.org/abs/0910.5479}{{\ttfamily arXiv:0910.5479 [hep-th]}}.

\bibitem{Feng:2004uq}
B.~Feng, Y.-H. He, and F.~Lam, ``{On correspondences between toric
  singularities and (p,q) webs},''
  \href{http://dx.doi.org/10.1016/j.nuclphysb.2004.08.048}{{\em Nucl. Phys. B}
  {\bfseries 701} (2004) 334--356},
  \href{http://arxiv.org/abs/hep-th/0403133}{{\ttfamily arXiv:hep-th/0403133}}.

\bibitem{Gulotta:2008ef}
D.~R. Gulotta, ``{Properly ordered dimers, R-charges, and an efficient inverse
  algorithm},'' \href{http://dx.doi.org/10.1088/1126-6708/2008/10/014}{{\em
  JHEP} {\bfseries 10} (2008) 014},
  \href{http://arxiv.org/abs/0807.3012}{{\ttfamily arXiv:0807.3012 [hep-th]}}.

\end{thebibliography}\endgroup

\end{document}